\documentclass[11pt,a4paper]{article}

\usepackage[T1]{fontenc}
\usepackage[utf8]{inputenc}
\usepackage{lmodern}
\usepackage[margin=2.5cm]{geometry}
\usepackage{amsmath,amssymb,bm}
\usepackage{graphicx}
\graphicspath{{figs/}{./}}
\usepackage{booktabs}
\usepackage[numbers,sort&compress]{natbib}
\usepackage[colorlinks=true,linkcolor=blue!50!black,citecolor=green!35!black,urlcolor=blue!60!black]{hyperref}
\usepackage{xcolor}
\usepackage[labelfont=bf]{caption}
\usepackage{tikz}
\usetikzlibrary{arrows.meta}

\title{Magnetic control of hydrodynamic spin coupling between rotors:\\
Ferrohydrodynamic effects in the two-rotor problem}
\author{E.~E.~Tzirtzilakis
\\[2pt]{\small Department of Civil Engineering, University of the Peloponnese,}
\\[2pt]{\small Patras 26334, Greece (etzirtzilakis@go.uop.gr)}}
\date{}

\begin{document}
\maketitle

\begin{abstract}
The hydrodynamic spin coupling between two rotors (``fluid gears'') is
investigated numerically for a magnetic fluid in non-uniform magnetic fields.
An active cylinder drives a torque-free passive one inside a circular corral,
with one rotor heated. The model combines magnetohydrodynamics (MHD) and
ferrohydrodynamics (FHD), with the full Kelvin force and temperature-dependent
Langevin magnetisation. For the oil-based ferrofluid EFH1 the Lorentz force is
negligible, and the Kelvin force drives thermomagnetic convection from the
heated rotor. Steady torque-free states are computed with Taylor--Hood finite
elements, the gear ratio $\Gamma$ being an additional unknown. Without a field
the results agree within 1--8\% with published planar simulations up to
$\mathrm{Re}\approx50$. About 17\,000 steady states cover 34 arrangements of
wires, NdFeB rods and magnet cores, gaps $G=0.1$--$0.7$,
$\mathrm{Re}=10$--$40$ and heating of 5--20\,K. Because the coupling is weak,
fields of 1--2\,mT change $\Gamma$ by an amount comparable to $\Gamma$ itself.
At 2\,mT it is reversed, strengthened or weakened in nine cases out of ten.
The outcome is set by the source position: at the experimental gap and
$\mathrm{Re}=20$ a source below the corral or behind the active rotor
strengthens the gear-like counter-rotation up to 2.7-fold, and a source above
the corral reverses it into belt-like corotation. Several sources do not act
additively, and a magnet core inside the active rotor raises $|\Gamma|$ at
small gaps to seven times the largest hydrodynamic value. At fixed geometry
and Reynolds number, field strength and heating act approximately through a
single thermomagnetic source parameter.
\end{abstract}

\section{Introduction}\label{sec:intro}

Hydrodynamic interactions between immersed bodies underlie sedimentation,
swarming and collective locomotion, and they have been studied extensively for
bodies that mainly translate. When the dominant motion is rotation, the
interaction takes a different form. A spinning body sets up a swirling flow,
and a nearby body responds to the shear of that flow. The rotation of one body
can therefore drive the rotation of another without mechanical contact
\citep{Smith2026spin,Pan2026planar}. This spin coupling is found over a wide
range of scales. Colloids spun by rotating magnetic or optical fields form
chiral fluids with odd free-surface flows \citep{Soni2019odd}, odd crystals
kneaded by motile dislocations \citep{Bililign2022}, pairs that orbit each other
through hydrodynamic spin--orbit coupling \citep{Modin2023spinorbit}, and
motile clusters of microrollers \citep{Driscoll2017rolling}. Spinner monolayers
can cavitate through collective flows \citep{Shen2023}. In biology, spinning
starfish embryos self-organise into living chiral crystals \citep{Tan2022chiral}.
At the millimetre scale, magnetised disks rotating at a liquid--air interface
self-assemble into ordered patterns through a repulsive hydrodynamic interaction
\citep{Grzybowski2000}, and particles spinning at intermediate Reynolds numbers
self-propel, flock and form chiral active phases \citep{Chen2025spinning}.

Rotationally coupled flows are also common in engineering. Pairs of
vertical-axis wind turbines interact aerodynamically, and suitable co- or
counter-rotating arrangements improve the performance of both rotors
\citep{Brownstein2019}. Rotating cylinders in closed containers are used to stir
and mix viscous fluids, as in two- and four-roll mills and vortex-mixing flows
\citep{Jana1994,Lagnado1990,Price2003,Hills2002}. They are also used to enhance
convective heat transfer, including in electrically conducting fluids and
ferrofluids under applied magnetic fields
\citep{Chatterjee2014,Chatterjee2016,Selimefendigil2015,Kim2018}. Transmitting
torque without mechanical contact is of independent technological interest, as
shown by magnetic gears \citep{Atallah2001} and magnetorheological fluid devices
\citep{Carlson2000}. A purely hydrodynamic coupling between rotors (a
``fluid gear'') adds a further option, because the sense of the transmitted
rotation is selected by the flow and not by a fixed mechanism
\citep{Smith2026spin}. Magnetic fluids are a natural way to make this coupling
controllable. In ferrofluids the field acts on the magnetisation of the fluid
\citep{Neuringer1964,Rosensweig1985book,Rinaldi2005,TorresDiaz2014}. In liquid
metals it acts through the Lorentz force on induced currents
\citep{Davidson2017book,Muller2001book}. In biomagnetic fluids, such as blood,
both mechanisms can act \citep{Tzirtzilakis2005model}.

Flows driven by rotating cylinders are classical archetypes in fluid mechanics.
In all of the studies below, every rotation rate is prescribed, and the
interest lies in the flow that the rotors generate. The fluid between two
concentric cylinders, the Taylor--Couette system, has been a paradigm of
hydrodynamic stability and turbulence since the work of
\citet{Taylor1923}. It has been reviewed for soft-matter applications
\citep{Fardin2014}, extended to eccentric geometries with axial through-flow
\citep{Leclercq2014}, and pushed into the ultimate turbulent regime
\citep{Ezeta2018}.

When several rotors are placed inside a bounded domain, the emphasis shifts to
streamline topology and mixing. \citet{Jana1994} combined experiments and
computations to study mixing in the vortex mixing flow and in multicellular
cavity flows. \citet{Lagnado1990} visualised the three-dimensional flow in a
four-roll mill. \citet{Hills2002} computed the two-dimensional flow in a
rectangular box containing two independently rotating circular cylinders. He
mapped the streamline topology and its transitions over a seven-parameter space
of geometry, rotation rates and Reynolds number, and confirmed the main
predicted features with a preliminary experiment. \citet{Price2003} studied the
creeping flow of a two-roll mill inside a cylindrical domain for an arbitrary
speed ratio. They identified the full sequence of topological transitions
between co- and counter-rotation, and suggested that varying the speed ratio
continuously could give an efficient chaotic mixer.

In periodic arrays of rotors, known as cellular flows, attention moves to the
transport of suspended objects. Examples are the stretch--coil transition of
fibres \citep{Young2007} and the L\'evy walks of elongated structures crossing
the vortical cells \citep{Hu2021}. Inertia enters such flows through the
centrifugal action of spinning bodies. This is well documented for a single
rotating sphere at low and moderate Reynolds numbers \citep{Dennis1980} and at
high rotation rates \citep{Bowden1963}.

A second group of studies lets at least one body respond freely to the flow.
In the Stokes regime, \citet{Guo2024rotors} considered microcylinders rotating
under prescribed torques inside a circular confinement. A single rotating
cylinder near a plane wall generates no net force in a Newtonian fluid. A pair
of cylinders, however, can form four non-trivial hydrodynamic bound states.
These states result from the competition between the interactions within the
pair and the interactions of each cylinder with the confining wall, and a
recirculation region appears between widely separated rotors. \citet{Dormy2024}
analysed the Stokes flow driven by two equal rotating cylinders enclosed in an
outer cylinder of radius $R_0$, with emphasis on the small-gap (lubrication)
limit. For counter-rotation, their numerical results indicate that the net force
decays like $(\log R_0)^{-1}$, and they give an exact solution in the contact
limit. For co-rotation, Watson's theory agrees well with their finite-element
solutions. Spin-coupled rotors have also been studied as free particles.
\citet{Ouyang2017rotator} simulated a self-rotating rotator and one or two
passive particles in a two-dimensional confined cavity. A single passive particle
gradually departs from the rotator while revolving around it and spinning about
its own axis. \citet{Modin2023spinorbit} showed experimentally that asynchronously
spinning optically driven micro-rotors orbit each other at an angular velocity
set by their spins. They explained this hydrodynamic spin--orbit coupling with
an analytical model in the Stokes limit.

The problem that motivates the present work was introduced by
\citet{Smith2026spin}, with a preliminary account in the APS Gallery of Fluid
Motion under the title ``Fluidic gears and pulleys'' \citep{Smith2025gallery}
and a commentary in \citet{Wells2026physicstoday}. Their set-up has two parts.
An \emph{active} cylinder (radius $r$) is driven at a prescribed angular
velocity $\Omega$. A \emph{passive} cylinder of the same size, mounted on air
bearings, rotates freely at the angular velocity $\omega$ selected by the
hydrodynamic torque. Both rotors are placed inside a circular corral of radius
$R$ and immersed in a glycerol--water mixture, with aspect ratio $h/r=8$, so
that the coupling is essentially quasi-two-dimensional. The state is described
by the Reynolds number $\mathrm{Re}=\Omega r^2/\nu$, the normalised gap
$G=g/(2R-4r)\in[0,1]$, the corral size $C=R/r$ and the gear ratio
$\Gamma=\omega/\Omega$. The authors measured $\Gamma$ for 1350 conditions with
$C\in(2,10)$. Gear-like counter-rotation ($\Gamma<0$) occurs at moderate gaps and low
$\mathrm{Re}$, but co-rotation ($\Gamma>0$) is equally widespread. It is reached
by three distinct routes. The first is a \emph{geometric} route at very small
gaps, where the inner shear zone is pinched. The second is a \emph{topological} route at
wide gaps, where an intervening recirculation reverses the flow washing the passive
rotor. The third is an \emph{inertial} route at higher $\mathrm{Re}$, tentatively
attributed to centrifugal strengthening of the outer flow. For $G=0.3$ this transition
occurs at $\mathrm{Re}\approx55$, a value reproduced by an empirical shear
model.
All three routes were interpreted as a competition between inner (gear-like)
and outer (belt-like) shear on the passive rotor, and no hysteresis was
observed.

\citet{Pan2026planar} performed the first finite-$\mathrm{Re}$ direct numerical
simulations of this problem, for strictly two-dimensional flow with $C=3$, $4.5$
and $6$. Their simulations recover the main features of the experimental gap
route at $\mathrm{Re}=20$: an intermediate band of counter-rotation, a
transition to co-rotation at wide gaps, gear ratios of order $10^{-2}$, and the
observed sequence of vortex attachment, detachment and merger.
The main discrepancy is the high-$\mathrm{Re}$ boundary. At $G\approx0.3$ the
planar gear ratio approaches zero from the counter-rotating side but does not
change sign up to $\mathrm{Re}=400$. At $G=0.22$ the spin reverses near
$\mathrm{Re}\approx44$, through a redistribution of the integrated torque rather
than through the experimental shear-competition mechanism. The authors
attribute the remaining mismatch to finite-depth secondary motion, end-wall
stresses and the geometry of the apparatus.

Two features of the problem are important for what follows. First, the
coupling is weak: in both the experiments and the planar simulations
$|\Gamma|=O(10^{-2})$ \citep{Smith2026spin,Pan2026planar}. Second, the sense of
rotation of the passive rotor is decided by a delicate balance of opposing
shear contributions. A mechanism that changes the vorticity distribution
around the passive rotor can therefore change both the magnitude and the sign
of $\Gamma$.

Magnetic fields act on rotating flows in several distinct ways, and the
literature falls into the following groups. Several numerical studies add a
magnetic field to enclosures containing rotating cylinders, almost always with
a thermal objective. \citet{Chatterjee2014} studied two-dimensional MHD mixed
convection in a cooled square enclosure containing two heated cylinders that
are stationary, co-rotating or counter-rotating, under a horizontal field
($10^3\le\mathrm{Ra}\le10^5$, $0\le\mathrm{Ha}\le50$, rotational
$\mathrm{Re}=100$). \citet{Chatterjee2016} considered a lid-driven,
differentially heated enclosure in which one of two cylinders rotates
($0\le\mathrm{Ha}\le50$, $1\le\mathrm{Ri}\le10$, $\Pr=0.02$--$7$). They found
that heat transfer depends strongly on the rotational speed.
\citet{Selimefendigil2015} examined ferrofluid mixed convection in a lid-driven
cavity with two rotating cylinders and a magnetic dipole below the heated
wall. The angular velocity, velocity ratio and diameter ratio of the cylinders
had a pronounced effect on heat transfer. \citet{Kim2018} analysed the thermal-flow characteristics of a
ferrofluid in a rotating eccentric cylinder under an external magnetic force,
reporting local Nusselt numbers, magnetophoretic forces and velocity
distributions for various eccentricities. In all of these studies the
rotation rates are \emph{imposed}, and the outputs are Nusselt numbers and flow
patterns. The torque balance of a freely rotating cylinder is not considered.

On the other hand, the governing equations of ferrohydrodynamics (FHD) were formulated by \citet{Neuringer1964} and are treated comprehensively by \citet{Rosensweig1985book}. Thermomagnetic convection, driven by temperature-dependent magnetisation, was analysed by \citet{Finlayson1970} and used in a magnetic-fluid cooling device by \citet{Matsuki1977}. Thermomagnetic convection has also been used to produce rotation directly. \citet{Zhou2009engine} built a miniature engine in which a rotor immersed in a kerosene-based magnetic fluid is driven by the thermomagnetic convection of the fluid in the field of a permanent NdFeB
magnet, and its speed is controlled by the heat load and the field. \citet{Liu2024liquidwheel} spun a ferrofluid ``liquid wheel'' by light-driven surface thermomagnetic convection. A static field hinders the rotation of the magnetic particles relative to the fluid vorticity, which produces the rotational viscosity, or magnetoviscous effect \citep{McTague1969,Shliomis1972,Odenbach2002book}. In an alternating field the rotational viscosity can become negative \citep{Shliomis1994,Bacri1995,Rosensweig1996}. A rotating uniform field drives bulk ``spin-up'' flow. \citet{Chaves2008} measured this flow in cylindrical
containers by ultrasound velocimetry and torque measurements and compared it with an asymptotic spin-diffusion theory. \citet{Zahn1995} used a uniform,
sinusoidally time-varying field to pump ferrofluid. In the Taylor--Couette geometry, \citet{Altmeyer2010} showed numerically that a transverse homogeneous field modulates Taylor and spiral vortex flows into wavy Taylor and wavy spiral vortex flows, the magnetically generated wavy Taylor vortices being stationary
(pinned by the field). Reviews of ferrofluid rheology and applications are given by \citet{Rinaldi2005} and \citet{TorresDiaz2014}.

As far as magnetohydrodynamics is concerned, in electrically conducting fluids such as liquid
metals and electrolytes, a field transverse to a fluid layer damps velocity
gradients along the field lines, and the flow becomes quasi-two-dimensional
\citep{Sommeria1982}. The core flow then feels a linear friction from the
Hartmann layers at the confining walls. \citet{Potherat2000} refined this
picture into an effective two-dimensional model that includes inertial effects
in the Hartmann layers. This framework has been tested in electrically driven
vortex flows in liquid-metal and electrolyte layers
\citep{Sommeria1986,Figueroa2009}. For a single cylinder,
\citet{Mutschke1997} showed numerically that a field aligned with the free
stream suppresses two-dimensional vortex shedding, while three-dimensional
instabilities can appear first. \citet{Grigoriadis2010} developed an MHD
immersed-boundary method and derived asymptotic laws for the lift, drag and
recirculation length of a cylinder as functions of the field strength. Rotating
MHD flows between cylinders are also relevant in astrophysics: experimental
evidence for the magnetorotational instability has been obtained in a
Taylor--Couette flow under a helical field \citep{Stefani2006}. Textbook
accounts are given by \citet{Davidson2017book} and \citet{Muller2001book}, and
the properties of the eutectic GaInSn alloy commonly used in laboratory MHD are
reported by \citet{Plevachuk2014}.

Biomagnetic fluid dynamics (BFD) unifies the two mechanisms. It is a model in which the fluid is both electrically
conducting and magnetisable, so that the Lorentz force and the magnetisation (Kelvin) force act together \citep{Tzirtzilakis2005model}. This formulation has
been applied to ducts and channels under uniform and localised fields     \citep{Tzirtzilakis2004duct,Tzirtzilakis2005localized,Papadopoulos2004curved,Tzirtzilakis2006turbulent}, stenoses and aneurysms
  \citep{Tzirtzilakis2008stenosis,Raptis2014finite,Tzirtzilakis2015aneurysm}, a lid-driven cavity \citep{Tzirtzilakis2013cavity},
  boundary layers over stretching sheets and plates \citep{Tzirtzilakis2003sheet,Tzirtzilakis2010sheet3D,Tzirtzilakis2010plate,Murtaza2017},
  stretched, swirling and rotating cylinders \citep{Alam2022interactions,Alam2023swirling,Alam2024rotating}, and targeted magnetic drug delivery \citep{Gul2022simulation,Gul2022twophase}. More recently, the micropolar description with micromagnetorotation (MMR) has been used to resolve the difference between fluid vorticity and particle microrotation in conducting fluids carrying magnetic particles \citep{Aslani2020micromagnetorotation,Aslani2024vorticity,Aslani2026stenosis}.

The numerical approaches used for these problems depend on the flow regime and on whether the bodies move. In the Stokes regime, boundary-based and
analytical methods prevail. Indicatively, \citet{Guo2024rotors} combined a high-order boundary-integral method with analytical solutions in bipolar coordinates, \citet{Dormy2024} used finite-element solutions together with lubrication theory, \citet{Price2003} computed the two-roll-mill flow with a least-squares approach, \citet{Modin2023spinorbit} relied on an analytical Stokes model for pairs of spheres, and \citet{Driscoll2017rolling} used Stokesian dynamics for large collections of microrollers.

At finite Reynolds numbers, both fixed-grid and body-fitted formulations have
been used. \citet{Hills2002} used a finite-difference scheme on a Cartesian
grid with variable spacing. To make the streamfunction determinate in the
multiply connected domain, he imposed that the pressure be single-valued
around each cylinder. The same condition is needed in any
stream-function--vorticity formulation of the two-rotor problem.
\citet{Pan2026planar} employed a distributed Lagrange multiplier/fictitious
domain (DLM/FD) method. The corral is embedded in a square domain, and rigid
motion and the corral no-slip condition are enforced by two Lagrange
multipliers. The velocity is discretised with $P_1$-iso-$P_2$ and the pressure
with $P_1$ finite elements on fixed Cartesian meshes, and the equations are
advanced by a first-order Lie operator splitting. This choice avoids body-fitted
remeshing when the passive body is allowed to move. Related fixed-grid
approaches include the immersed boundary method
\citep{Peskin2002,Griffith2020}, which has been extended to MHD flow past
cylinders \citep{Grigoriadis2010}. Other examples are the volume-penalisation
Navier--Stokes simulations of spinning particles by \citet{Chen2025spinning}
and the lattice Boltzmann simulations of spinner monolayers by
\citet{Shen2023}.

The magnetic-enclosure studies use conventional discretisations: the finite
volume method \citep{Chatterjee2014}, the finite element method
\citep{Selimefendigil2015} and a commercial finite-element code
\citep{Kim2018}. For ferrofluid
Taylor--Couette flow, \citet{Altmeyer2010} combined finite differences with a
Galerkin method. For combined MHD--FHD flows, indicative tools used include stream-function--vorticity methodologies \citep{Tzirtzilakis2008numerical}, finite element \citep{Raptis2014finite} and finite volume \citep{Chrimatopoulos2024finite,Katsoudas2025finite} solvers, the latter benchmarked for MHD flow over a backward-facing step, and OpenFOAM solvers for MHD micropolar flows with and without micromagnetorotation \citep{Aslani2026openfoam}.
On the experimental side, the torque on the passive rotor is inferred from its
measured rotation \citep{Smith2026spin}. Flow-based estimates of wall shear
stress by high-resolution particle image velocimetry
\citep{Kahler2006,Andre2015} have been suggested as a route to direct torque
measurement.

To the best of the authors' knowledge, two lines of work have remained separate. The hydrodynamic spin coupling of a torque-free rotor has been studied only for
non-magnetic Newtonian fluids \citep{Smith2026spin,Pan2026planar}. Conversely, studies of rotating cylinders under magnetic fields prescribe every rotation rate and focus on heat transfer \citep{Chatterjee2014,Chatterjee2016,Selimefendigil2015,Kim2018}. The thermomagnetic engines of \citet{Zhou2009engine} and \citet{Liu2024liquidwheel} show that thermomagnetic convection can drive rotation, but they do not concern the coupling between an active and a torque-free passive rotor. It is therefore not known whether, and by which mechanism, a magnetic field can strengthen, weaken or reverse the hydrodynamic coupling between two rotors, nor how the answer depends on the position and arrangement of the field sources. The present contribution is confined to this question, the magnetic control of the hydrodynamic spin coupling between an active and a torque-free passive solid rotor in a corral.

In the present work, the two-rotor problem of \citet{Smith2026spin} is extended to magnetic fluids, using the combined MHD--FHD formulation of \citet{Tzirtzilakis2005model}. The structure of the torque balance is taken as the starting point. In the Stokes limit, a torque-free cylinder rotates at half the
mean ambient vorticity over its cross-section. The flow around an isolated rotor is irrotational, so the coupling in the purely hydrodynamic problem comes only from confinement, finite size and inertia. This explains why $|\Gamma|$ is small. It follows that the coupling can be changed appreciably by a body force that modifies the distribution of vorticity near the passive rotor. In a heated ferrofluid such a force is provided by the Kelvin force, which drives thermomagnetic convection. The full Kelvin force and a magnetisation that depends on both the field and the temperature are therefore included from the outset. The weakness of the hydrodynamic coupling is then an opportunity: a modest field-induced flow near the passive rotor can change $\Gamma$ by an amount comparable to $\Gamma$ itself. On this basis the present work makes the following contributions.
\begin{enumerate}
  \item \emph{Formulation.} The two-rotor problem with a torque-free passive
    rotor is formulated for a fluid that is both electrically conducting and
    magnetisable, with a Langevin magnetisation that depends on temperature
    and with the full Kelvin force. For an oil-based ferrofluid the Lorentz
    terms are negligible, and the field acts through the Kelvin force alone,
    which drives thermomagnetic convection from the heated rotor. A uniform in-plane field exerts no Kelvin force, and an
    axial field acts on a conducting fluid only through Hartmann friction
    on the end walls (\S\ref{sec:uniform}).
  \item \emph{Method.} Steady torque-free states are computed directly from the
    steady equations, with the gear ratio as an additional unknown fixed by
    the torque-free condition, isoparametric Taylor--Hood elements and
    consistent (residual-based) torques and heat fluxes. Without a field, this
    body-fitted formulation reproduces the planar results of
    \citet{Pan2026planar}, obtained with an independent fictitious-domain
    method, to within 1--8\% up to $\mathrm{Re}\approx50$.
  \item \emph{A systematic map of the magnetic control.} For the ferrofluid
    EFH1, heated by 5--20\,K at one rotor, a database of 1\,526
    continuation paths and about 17\,000 steady states is computed. It covers 34
    arrangements of field sources of practical strength, gaps
    $G=0.1$--0.7 and $\mathrm{Re}=10$--40. The sources are straight wires and
    coil sides, external NdFeB rods, singly, in pairs and in crosses of
    four, and a stationary NdFeB core inside either rotor. Fields of
    1--2\,mT change $\Gamma$ by an amount comparable to $\Gamma$ itself, and
    at 2\,mT the coupling is reversed, strengthened or weakened in nine
    cases out of ten.
  \item \emph{Mechanism and governing parameter.} The position of the source relative to the sense of rotation decides the outcome. A source below the corral or behind the active rotor strengthens the gear-like counter-rotation up to 2.7-fold at the experimental gap and $\mathrm{Re}=20$, and a source above the corral reverses it into belt-like corotation. The field thus controls the competition between inner and outer shear identified by \citet{Smith2026spin}. It opens a
    thermomagnetic route between the two modes that, unlike the geometric, topological and inertial routes, can be followed in both directions at fixed geometry and Reynolds number. The computed flow fields of all groups of sources fall into four classes: isolation of the passive rotor, enclosure of both rotors, a field-induced cell that reverses the gap flow, and a balance that decouples the rotors. The field therefore reproduces both the belt-like and the topological route to corotation. The effects of several sources do not add. At $\mathrm{Re}=10$, two crosses of four wires that differ only in two current senses impose one or the other sense of rotation at almost every gap. A core inside the active rotor raises $|\Gamma|$ to seven times the largest hydrodynamic value at small gaps, without extra driving torque at the experimental gap, but decouples the rotors at wide gaps. For fixed geometry and Reynolds number, the field strength and the
    heating act approximately through a single thermomagnetic source parameter.
\end{enumerate}

These results are steady solution branches of a planar model with the applied field and without gravity. Their stability is not examined, and the magnetostatic self-field of the ferrofluid and buoyancy are not included. The consequences for a laboratory realisation are estimated in \S\ref{sec:parameters} and \S\ref{sec:limitations}.

The paper is organised as follows. The mathematical formulation, the magnetisation law and the physical parameters of the ferrofluid and of the field sources are presented in Section~\ref{sec:formulation}. The finite-element method is described in Section~\ref{sec:numerics}. In Section~\ref{sec:results} the solver is validated against the zero-field results of \citet{Pan2026planar}, and the effect of the magnetic field on the gear ratio, on the torque needed to drive the active rotor and on the heat transfer is presented. The conclusions and the outlook are summarised in Section~\ref{sec:conclusions}.

\section{Mathematical formulation}\label{sec:formulation}

\subsection{Physical problem}\label{sec:problem}
\begin{figure}[t]
\centering
\begin{tikzpicture}[scale=0.55,>=Stealth]
  \draw[very thick,gray!70] (0,0) circle (4.5);
  \fill[blue!12,draw=blue!60!black,thick] (-1.75,0) circle (1);
  \fill[red!10,draw=red!60!black,thick] (1.75,0) circle (1);
  \draw[->,thick,blue!60!black] (-1.75,0) ++(150:1.35) arc (150:30:1.35);
  \draw[<->,thick,red!60!black] (1.75,0) ++(150:1.35) arc (150:30:1.35);
  \node[blue!50!black] at (-1.75,1.75) {$\bar\Omega$};
  \node[red!50!black] at (1.75,1.75) {$\bar\omega$};
  \node at (-1.75,0) {$A$};
  \node at (1.75,0) {$P$};
  \draw[<->] (-0.75,-1.35) -- (0.75,-1.35) node[midway,below]{$\bar g$};
  \draw[<->] (-1.75,-2.25) -- (1.75,-2.25) node[midway,below]{$\bar D$};
  \draw[->] (0,0) -- (0,4.5) node[midway,left]{$\bar R$};
  \draw[->,gray] (-4.9,0) -- (5.6,0) node[right]{$\bar x$};
  \draw[->,gray] (0,-4.9) -- (0,5.2) node[above]{$\bar y$};
  \foreach \p/\l/\a in {{(0,-5.5)}/B/right, {(-5.5,0)}/L/above, {(5.5,0)}/R/above, {(0,5.5)}/T/right}{
    \draw[thick,orange!80!black,fill=orange!20] \p circle (0.22);
    \node[\a=3pt,orange!60!black] at \p {\small \l};}
  \draw[densely dotted,orange!80!black] (0,0) circle (5.5);
  \node[orange!60!black] at (4.3,-4.6) {\small $|\bar{\bm x}|=\bar d_s$};
  \node[gray!60!black] at (-3.6,3.5) {corral};
\end{tikzpicture}
\caption{Geometry of the two-rotor problem (drawn to scale for $C=4.5$,
$G=0.3$). $A$: active rotor with prescribed $\bar\Omega$. $P$: passive,
torque-free rotor. Circles: positions of the external field sources (wires or
NdFeB rods parallel to the rotor axes) at distance $\bar d_s$ from the corral
centre: T (above), B (below), L (left of $A$) and R (right of $P$). Pairs
occupy T+B or L+R, and crosses all four positions. Wires:
$\bar d_s=5\bar r$. Rods: $\bar d_s=6\bar r$.}
\label{fig:geometry}
\end{figure}
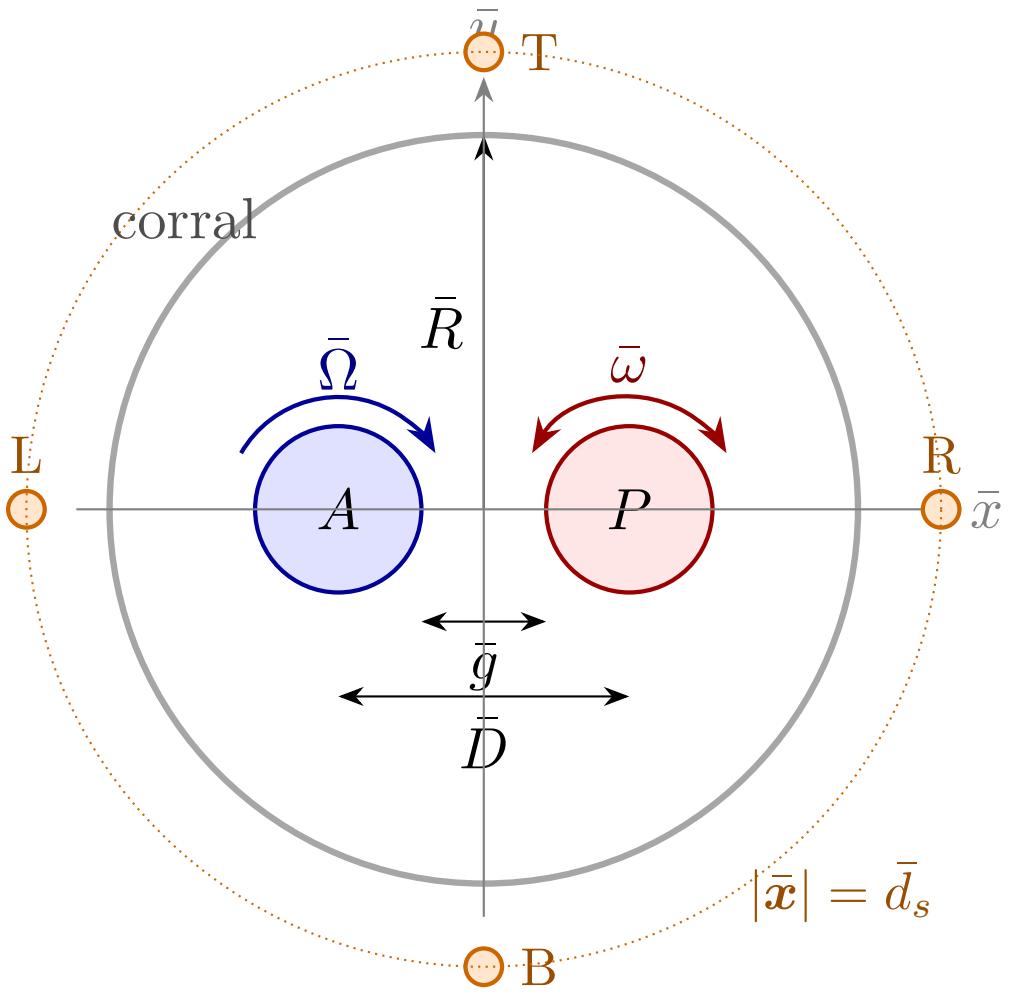

The two-rotor problem of \citet{Smith2026spin} is considered in the planar
setting of \citet{Pan2026planar}. Two identical circular cylinders of radius
$\bar r$ are placed inside a stationary circular corral of radius $\bar R$.
Their centres lie on a diameter of the corral, symmetrically about its centre,
at a surface-to-surface distance $\bar g$ (Fig.~\ref{fig:geometry}). The
\emph{active} rotor $A$ is driven at a prescribed angular velocity $\bar\Omega$.
The \emph{passive} rotor $P$ is free to rotate about its fixed axis, and its
angular velocity $\bar\omega$ is determined by the hydrodynamic torque. The
rotors and the corral are non-magnetic and electrically insulating.

The space between them is filled with a biomagnetic fluid in the sense of
\citet{Tzirtzilakis2005model}: a homogeneous, Newtonian, incompressible fluid
that is \emph{both} electrically conducting and magnetisable. Depending on the
material parameters, this description covers a ferrofluid (dominant
magnetisation), a liquid metal (dominant conductivity) or blood (both
mechanisms). Following \citet{Tzirtzilakis2013cavity}, the flow is assumed to
be laminar and two-dimensional, with three further assumptions. The
increase of viscosity due to the magnetic field is neglected. The magnetisation is in
equilibrium and parallel to the field ($\bar{\bm M}\parallel\bar{\bm H}$). For
in-plane magnetic fields, the electrical problem is axially homogeneous, with no axial
voltage gradient or end-wall effects (\S\ref{sec:2D}).
Two kinds of field source are considered, both parallel to the rotor axes and lying
outside the fluid. The first kind is the \emph{line current} (a straight
wire): a single wire, a pair of wires carrying equal currents in the same or in
opposite directions, or a cross of four wires, one at each of the positions T, B, L
and R defined below. The second kind is the \emph{line dipole}, a long permanent
magnet of circular cross-section magnetised perpendicular to its axis (a diametrically
magnetised NdFeB rod): a single rod, a pair of identical rods magnetised parallel or
antiparallel, or a cross of four rods, one at each of T, B, L and R.
The sources are placed at distance $\bar d_s$ from the corral centre, at four
positions (Fig.~\ref{fig:geometry}): above and below the corral on the
perpendicular bisector of the line of centres (positions T and B), and on the
line of centres, to the left of $A$ (L) and to the right of $P$ (R). A single
source occupies one of these positions, a pair occupies T and B or L and R,
and a cross of four sources occupies all four. In a further group of cases
the dipole is a stationary magnet core inside one of the rotors
(\S\ref{sec:parameters}).
The case of a uniform field is treated as a limit (\S\ref{sec:uniform}). The magnetic Reynolds number
$\bar\mu_0\bar\sigma\bar\Omega\bar r^{2}$ is assumed small, so that the applied
field is not modified by the flow. As in the BFD model, the field is also not
modified by the magnetisation of the fluid. This prescribed-field
approximation is justified for weakly magnetisable fluids. The ferrofluid
considered here is not weakly magnetisable, and the consequences for the
quantitative results are discussed in \S\ref{sec:parameters} and
\S\ref{sec:limitations}.
The rotor axes are taken to be vertical, i.e.\ parallel to gravity, as in
the apparatus of \citet{Smith2026spin}, where the rotors hang from the lid of
the tank and nearly span its depth. The role of gravity is discussed at the
end of \S\ref{sec:2D}.

\subsection{Governing equations}\label{sec:equations}
According to the BFD model of \citet{Tzirtzilakis2005model}, the flow of an
incompressible, Newtonian, electrically conducting and magnetisable fluid is
governed by the continuity, momentum and energy equations:
\begin{align}
&\bar\nabla\cdot\bar{\bm q}=0, \label{eq:cont}\\[2pt]
&\bar\rho\,\frac{D\bar{\bm q}}{D\bar t}=-\bar\nabla\bar p+\bar\mu\,\bar\nabla^{2}\bar{\bm q}
 +\bar\mu_0\bar M\,\bar\nabla\bar H+\bar{\bm J}\times\bar{\bm B}, \label{eq:mom}\\[2pt]
&\bar\rho\bar C_p\,\frac{D\bar T}{D\bar t}
 +\bar\mu_0\bar T\,\frac{\partial\bar M}{\partial\bar T}\,\frac{D\bar H}{D\bar t}
 -\frac{\bar{\bm J}\cdot\bar{\bm J}}{\bar\sigma}=\bar k\,\bar\nabla^{2}\bar T+\bar\mu\,\bar\Phi, \label{eq:energy}
\end{align}
with the quasi-static electrical relations
\begin{equation}
\bar{\bm J}=\bar\sigma\,(-\bar\nabla\bar\varphi_e+\bar{\bm q}\times\bar{\bm B}),
\qquad \bar\nabla\cdot\bar{\bm J}=0,
\label{eq:ohm}
\end{equation}
where bars denote dimensional quantities and $\bar\varphi_e$ is the electric potential. Under the prescribed-field
approximation, $\bar{\bm H}$ and $\bar{\bm B}$ denote the applied fields,
which satisfy in the fluid
\begin{equation}
\bar\nabla\times\bar{\bm H}=\bm0,\qquad
\bar\nabla\cdot\bar{\bm B}=0,\qquad \bar{\bm B}=\bar\mu_0\bar{\bm H}.
\label{eq:maxwell}
\end{equation}
Amp\`ere's law applies to the total field,
$\bar\nabla\times(\bar{\bm H}+\bar{\bm H}_{\mathrm{ind}})=\bar{\bm J}$.
The induced correction $\bar{\bm H}_{\mathrm{ind}}$ is neglected in the
force terms at small magnetic Reynolds number.
Here $\bar{\bm q}=(\bar u,\bar v)$ is the velocity, $\bar p$ the pressure,
$\bar T$ the temperature and $\bar\Phi$ the viscous dissipation function. The
density, dynamic viscosity, electrical conductivity, specific heat and thermal
conductivity of the fluid are denoted by $\bar\rho$, $\bar\mu$, $\bar\sigma$,
$\bar C_p$ and $\bar k$, and $\bar\mu_0$ is the magnetic permeability of
vacuum. The applied magnetic field strength intensity is
$\bar{\bm H}=(\bar H_x,\bar H_y)$ and $\bar H=|\bar{\bm H}|$ its magnitude.
Finally, $\bar{\bm B}$ is the magnetic flux density and $\bar M$ the magnitude
of the magnetisation.

The term $\bar\mu_0\bar M\bar\nabla\bar H$ is the magnetisation (Kelvin) force of FHD, and $\bar{\bm J}\times\bar{\bm B}$ is the Lorentz force of MHD. In the general derivation of \citet{Tzirtzilakis2005model}, before neglecting the induced magnetic field, the magnetic force $\bar\mu_0(\bar{\bm M}\cdot\bar\nabla)\bar{\bm H}$ for
$\bar{\bm M}\parallel\bar{\bm H}$ is shown to equal $\bar\mu_0\bar M\bar\nabla\bar H+(\bar M/\bar H)(\bar{\bm J}\times\bar{\bm B})$. Equation~\eqref{eq:mom} follows by adding the Lorentz force and neglecting the second term for $\bar M/\bar H\ll1$. For the fluids considered here this approximation is not restrictive. In a ferrofluid $\bar M/\bar H$ may be of order one, but the conductivity, and with it $\bar{\bm J}$, is negligible. In a liquid metal $\bar M\approx0$. The term $\bar\mu_0\bar T(\partial\bar M/\partial\bar T)(D\bar H/D\bar t)$ in \eqref{eq:energy} is the magnetocaloric power and $\bar J^{2}/\bar\sigma$ is the Joule heating \citep{Tzirtzilakis2005model}.

In the driven-cavity problem of \citet{Tzirtzilakis2013cavity} the magnetisation varies linearly with the field, $\bar M=\bar\chi\bar H$. Here the
Langevin function is adopted. Among the magnetisation laws listed by \citet{Tzirtzilakis2005model}, it is the most accurate. It describes both
ferrofluids and the magnetisation of red blood cells, and it accounts for
saturation \citep[Eq.~(9)]{Tzirtzilakis2005model}:
\begin{equation}
\bar M=\bar M_s\,\mathcal L(\xi),\qquad
\mathcal L(\xi)=\coth\xi-\frac1\xi,\qquad
\xi=\frac{\bar\mu_0\bar m\bar H}{\bar\kappa\bar T},\qquad \bar M_s=\bar m\bar N.
\label{eq:langevin}
\end{equation}
Here $\bar m$ is the magnetic moment of a particle, $\bar N$ the number of
particles per unit volume, $\bar\kappa$ the Boltzmann constant and $\bar M_s$
the saturation magnetisation. Two limits are useful. For weak fields, $\xi\ll1$,
Eq.~\eqref{eq:langevin} gives $\mathcal L\simeq\xi/3$, which recovers the linear law
of \citet{Tzirtzilakis2013cavity} with the Curie susceptibility
$\bar\chi=\bar\mu_0\bar m\bar M_s/(3\bar\kappa\bar T)$. For strong fields,
$\xi\gg1$, the magnetisation saturates, $\bar M\to\bar M_s$.
Because $\xi$ contains $\bar T$, the Langevin law also supplies the temperature
dependence needed in the magnetocaloric term. Differentiating at constant
$\bar H$ gives the closed form
\begin{equation}
\bar T\,\frac{\partial\bar M}{\partial\bar T}=-\bar M_s\,\xi\,\mathcal L'(\xi),\qquad
\mathcal L'(\xi)=\frac{1}{\xi^{2}}-\frac{1}{\sinh^{2}\xi}.
\label{eq:dMdT}
\end{equation}
In Eq.~\eqref{eq:dMdT} $\bar M_s$ is taken independent of the temperature, so that the whole temperature dependence of the magnetisation is supplied by the Langevin argument. The thermal dilution of the particles and the temperature dependence of their domain magnetisation, which are neglected, are estimated in \S\ref{sec:parameters}.

As in \citet[Eq.~(6)]{Tzirtzilakis2013cavity}, the field generated by the line current at $(\bar a,\bar b)$ has the components
\begin{equation}
\bar H_x=\frac{\gamma}{2\pi}\,\frac{\bar y-\bar b}{(\bar x-\bar a)^{2}+(\bar y-\bar b)^{2}},\qquad
\bar H_y=-\frac{\gamma}{2\pi}\,\frac{\bar x-\bar a}{(\bar x-\bar a)^{2}+(\bar y-\bar b)^{2}},
\label{eq:Hcomp}
\end{equation}
and the magnitude
\begin{equation}
\bar H(\bar x,\bar y)=\frac{\gamma}{2\pi}\,\frac{1}{\sqrt{(\bar x-\bar a)^{2}+(\bar y-\bar b)^{2}}}.
\label{eq:Hmag}
\end{equation}
The field lies in the plane of the flow, is curl-free and divergence-free in
the fluid, and is azimuthal about the wire.

For a long cylindrical magnet of radius $\bar R_m$ and remanence $\bar B_r$,
uniformly magnetised along the unit vector $\hat{\bm m}=(m_x,m_y)$
perpendicular to its axis, the exterior field is that of a two-dimensional
line dipole \citep{Furlani2007targeting,Bjork2010halbach}. With
$X=\bar x-\bar a$, $Y=\bar y-\bar b$ and $\bar\rho^{2}=X^{2}+Y^{2}$, it reads
\begin{equation}
\begin{gathered}
\bar{\bm H}=\frac{\bar K}{\bar\rho^{2}}\big[2(\hat{\bm m}\cdot\hat{\bm\rho})\hat{\bm\rho}-\hat{\bm m}\big],\\[2pt]
\bar H_x=\bar K\,\frac{m_x(X^{2}-Y^{2})+2m_yXY}{\bar\rho^{4}},\qquad
\bar H_y=\bar K\,\frac{2m_xXY+m_y(Y^{2}-X^{2})}{\bar\rho^{4}},
\end{gathered}
\label{eq:Hdip}
\end{equation}
with $\bar K=\bar B_r\bar R_m^{2}/(2\bar\mu_0)$. Its magnitude,
\begin{equation}
\bar H(\bar x,\bar y)=\frac{\bar K}{\bar\rho^{2}}=\frac{\bar B_r}{2\bar\mu_0}\,\frac{\bar R_m^{2}}{(\bar x-\bar a)^{2}+(\bar y-\bar b)^{2}},
\label{eq:Hdipmag}
\end{equation}
depends only on the distance from the magnet axis. On the magnet surface it
equals $\bar B_r/(2\bar\mu_0)$ at every angle. For both sources, therefore,
$\bar H$ is a function of the distance from the source alone, and
$\bar\nabla\bar H$ points towards the source. The direction of
$\hat{\bm m}$ enters only through the Lorentz terms.

\subsection{Two-dimensional equations}\label{sec:2D}
For the in-plane-field cases an axially homogeneous model is used:
$\bar{\bm q}=(\bar u,\bar v,0)$ and
$\bar{\bm B}=\bar\mu_0(\bar H_x,\bar H_y,0)$ are independent of $\bar z$.
The electric potential is taken as $\bar\varphi_e=\bar\varphi_e(\bar x,\bar y)$, with
no imposed axial electric field, and electrical end effects are excluded. Since
$\bar{\bm q}\times\bar{\bm B}$ has only an axial component, charge conservation
in Eq.~\eqref{eq:ohm}, with constant $\bar\sigma$, reduces to Laplace's
equation in the fluid, with homogeneous Neumann conditions on the insulating
rotor and corral boundaries:
\begin{equation}
\bar\nabla_{xy}^{2}\bar\varphi_e=0,\qquad
\frac{\partial\bar\varphi_e}{\partial\bar n}=0
\quad\text{on the boundaries}.
\label{eq:potential2D}
\end{equation}
The connected fluid domain therefore has a constant electric potential.
The current density is normal to the plane:
\begin{equation}
\bar{\bm J}=\bar J_z\,\hat{\bm z},\qquad
\bar J_z=\bar\sigma\,(\bar u\bar B_y-\bar v\bar B_x).
\label{eq:Jz}
\end{equation}
Thus the electric potential is eliminated analytically, and no additional
electrical equation is solved. Charge conservation is satisfied identically,
$\bar\nabla\cdot\bar{\bm J}=\partial\bar J_z/\partial\bar z=0$.
Equations~\eqref{eq:cont}--\eqref{eq:energy} then take the form
\citep{Tzirtzilakis2013cavity}
\begin{align}
&\frac{\partial\bar u}{\partial\bar x}+\frac{\partial\bar v}{\partial\bar y}=0, \label{eq:cont2}\\[2pt]
&\bar\rho\Big(\frac{\partial\bar u}{\partial\bar t}+\bar u\frac{\partial\bar u}{\partial\bar x}+\bar v\frac{\partial\bar u}{\partial\bar y}\Big)
=-\frac{\partial\bar p}{\partial\bar x}+\bar\mu_0\bar M\frac{\partial\bar H}{\partial\bar x}
-\bar\sigma\bar B_y^{2}\bar u+\bar\sigma\bar B_x\bar B_y\bar v
+\bar\mu\Big(\frac{\partial^{2}\bar u}{\partial\bar x^{2}}+\frac{\partial^{2}\bar u}{\partial\bar y^{2}}\Big), \label{eq:xmom}\\[2pt]
&\bar\rho\Big(\frac{\partial\bar v}{\partial\bar t}+\bar u\frac{\partial\bar v}{\partial\bar x}+\bar v\frac{\partial\bar v}{\partial\bar y}\Big)
=-\frac{\partial\bar p}{\partial\bar y}+\bar\mu_0\bar M\frac{\partial\bar H}{\partial\bar y}
-\bar\sigma\bar B_x^{2}\bar v+\bar\sigma\bar B_x\bar B_y\bar u
+\bar\mu\Big(\frac{\partial^{2}\bar v}{\partial\bar x^{2}}+\frac{\partial^{2}\bar v}{\partial\bar y^{2}}\Big), \label{eq:ymom}\\[2pt]
&\bar\rho\bar C_p\Big(\frac{\partial\bar T}{\partial\bar t}+\bar u\frac{\partial\bar T}{\partial\bar x}+\bar v\frac{\partial\bar T}{\partial\bar y}\Big)
-\bar\mu_0\bar M_s\,\xi\mathcal L'(\xi)\Big(\bar u\frac{\partial\bar H}{\partial\bar x}+\bar v\frac{\partial\bar H}{\partial\bar y}\Big)
-\frac{\bar J_z^{2}}{\bar\sigma}
=\bar k\Big(\frac{\partial^{2}\bar T}{\partial\bar x^{2}}+\frac{\partial^{2}\bar T}{\partial\bar y^{2}}\Big)+\bar\mu\bar\Phi, \label{eq:energy2}
\end{align}
with
$\bar\Phi=2\big[(\partial_{\bar x}\bar u)^{2}+(\partial_{\bar y}\bar v)^{2}\big]+(\partial_{\bar y}\bar u+\partial_{\bar x}\bar v)^{2}$.
Since the applied field is steady, $D\bar H/D\bar t=\bar{\bm q}\cdot\bar\nabla\bar H$,
and Eq.~\eqref{eq:dMdT} has been used in the energy equation. The momentum
equations~\eqref{eq:xmom}--\eqref{eq:ymom} have the same form as Eqs.~(2)--(3)
of \citet{Tzirtzilakis2013cavity}, with the time derivative retained and
$\bar M$ given by Eq.~\eqref{eq:langevin} in place of $\bar\chi\bar H$.

Gravity has been omitted from Eqs.~\eqref{eq:mom} and \eqref{eq:xmom}--\eqref{eq:ymom}. With vertical rotor
axes, the Boussinesq buoyancy force $-\bar\rho\bar\beta_T(\bar T-\bar T_0)\bar{\bm g}$, where $\bar\beta_T$ is the thermal expansion coefficient, is parallel to the axes. In an axially homogeneous flow it enters only the equation for an axial velocity $\bar w(\bar x,\bar y)$, which is driven by the temperature field and by a
uniform axial pressure gradient fixed by zero net axial flux. Since all fields are independent of $\bar z$, $\bar w$ does not appear in the in-plane
momentum equations or in the energy equation, and it exerts no torque about the rotor axes. Within the axially homogeneous model, therefore, buoyancy
produces an axial flow but does not affect $(\bar u,\bar v,\bar T)$ or $\Gamma$, and the gravity-free planar problem is exact. The axial flow is not suppressed, however. In a container of finite height it must turn round near the ends, where it has in-plane components, and it may itself become unstable. With horizontal axes buoyancy would act in the plane of the flow. Estimates of both effects for the laboratory parameters are given in \S\ref{sec:parameters}. They restrict the identification of the planar problem with a laboratory set-up.

\subsection{Boundary conditions and torque balance}\label{sec:bc}
The origin is placed at the centre of the corral, and the rotor centres are
$\bar{\bm x}_A=(-\bar D/2,0)$ and $\bar{\bm x}_P=(\bar D/2,0)$, with
$\bar D=2\bar r+\bar g$. The velocity boundary conditions are no slip on all
solid surfaces:
\begin{equation}
\begin{aligned}
&\text{corral } (|\bar{\bm x}|=\bar R): && \bar u=\bar v=0,\\
&\text{active rotor } (|\bar{\bm x}-\bar{\bm x}_A|=\bar r): && \bar u=-\bar\Omega\,\bar y,\quad \bar v=\bar\Omega\,(\bar x-\bar x_A),\\
&\text{passive rotor } (|\bar{\bm x}-\bar{\bm x}_P|=\bar r): && \bar u=-\bar\omega\,\bar y,\quad \bar v=\bar\omega\,(\bar x-\bar x_P).
\end{aligned}
\label{eq:bc}
\end{equation}

For the temperature, one rotor (the \emph{heated}
rotor, either $A$ or $P$, \S\ref{sec:parameters}) is kept at $\bar T_A$ and the
corral at $\bar T_0<\bar T_A$. The other rotor is taken as adiabatic.

The angular velocity of the passive rotor is not prescribed. It follows from
the angular momentum balance
\begin{equation}
\bar I_P\,\frac{d\bar\omega}{d\bar t}=\bar h\,\bar{\mathcal T}_P,\qquad
\bar{\mathcal T}_P=\bar r^{2}\!\int_{0}^{2\pi}\bar\tau_{\rho\theta}\big|_{\rho=\bar r}\,d\theta,\qquad
\bar\tau_{\rho\theta}=\bar\mu\Big(\frac{\partial\bar u_\theta}{\partial\bar\rho}-\frac{\bar u_\theta}{\bar\rho}\Big),
\label{eq:torque}
\end{equation}
where $(\bar\rho,\theta)$ are polar coordinates centred at $\bar{\bm x}_P$,
$\bar u_\theta$ is the azimuthal velocity, $\bar{\mathcal T}_P$ the torque per
unit length, $\bar h$ the rotor length and $\bar I_P$ the moment of inertia.
The torque is purely viscous, for three reasons. $(i)$ On a circular rotor the pressure traction is radial and exerts no torque about the centre. $(ii)$ The rotor is non-magnetic and insulating, so no body force acts inside it. $(iii)$ The magnetisation drops to zero across the rotor surface, and the singular part of the Kelvin force there is normal to the surface, because the tangential derivative of $\bar H$ is continuous.

A steady terminal state is necessarily torque-free, $\bar{\mathcal T}_P=0$, and the gear ratio is $\Gamma=\bar\omega/\bar\Omega$ \citep{Smith2026spin}.

\subsection{Transformation of equations}\label{sec:nondim}
Following \citet{Tzirtzilakis2013cavity}, but with the rotor radius as the
length scale and the rotor surface speed as the velocity scale (so that the
Reynolds number coincides with that of \citet{Smith2026spin}), dimensionless
variables are introduced as:
\begin{equation}
x=\frac{\bar x}{\bar r},\quad y=\frac{\bar y}{\bar r},\quad
u=\frac{\bar u}{\bar\Omega\bar r},\quad v=\frac{\bar v}{\bar\Omega\bar r},\quad
t=\bar\Omega\bar t,\quad
p=\frac{\bar p}{\bar\rho\bar\Omega^{2}\bar r^{2}},
\label{eq:nd1}
\end{equation}
\begin{equation}
H_x=\frac{\bar H_x}{\bar H_0},\quad H_y=\frac{\bar H_y}{\bar H_0},\quad
H=\frac{\bar H}{\bar H_0},\quad M=\frac{\bar M}{\bar M_s},\quad
\theta=\frac{\bar T-\bar T_0}{\bar T_A-\bar T_0},
\label{eq:nd2}
\end{equation}
where $\bar H_0$ is a reference magnetic field strength intensity. For an
external source it is the magnetic field strength intensity that \emph{one}
source produces at the corral centre, i.e.\ at the distance $\bar d_s$ from its
axis. For a magnet core inside a rotor it is the magnetic field strength
intensity at the distance $\bar d_s=3.5\bar r$ from the core axis
(\S\ref{sec:parameters}). With $d_s=\bar d_s/\bar r$, $(a,b)=(\bar a,\bar b)/\bar r$
and $\rho_s^{2}=(x-a)^{2}+(y-b)^{2}$, the dimensionless field of a wire is
\begin{equation}
H_x=s\,d_s\,\frac{y-b}{\rho_s^{2}},\qquad
H_y=-s\,d_s\,\frac{x-a}{\rho_s^{2}},\qquad
H=\frac{d_s}{\rho_s},
\label{eq:Hnd}
\end{equation}
where $s=\pm1$ gives the sense of the current, and that of a magnet is
\begin{equation}
\begin{gathered}
H_x=d_s^{2}\,\frac{m_x\big[(x-a)^{2}-(y-b)^{2}\big]+2m_y(x-a)(y-b)}{\rho_s^{4}},\\[2pt]
H_y=d_s^{2}\,\frac{2m_x(x-a)(y-b)+m_y\big[(y-b)^{2}-(x-a)^{2}\big]}{\rho_s^{4}},\qquad
H=\frac{d_s^{2}}{\rho_s^{2}}.
\end{gathered}
\label{eq:Hdipnd}
\end{equation}
For several sources the vector fields are added, $\bm H=\sum_k\bm H_k$, and
$H=|\bm H|$. The magnitudes then no longer add, and the result depends on the
current senses or on the orientations of the magnets
(\S\ref{sec:final}).
Substitution into
Eqs.~\eqref{eq:cont2}--\eqref{eq:energy2} gives
\begin{align}
&\frac{\partial u}{\partial x}+\frac{\partial v}{\partial y}=0, \label{eq:ndcont}\\[2pt]
&\frac{\partial u}{\partial t}+u\frac{\partial u}{\partial x}+v\frac{\partial u}{\partial y}
=-\frac{\partial p}{\partial x}+\mathrm{Mn_F}\,M\,\frac{\partial H}{\partial x}
-N\big(uH_y^{2}-vH_xH_y\big)
+\frac{1}{\mathrm{Re}}\Big(\frac{\partial^{2}u}{\partial x^{2}}+\frac{\partial^{2}u}{\partial y^{2}}\Big), \label{eq:ndx}\\[2pt]
&\frac{\partial v}{\partial t}+u\frac{\partial v}{\partial x}+v\frac{\partial v}{\partial y}
=-\frac{\partial p}{\partial y}+\mathrm{Mn_F}\,M\,\frac{\partial H}{\partial y}
-N\big(vH_x^{2}-uH_xH_y\big)
+\frac{1}{\mathrm{Re}}\Big(\frac{\partial^{2}v}{\partial x^{2}}+\frac{\partial^{2}v}{\partial y^{2}}\Big), \label{eq:ndy}\\[2pt]
&\frac{\partial\theta}{\partial t}+u\frac{\partial\theta}{\partial x}+v\frac{\partial\theta}{\partial y}
-\mathrm{Mn_F}\,\mathrm{Ec}\;\xi\mathcal L'(\xi)\Big(u\frac{\partial H}{\partial x}+v\frac{\partial H}{\partial y}\Big)
-N\,\mathrm{Ec}\,\big(uH_y-vH_x\big)^{2} \nonumber\\
&\hspace{6.2cm}=\frac{1}{\mathrm{Pr}\,\mathrm{Re}}\Big(\frac{\partial^{2}\theta}{\partial x^{2}}+\frac{\partial^{2}\theta}{\partial y^{2}}\Big)
+\frac{\mathrm{Ec}}{\mathrm{Re}}\,\Phi, \label{eq:ndT}
\end{align}
where the dimensionless magnetisation and Langevin argument are
\begin{equation}
M=\mathcal L(\xi),\qquad \xi=\frac{\alpha\,H}{1+\varepsilon\,\theta}.
\label{eq:ndLangevin}
\end{equation}
The dimensionless parameters are listed below.

\begin{itemize}
  \item Hydrodynamics:
    \begin{equation}
    \mathrm{Re}=\frac{\bar\Omega\bar r^{2}}{\bar\nu}\quad\text{(Reynolds number)},
    \end{equation}
    and the geometric parameters
    \begin{equation}
    G=\frac{\bar g}{2\bar R-4\bar r},\qquad C=\frac{\bar R}{\bar r},\qquad
    D=2+G\,(2C-4).
    \end{equation}
  \item MHD:
    \begin{equation}
    N=\frac{\bar\sigma\bar\mu_0^{2}\bar H_0^{2}}{\bar\rho\bar\Omega}=\frac{\mathrm{Ha}^{2}}{\mathrm{Re}},\qquad
    \mathrm{Ha}=\bar\mu_0\bar H_0\bar r\sqrt{\frac{\bar\sigma}{\bar\mu}}
    \quad\text{(Stuart and Hartmann numbers)}.
    \end{equation}
  \item FHD:
    \begin{equation}
    \mathrm{Mn_F}=\frac{\bar\mu_0\bar M_s\bar H_0}{\bar\rho\bar\Omega^{2}\bar r^{2}}
    \quad\text{(magnetic number)},\qquad
    \alpha=\frac{\bar\mu_0\bar m\bar H_0}{\bar\kappa\bar T_0}
    \quad\text{(Langevin parameter)}.
    \end{equation}
  \item Heat transfer:
    \begin{equation}
    \mathrm{Pr}=\frac{\bar\mu\bar C_p}{\bar k},\qquad
    \mathrm{Ec}=\frac{\bar\Omega^{2}\bar r^{2}}{\bar C_p(\bar T_A-\bar T_0)},\qquad
    \varepsilon=\frac{\bar T_A-\bar T_0}{\bar T_0}.
    \end{equation}
\end{itemize}

The magnetic number is defined with the saturation magnetisation, so $M\le1$
everywhere. In the weak-field regime $\xi\ll1$,
\begin{equation}
M\simeq\frac{\alpha H}{3(1+\varepsilon\theta)},\qquad
\mathrm{Mn_F}\,M\,\frac{\partial H}{\partial x_i}
\simeq\frac{\mathrm{Mn_F^{lin}}}{1+\varepsilon\theta}\,
H\,\frac{\partial H}{\partial x_i},
\label{eq:weakfield}
\end{equation}
where
$\mathrm{Mn_F^{lin}}=\alpha\,\mathrm{Mn_F}/3
=\bar\mu_0\bar\chi_0\bar H_0^{2}/(\bar\rho\bar\Omega^{2}\bar r^{2})$
and $\bar\chi_0=\bar\mu_0\bar m\bar M_s/(3\bar\kappa\bar T_0)$ is the
susceptibility at the reference temperature. The factor
$(1+\varepsilon\theta)^{-1}$ retains the Curie temperature dependence.
Apart from this factor, Eq.~\eqref{eq:weakfield} has the form of the linear
force and magnetic number of Eqs.~(11)--(12) of \citet{Tzirtzilakis2013cavity}.
As in that work, $N=0$ gives pure FHD, $\mathrm{Mn_F}=0$ gives pure MHD, and
$N=\mathrm{Mn_F}=0$ recovers the hydrodynamic problem of \citet{Pan2026planar}.

The dimensionless boundary conditions are
\begin{equation}
\begin{aligned}
&|\bm x|=C: && u=v=0,\quad \theta=0,\\
&|\bm x-\bm x_A|=1: && u=-y,\quad v=x+D/2,\\
&|\bm x-\bm x_P|=1: && u=-\Gamma(t)\,y,\quad v=\Gamma(t)\,(x-D/2),
\end{aligned}
\label{eq:ndbc}
\end{equation}
with $\theta=1$ on the heated rotor and $\partial\theta/\partial n=0$ on the
other,
and the torque balance~\eqref{eq:torque} becomes
\begin{equation}
\frac{d\Gamma}{dt}=\frac{2}{\pi}\,\frac{\bar\rho}{\bar\rho_s}\,\frac{1}{\mathrm{Re}}\;\mathcal T_P,\qquad
\mathcal T_P=\int_{0}^{2\pi}\Big(\frac{\partial u_\theta}{\partial\rho}-\frac{u_\theta}{\rho}\Big)\Big|_{\rho=1}d\theta,
\label{eq:ndtorque}
\end{equation}
where $\bar\rho_s$ is the density of the rotor and
$\bar I_P=\tfrac12\pi\bar\rho_s\bar r^{4}\bar h$. The terminal gear ratio is
the value of $\Gamma$ for which $\mathcal T_P=0$. Since it is a zero-torque
state, it does not depend on $\bar\rho/\bar\rho_s$ when the steady state is
unique \citep{Pan2026planar}. Whether a given torque-free steady state is
reached in time, i.e.\ its stability, is a separate question, whose answer
may depend on $\bar\rho/\bar\rho_s$. It is not examined in this work
(\S\ref{sec:limitations}).

\subsection{Uniform and axial fields}\label{sec:uniform}
For a uniform in-plane field $\bm H=(\cos\beta,\sin\beta)$, at angle $\beta$ to the line of centres,
$\nabla H=0$, so the FHD terms and the magnetocaloric term vanish
identically. Only the Lorentz terms of Eqs.~\eqref{eq:ndx}--\eqref{eq:ndy}
remain, with constant $H_x$ and $H_y$. They act as an anisotropic friction
that damps only the velocity component normal to the field.

For a uniform axial field $\bar{\bm B}=\bar B_0\hat{\bm z}$
parallel to the rotor axes and a strictly two-dimensional flow, the
electromotive force $\bar{\bm q}\times\bar{\bm B}=-\bar B_0\bar\nabla\bar\psi$
is a gradient, where $\bar\psi$ is the stream function. It is balanced by the
electrostatic potential, so no current flows. The effect of an axial field
therefore arises only through the Hartmann layers on the end walls. For a
layer of height $\bar h$ bounded by two insulating walls normal to the field,
the depth-averaged quasi-two-dimensional model
\citep{Sommeria1982,Potherat2000} adds to the right-hand side of
Eqs.~\eqref{eq:ndx}--\eqref{eq:ndy} the linear friction
\begin{equation}
-\lambda_H\,(u,v),\qquad
\lambda_H=\frac{2\,\mathrm{Ha}_h}{\mathrm{Re}}\Big(\frac{\bar r}{\bar h}\Big)^{2},\qquad
\mathrm{Ha}_h=\bar B_0\bar h\sqrt{\frac{\bar\sigma}{\bar\mu}} .
\label{eq:hartmann}
\end{equation}
This model requires $\mathrm{Ha}_h\gg1$. It introduces the screening length
$\bar\ell=\bar h/\sqrt{2\mathrm{Ha}_h}$, over which the swirl of the active
rotor decays.

\subsection{Final mathematical formulation of the problem}\label{sec:final}
For the realistic case scenario in which the oil-based ferrofluid EFH1
(\S\ref{sec:parameters}) is adopted, $N<10^{-14}$ and
$\mathrm{Ec}\approx2\times10^{-8}$. Thus, the Lorentz terms of the momentum
equations, as well as the Joule, viscous-dissipation and magnetocaloric terms
of the energy equation, can be neglected.

Moreover, steady torque-free states are considered (\S\ref{sec:numerics}), and
the governing equations \eqref{eq:ndcont}--\eqref{eq:ndT} are finally reduced
to:
\begin{align}
&\frac{\partial u}{\partial x}+\frac{\partial v}{\partial y}=0, \label{eq:fcont}\\[2pt]
&u\frac{\partial u}{\partial x}+v\frac{\partial u}{\partial y}
=-\frac{\partial p}{\partial x}
+\mathrm{Mn_F}\,\mathcal L(\xi)\,\frac{\partial H}{\partial x}
+\frac{1}{\mathrm{Re}}\Big(\frac{\partial^{2}u}{\partial x^{2}}+\frac{\partial^{2}u}{\partial y^{2}}\Big), \label{eq:fx}\\[2pt]
&u\frac{\partial v}{\partial x}+v\frac{\partial v}{\partial y}
=-\frac{\partial p}{\partial y}
+\mathrm{Mn_F}\,\mathcal L(\xi)\,\frac{\partial H}{\partial y}
+\frac{1}{\mathrm{Re}}\Big(\frac{\partial^{2}v}{\partial x^{2}}+\frac{\partial^{2}v}{\partial y^{2}}\Big), \label{eq:fy}\\[2pt]
&u\frac{\partial\theta}{\partial x}+v\frac{\partial\theta}{\partial y}
=\frac{1}{\mathrm{Re}\,\mathrm{Pr}}\Big(\frac{\partial^{2}\theta}{\partial x^{2}}+\frac{\partial^{2}\theta}{\partial y^{2}}\Big), \label{eq:fT}
\end{align}
with
\begin{equation}
\xi=\frac{\alpha H}{1+\varepsilon\theta},\qquad
\mathcal L(\xi)=\coth\xi-\frac1\xi .
\label{eq:fxi}
\end{equation}

The temperature enters the momentum equations through the Langevin argument
$\xi$, i.e.\ through the temperature dependence of the magnetisation, which
couples Eqs.~\eqref{eq:fx}--\eqref{eq:fy} to the energy
equation~\eqref{eq:fT}.

The dimensionless magnetic field strength intensity $H(x,y)$ is given by
Eqs.~\eqref{eq:Hnd}--\eqref{eq:Hdipnd}. For a single wire $H=d_s/\rho_s$,
whereas for a single magnet, outside the corral or as a core inside a rotor,
$H=d_s^{2}/\rho_s^{2}$.

For the pairs at $(0,\mp d_s)$, the distances from the lower and the upper
source are denoted by $\rho_{1}$ and $\rho_{2}$, with $\rho=(x^{2}+y^{2})^{1/2}$ and
$z=x+\mathrm{i}y$. The superposition gives the closed forms
\begin{equation}
H_{\uparrow\downarrow}=\frac{2d_s^{2}}{\rho_1\rho_2},\qquad
H_{\uparrow\uparrow}=\frac{2d_s\,\rho}{\rho_1\rho_2},\qquad
H_{\mathrm{dip}}=\frac{2d_s^{2}\,|z^{2}-d_s^{2}|}{\rho_1^{2}\rho_2^{2}},
\label{eq:Hpairs}
\end{equation}
for two wires with opposite ($\uparrow\downarrow$) and equal
($\uparrow\uparrow$) currents and for two identical rods magnetised in the
same direction. For parallel magnets the magnitude does not depend on the
common direction of magnetisation. All three fields are symmetric about both
axes. The qualitative difference between the two wire pairs is worth noting.
For opposite currents, as in the two sides of a long loop, the fields add
in the gap between the rotors ($H=2$ at the centre). For equal currents they cancel
there ($H=0$ at the centre).
The pairs at $(\mp d_s,0)$ follow by a rotation through $90^\circ$. For the
crosses of four sources the fields are superposed in the solver
(\S\ref{sec:numerics}). The eight arrangements of the current senses, or of
the magnetisations, that are distinct up to a global sign are considered for
both wires and rods. Their field magnitude is shown in Fig.~\ref{fig:Hcrosses}.
At the corral centre $H$ ranges from 0 to $2\sqrt2$ for the crosses of wires
and from 0 to 4 for the crosses of rods, depending on the senses. In every multi-source arrangement each source alone
gives $H=1$ at the corral centre. For several sources only the relative
senses of the currents or magnetisations matter, since a global change of
sign leaves $H$ unchanged.

\paragraph{Boundary conditions.} Following the notation of
\citet{Tzirtzilakis2013cavity}, the boundary conditions are
\begin{equation}
\left.
\begin{aligned}
&\text{Corral wall } (x^{2}+y^{2}=C^{2}): && u=0,\quad v=0,\quad \theta=0.\\
&\text{Active rotor } \big((x+D/2)^{2}+y^{2}=1\big): && u=-y,\quad v=x+D/2,\quad \theta=1\ \text{or}\ \partial\theta/\partial n=0.\\
&\text{Passive rotor } \big((x-D/2)^{2}+y^{2}=1\big): && u=-\Gamma y,\quad v=\Gamma\,(x-D/2),\quad \partial\theta/\partial n=0\ \text{or}\ \theta=1.
\end{aligned}
\right\}
\label{eq:fbc}
\end{equation}
The heated rotor has $\theta=1$ and the other rotor is adiabatic. The heated
rotor is $A$ for the external sources and for the core in $P$, and $P$ for the
core in $A$. The gear ratio $\Gamma$ is not prescribed. It is the additional
unknown fixed by the torque-free condition on the passive rotor,
\begin{equation}
\mathcal T_P(\Gamma)=\int_{0}^{2\pi}\Big(\frac{\partial u_\varphi}{\partial\rho}-\frac{u_\varphi}{\rho}\Big)\Big|_{\rho=1}d\varphi=0,
\label{eq:ftorque}
\end{equation}
where $(\rho,\varphi)$ are polar coordinates centred on $\bm x_P$ and
$u_\varphi$ is the azimuthal velocity.

\paragraph{Output quantities.} Besides $\Gamma$, two further quantities are
reported. The first is the torque exerted by the fluid on the active
rotor, $\mathcal T_A=\bar{\mathcal T}_A/(\bar\mu\bar\Omega\bar r^{2})$, given by
the integral of Eq.~\eqref{eq:ftorque} about $\bm x_A$. With this sign convention,
$\mathcal T_A<0$ when the fluid resists the imposed rotation. The mechanical power
supplied by the drive per unit length is
$\bar{\mathcal P}_A=-\bar\Omega\,\bar{\mathcal T}_A$, so
$\bar{\mathcal P}_A/(\bar\mu\bar\Omega^{2}\bar r^{2})=-\mathcal T_A$. The second
is the mean Nusselt number of the heated rotor,
\begin{equation}
\mathrm{Nu}=-\frac{1}{2\pi}\int_{0}^{2\pi}\frac{\partial\theta}{\partial\rho}\Big|_{\rho=1}d\varphi
=\frac{\bar Q}{2\pi\bar k\,(\bar T_A-\bar T_0)},
\label{eq:Nu}
\end{equation}
where $\bar Q$ is the heat transferred to the fluid per unit length and
$\rho$ is measured from the centre of the heated rotor.
For conduction alone through a concentric annulus,
$\mathrm{Nu}=1/\ln C=0.665$.

\paragraph{Thermomagnetic vorticity source.} The equations are solved in
primitive variables (\S\ref{sec:numerics}). For the interpretation of the
results, the action of the Kelvin force on the flow is expressed through the
vorticity $\zeta=\partial v/\partial x-\partial u/\partial y$. With
$M=\mathcal L(\xi)$ and $\xi$ given by Eq.~\eqref{eq:fxi}, the curl of the
Kelvin force in Eqs.~\eqref{eq:fx}--\eqref{eq:fy} is
\begin{equation}
\mathrm{Mn_F}\,\frac{\partial M}{\partial\theta}\Big(\frac{\partial\theta}{\partial x}\frac{\partial H}{\partial y}-\frac{\partial\theta}{\partial y}\frac{\partial H}{\partial x}\Big)
=-\mathrm{Mn_F}\,\frac{\varepsilon\,\xi\,\mathcal L'(\xi)}{1+\varepsilon\theta}\;\hat{\bm z}\cdot(\nabla\theta\times\nabla H).
\label{eq:fhdsource}
\end{equation}
This is the vorticity source of thermomagnetic convection
\citep{Finlayson1970}. It is non-zero wherever the temperature and field
gradients are not collinear, and it has the structure of the baroclinic source
of natural convection, with the field gradient, which points towards the field
source, in the role of gravity: fluid that is hotter, and therefore less
magnetised, is pushed away from the field source. Its strength is controlled
by $\mathrm{Mn_F}\,\varepsilon$ and by $\xi\mathcal L'(\xi)$, which peaks at
moderate $\xi$ and vanishes both for weak fields and at saturation. Through
the torque-free condition~\eqref{eq:ftorque}, the flow driven by this source
changes the gear ratio.

\subsection{Case scenarios}\label{sec:parameters}

Ferrotec EFH1, a light-mineral-oil-based magnetite ferrofluid
\citep{FerrotecEFH1}, is considered as the working fluid. It has been characterised in detail in a
ferrohydrodynamic pumping experiment \citep{Mao2011pumping} and used in
numerical studies of rotating and thermomagnetic flows
\citep{Kim2018,Lee2019thermomagnetic}. The property values adopted here are
listed in Table~\ref{tab:efh1}. The datasheet gives an initial susceptibility of 2.64 (SI). The measured value
$\chi_0=1.56$ \citep{Mao2011pumping} is consistent with the datasheet relative
permeability of about 2.6, which corresponds to $\chi_0\approx1.6$. The
measured value is therefore used. The thermal conductivity and specific heat are the nominal values used by
\citet{Kim2018} and \citet{Lee2019thermomagnetic}. No direct measurement for
EFH1 was found. 

Oil-based ferrofluids are electrical insulators, with conductivity of order
$10^{-8}$--$10^{-6}$\,S/m. For EFH1 the Lorentz terms are therefore negligible
($N<10^{-14}$ in all cases below for the value $\bar\sigma=10^{-8}$\,S/m
adopted here, and $N<10^{-12}$ even for $\bar\sigma=10^{-6}$\,S/m), and the
flow is governed by FHD alone.

\begin{table}[h]
\centering\small
\caption{Properties of the ferrofluid EFH1 used in the scenarios.}
\label{tab:efh1}
\footnotesize
\begin{tabular}{p{5.3cm}p{3.6cm}p{5.6cm}}
\toprule
Property & Value & Source\\
\midrule
Saturation magnetisation $\bar M_s$ & 35.0\,kA/m ($\bar\mu_0\bar M_s=44$\,mT) & \citet{FerrotecEFH1} (34.4\,kA/m measured, \citealp{Mao2011pumping})\\
Initial susceptibility $\chi_0$ & 1.56 & measured \citep{Mao2011pumping}\\
Magnetite volume fraction $\phi$ & 7.9\% & \citet{FerrotecEFH1}\\
Density $\bar\rho$ & 1210\,kg/m$^3$ (25\,\textdegree C) & \citet{FerrotecEFH1}\\
Viscosity $\bar\mu$ & 6\,mPa\,s (27\,\textdegree C) & \citet{FerrotecEFH1}\\
Thermal conductivity $\bar k$ & 0.19\,W/(m\,K) & \citet{Kim2018,Lee2019thermomagnetic}\\
Specific heat $\bar C_p$ & 1840\,J/(kg\,K) & \citet{Kim2018,Lee2019thermomagnetic}\\
Domain magnetisation of magnetite $\bar M_d$ & 446\,kA/m & \citet{Rosensweig1985book}\\
\midrule
Particle moment\newline $\bar m=3\bar\kappa\bar T_0\chi_0/(\bar\mu_0\bar M_s)$ & $4.41\times10^{-19}$\,A\,m$^2$ & derived\\
Magnetic diameter $(6\bar m/\pi\bar M_d)^{1/3}$ & 12.4\,nm (nominal 10\,nm) & derived\\
$\xi/\bar H=\bar\mu_0\bar m/(\bar\kappa\bar T_0)$ & $1.34\times10^{-4}$\,m/A & derived ($\bar T_0=300$\,K)\\
\bottomrule
\end{tabular}
\end{table}

The particle moment in Eq.~\eqref{eq:langevin} is chosen so that the
weak-field limit of the Langevin law reproduces the measured initial
susceptibility, $\chi_0=\bar\mu_0\bar m\bar M_s/(3\bar\kappa\bar T_0)$. This
gives a magnetic diameter of 12.4\,nm, somewhat larger than the nominal
10\,nm. A diameter inferred from the initial susceptibility is weighted
towards the larger particles of the distribution. With
$\bar\mu_0\bar m/(\bar\kappa\bar T_0)=1.34\times10^{-4}$\,m/A, the
magnetisation reaches half saturation ($\xi\approx1.8$) at
$\bar H\approx1.35\times10^{4}$\,A/m ($\bar\mu_0\bar H\approx17$\,mT).

The Langevin law with constant $\bar M_s$ captures only the $1/\bar T$
dependence of $\xi$. The decrease of the domain magnetisation with temperature
and the thermal dilution of the particles, which are neglected here, would add
further temperature sensitivity. The dilution alone,
$\bar M_s\propto\bar\rho$, adds to the pyromagnetic coefficient
$-\partial\bar M/\partial\bar T$ of Eq.~\eqref{eq:dMdT} the term
$\bar\beta_T\bar M_s\mathcal L(\xi)$. Relative to the retained term it is
$\bar\beta_T\bar T_0\,\mathcal L(\xi)/[\xi\mathcal L'(\xi)]$. With
$\bar\beta_T=8.6\times10^{-4}$\,K$^{-1}$ for EFH1
\citep{Lee2019thermomagnetic} this ratio is $0.26$ in the Curie regime, which
covers the wires and the rods up to about 4\,mT, and $0.29$, $0.40$ and $0.70$
at the reference points of the three cores ($\xi=0.90$, 2.03 and 3.61). The
decrease of the domain magnetisation adds a contribution of the same sign,
which is not quantified here. The thermomagnetic source of
Eq.~\eqref{eq:fhdsource} would therefore be larger than in the present model
by at least about a quarter for the external sources and by about 70\% at the reference point of
the strongest core, and by more where the field is closer to saturation. Its
effect on $\Gamma$ is not a uniform rescaling of the field
strength, because the ratio varies with $\xi$ across the fluid, and it has
not been computed.

The geometry of \S\ref{sec:benchmark} is scaled to a laboratory set-up with
rotor radius $\bar r=5$\,mm, $C=4.5$ (corral radius 22.5\,mm), $G=0.3$ and
rotor height $8\bar r$. The kinematic viscosity
$\bar\nu=4.96\times10^{-6}$\,m$^2$/s gives $\bar\Omega=0.198\,\mathrm{Re}$
rad/s, i.e.\ 38\,rpm at $\mathrm{Re}=20$ and 189\,rpm at $\mathrm{Re}=100$.
The reference temperature is $\bar T_0=300$\,K. In all magnetic cases
one rotor is heated by $\bar T_A-\bar T_0=10$\,K. This gives
$\varepsilon=0.033$, $\mathrm{Pr}=58$ and $\mathrm{Ec}\approx2\times10^{-8}$,
so viscous and magnetocaloric heating are negligible. The P\'eclet number
$\mathrm{Re}\,\mathrm{Pr}$ ranges from about 300 at $\mathrm{Re}=5$ to about
6000 at $\mathrm{Re}=100$, so the thermal boundary layers are thin and must be resolved by local mesh refinement.

\begin{figure}[t]
\centering
\includegraphics[width=\textwidth]{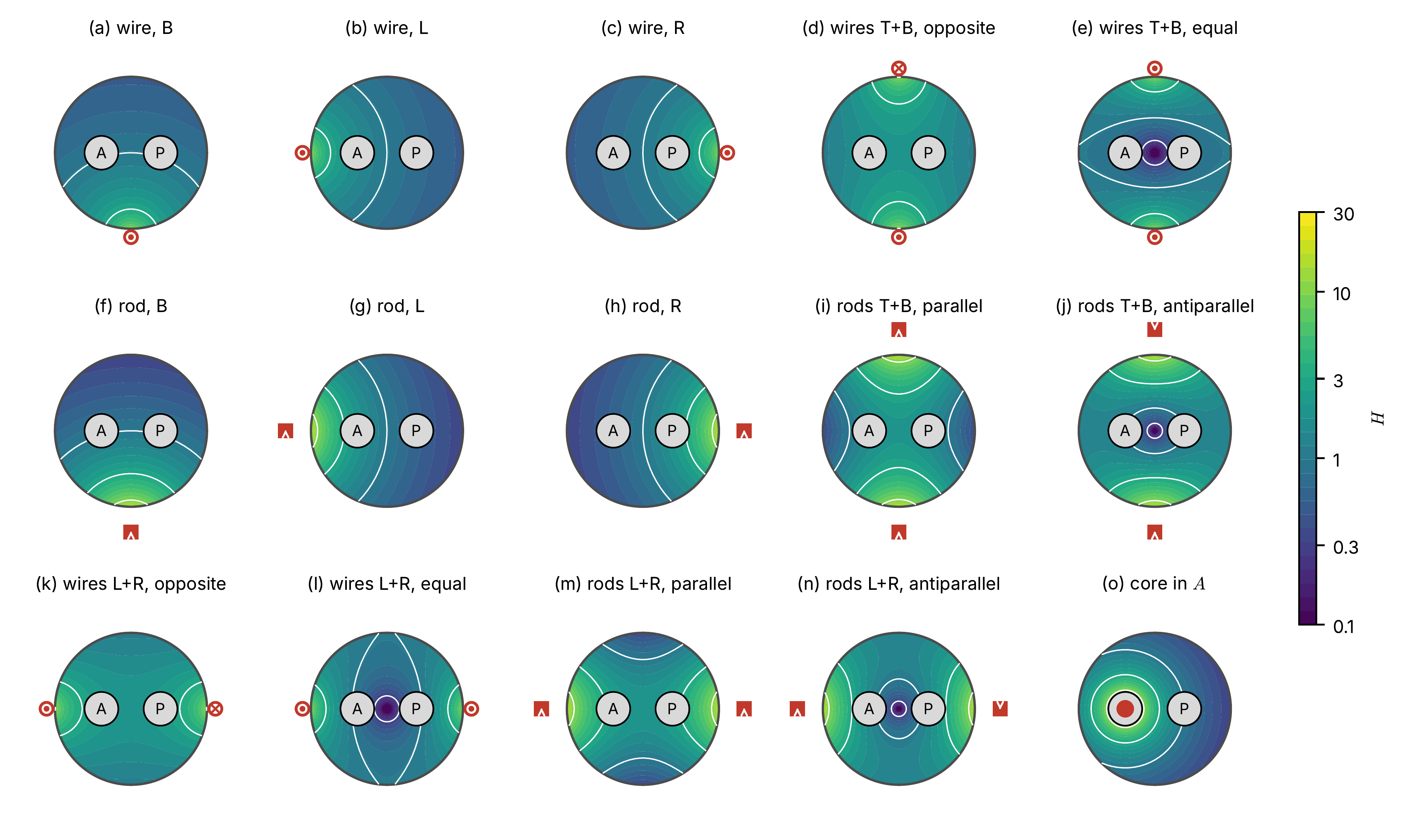}
\caption{Dimensionless magnetic field strength intensity $H$ in the fluid
($C=4.5$, $G=0.3$) for the single sources, the pairs and the core. Top row:
wires at $d_s=5$ (single wire at B, L, R, and pairs T+B with opposite and with
equal currents). Middle row: NdFeB rods at $d_s=6$ (single rod at B, L, R, and
pairs T+B magnetised parallel and antiparallel). Bottom row: pairs L+R of
wires and of rods, and a magnet core in rotor $A$ ($d_s=3.5$). Symbols at the
true source positions: wires, $\odot$ and $\otimes$ for the senses $+$ and
$-$ (as in Fig.~\ref{fig:magfieldsmulti}). Rods: squares with the arrow along
the magnetisation. Core: red disc. White lines: $H=0.3$, 1, 3 and 10. For a
single source $H=1$ at the corral centre (at 3.5 from the core for (o)). The
field of a single source at T is the mirror image of (a) and (f).}
\label{fig:Hconfigs}
\end{figure}

\begin{figure}[t]
\centering
\includegraphics[width=0.85\textwidth]{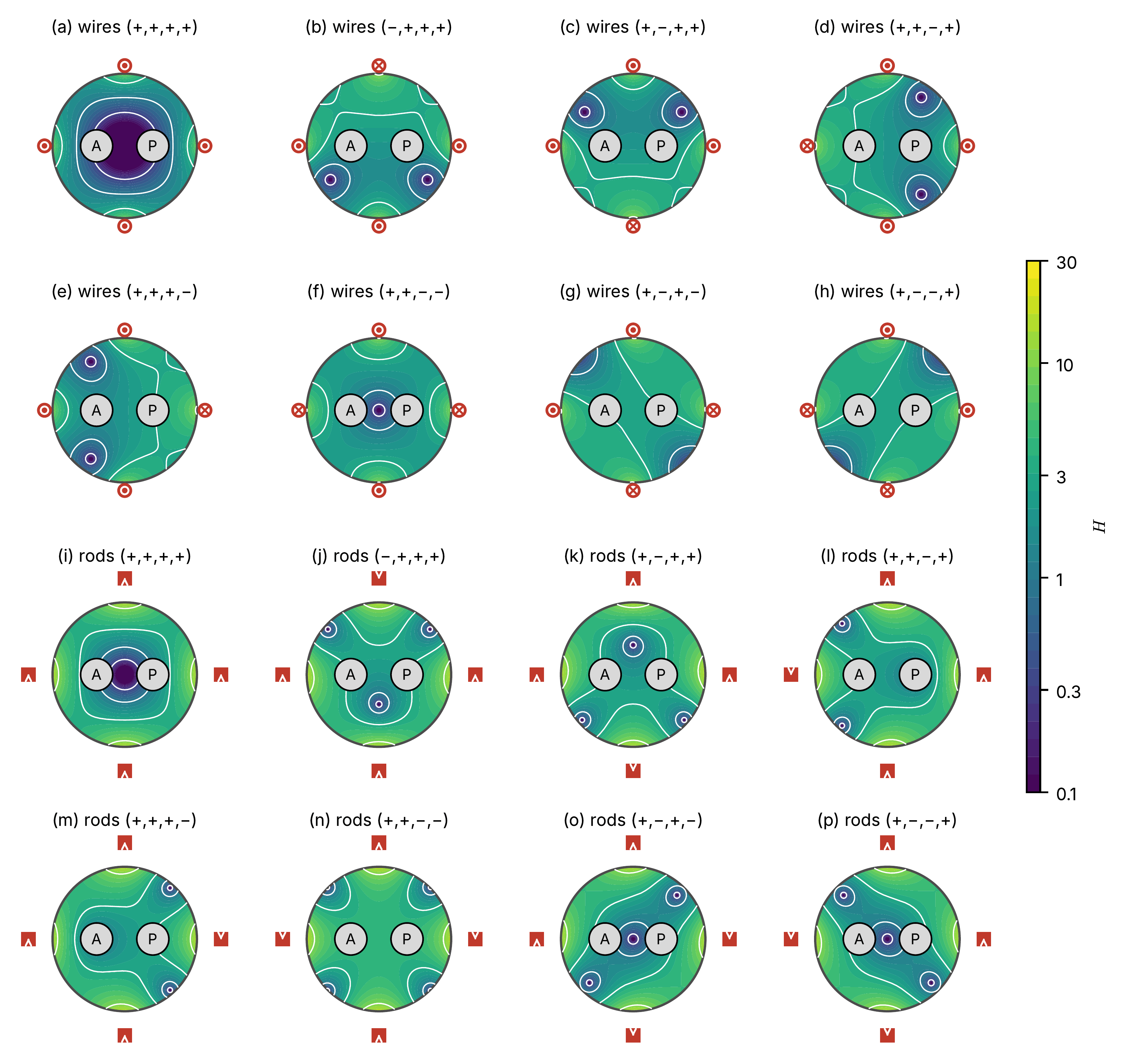}
\caption{Dimensionless magnetic field strength intensity $H$ in the fluid
($C=4.5$, $G=0.3$) for the crosses of four sources at (T, B, L, R): (a)--(h)
wires at $d_s=5$ and (i)--(p) NdFeB rods at $d_s=6$, for the eight
arrangements of the current senses or magnetisations that are distinct up to
a global sign (Table~\ref{tab:campaigns}). Each source alone gives $H=1$ at
the corral centre. Symbols as in Fig.~\ref{fig:Hconfigs}. White lines:
$H=0.3$, 1, 3 and 10.}
\label{fig:Hcrosses}
\end{figure}

The external magnetic field sources are placed at the positions T, B, L and R of
Fig.~\ref{fig:geometry}, singly, in pairs (T+B, L+R) or as crosses of four.
The magnitude $H$ of the dimensionless magnetic field strength intensity is
shown in Fig.~\ref{fig:Hconfigs} for the single sources at B, L and R, for the
eight pairs and for a core, and in Fig.~\ref{fig:Hcrosses} for the sixteen
crosses of four sources. The field of a single source at T is the mirror image
of that at B.

\begin{figure}[t]
\centering
\includegraphics[width=\textwidth]{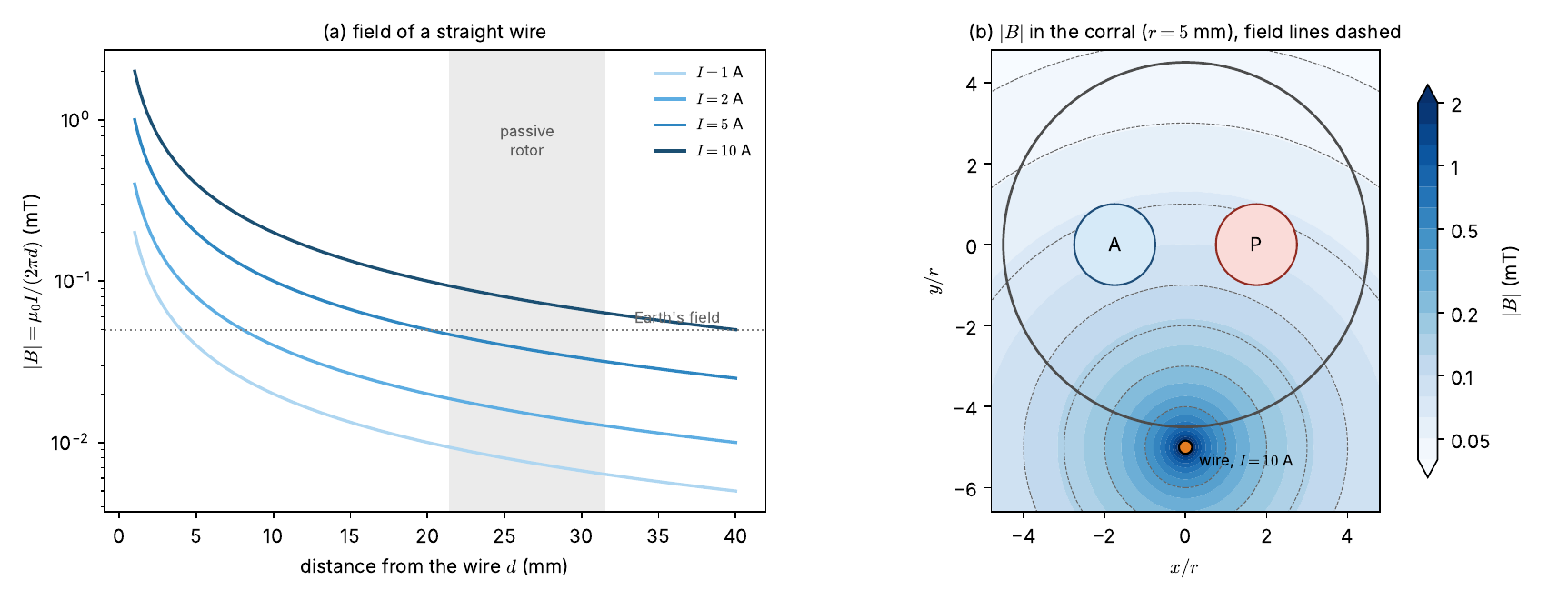}\\[4pt]
\includegraphics[width=\textwidth]{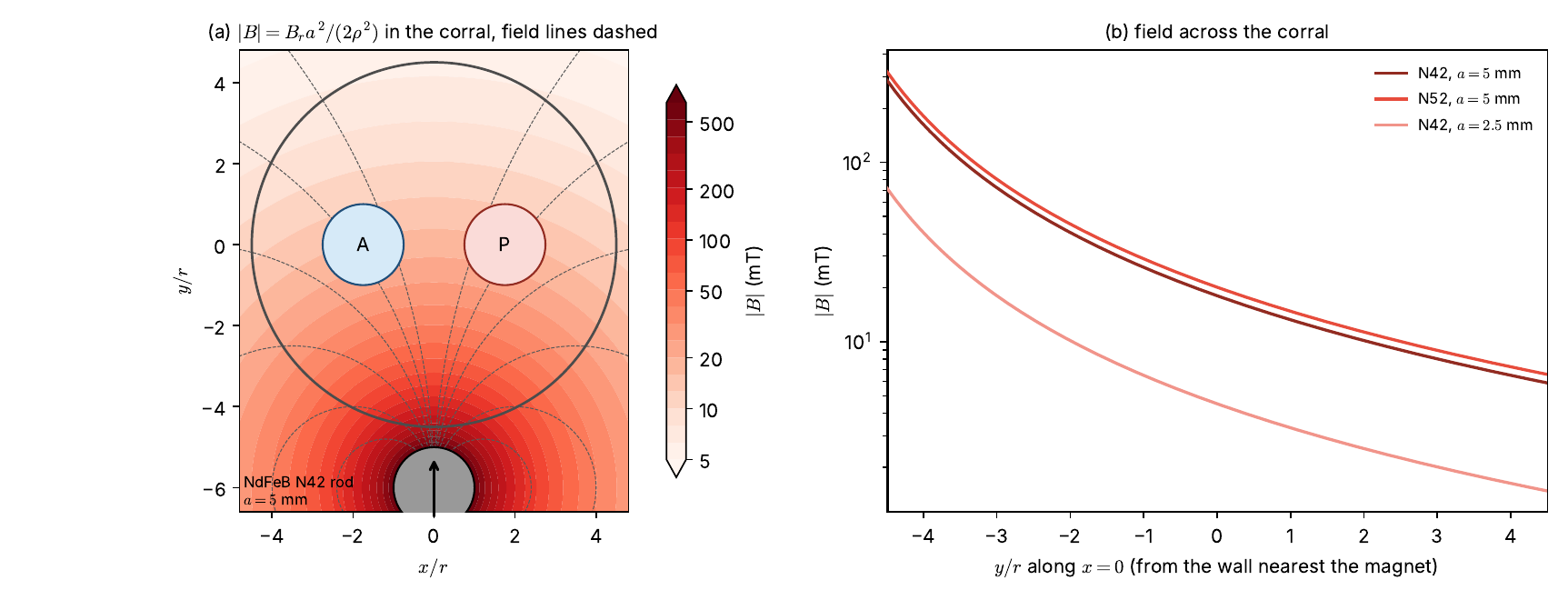}
\caption{Field sources for $\bar r=5$\,mm. Top: (a) field of a straight wire
for $I=1$--$10$\,A (the grey band marks the distances of the passive rotor).
(b) $|\bar B|$ in the corral for $I=10$\,A at position B. Bottom: (a)
$|\bar B|$ of a diametrically magnetised N42 rod ($\bar R_m=5$\,mm) at
position B. (b) Field along $x=0$ across the corral for three rods. Dashed
lines: field lines.}
\label{fig:fields}
\end{figure}

  The \emph{wires} are at $\bar d_s=5\bar r=25$\,mm from the
    corral centre, i.e.\ 2.5\,mm outside the corral wall. A single straight
    wire carries $I=1$--$10$\,A. A current of 10\,A is carried without
    difficulty by a copper conductor of 1.5--2.5\,mm$^2$ cross-section. At the
    corral centre such a wire gives only
    $\bar B_c=\bar\mu_0 I/(2\pi\bar d_s)=0.08$\,mT, comparable to the
    geomagnetic field of about 0.05\,mT (Fig.~\ref{fig:fields}a). Bundles of
    $N$ conductors, each carrying at most 10\,A, are therefore also considered. Such
    a bundle is the straight side of a long racetrack coil parallel to the
    rotor axes, with $NI$ up to 500 ampere-turns (50 turns of 10\,A) and
    $\bar B_c=4$\,mT. A pair with opposite currents corresponds to the
    two sides of one long loop around the corral. With equal currents it
    corresponds to two separate circuits. A cross of four wires combines the
    pairs T+B and L+R. The eight arrangements of the current senses that are
    distinct up to a global sign are listed in Table~\ref{tab:campaigns}.

The \emph{permanent magnets} are diametrically magnetised
    NdFeB rods of grade N42 ($\bar B_r\approx1.30$\,T, \citealp{ArnoldNdFeB}),
    with their axes at $\bar d_s=6\bar r=30$\,mm from the corral centre. Each
    produces the line-dipole field~\eqref{eq:Hdip}, with $|\bar B|=\bar B_r/2$
    on its surface (Fig.~\ref{fig:fields}b). At the corral centre
    $\bar B_c=\bar B_r\bar R_m^{2}/(2\bar d_s^{2})$, so the range
    $\bar B_c=0.25$--16\,mT considered below corresponds to rod radii
    $\bar R_m=0.6$--4.7\,mm. The rods of a pair are identical and
    magnetised in the same (parallel) or in opposite (antiparallel)
    directions. All rods, including the four rods of a cross, are magnetised
    along $\pm\hat{\bm y}$, i.e.\ perpendicular to the line of centres.
For a single external source, $\bar H_0=\bar B_c/\bar\mu_0$ is the reference
magnetic field strength intensity of \S\ref{sec:nondim}, and $\bar B_c$ the
corresponding magnetic flux density. In the arrangements with several sources,
each source alone has this strength, and $\bar B_0=\bar\mu_0\bar H_0$ refers
to one source. Across the fluid, $H$ then varies between
0.53 and 10 for a single wire and between 0.33 and 16 for a single rod. In
the crosses it reaches 11 (wires) and 16 (rods) near the corral wall and
vanishes at isolated points (Fig.~\ref{fig:Hcrosses}).

In a third group of scenarios the field source is a stationary, diametrically magnetised N42 core of radius
$\bar a_c=2$, $3$ or $4$\,mm. The core is mounted on the fixed axle inside one
of the rotors, which is then a non-magnetic sleeve of radius 5\,mm rotating
around it. The field is the line dipole~\eqref{eq:Hdipnd} centred on that
rotor: $|\bar B|=0.104$, $0.234$ and $0.416$\,T on the sleeve surface, and
8.5, 19.1 and 34.0\,mT at a distance of 17.5\,mm (the other rotor's centre
for $G=0.3$). In the scenario \emph{R1} the core is in the active rotor
$A$ and the passive rotor $P$ is heated. In \emph{R2} the core is in $P$ and $A$ is
heated.
In both cases the magnet and the heat source are on different rotors, so the
temperature and field gradients are not collinear.

The core is kept stationary for a practical reason. A magnetised rotor
immersed in EFH1 experiences a reluctance torque from the magnetisation of the
surrounding fluid, which is not axisymmetric because of the other rotor and
the corral. This torque was computed from the two-dimensional non-linear
magnetostatic problem with the Langevin law. For a 40-mm-high solid N42
passive rotor it reaches $4\times10^{-4}$\,N\,m, more than a thousand times
the viscous torque $|\bar{\mathcal T}_A|\,\bar h\approx3.4\times10^{-7}$\,N\,m
that drives the active rotor at $\mathrm{Re}=20$. A magnetised
passive rotor would therefore lock at a preferred angle. A non-magnetic sleeve
around a stationary core avoids this. Since $|\bar{\bm H}|$ of a line dipole is
axisymmetric, the field acting on the fluid is the same as that of a rotating
magnet.

At $\mathrm{Re}=20$ these scenarios have $\alpha=0.90$, $2.03$ and $3.61$,
$\mathrm{Mn_F}=624$, $1404$ and $2497$, and source strengths
$\mathrm{Mn_F}\varepsilon\,\xi\mathcal L'(\xi)=5.4$, $16$ and $22$, all
evaluated at the reference distance of 17.5\,mm.

\begin{figure}[t]
\centering
\includegraphics[width=0.66\textwidth]{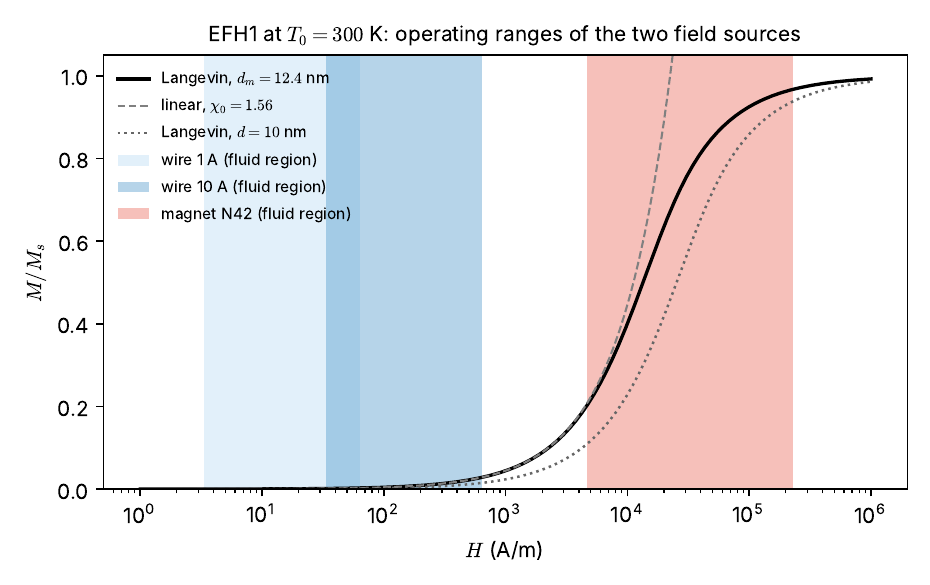}
\caption{Langevin magnetisation of EFH1 at $\bar T_0=300$\,K and the range of
$\bar H$ over the fluid domain for a single wire ($I=1$ and $10$\,A) and for
an N42 rod of radius 5\,mm at position B.}
\label{fig:langevin}
\end{figure}

These ranges are placed on the magnetisation curve in Fig.~\ref{fig:langevin}.
A single wire with $I\le10$\,A operates entirely in the linear (Curie)
regime: $\alpha=0.0085$, so $\xi<0.09$. A bundle of 500 ampere-turns has
$\alpha=0.43$ and reaches $\xi\approx4$ near the wall closest to it. The rods with
$\bar B_c=1$--$16$\,mT span the transition from the linear to the saturated part of
the Langevin curve, and the strongest rod ($\alpha=1.70$) reaches $\xi\approx27$
near the wall.

\paragraph{Dimensionless parameters.}
The resulting parameters at $\mathrm{Re}=20$ are listed in Table~\ref{tab:scenarios}.
Since $\bar H_0$ is fixed by the source,
$\mathrm{Mn_F}\propto\bar\Omega^{-2}\propto\mathrm{Re}^{-2}$. The same source
therefore acts sixteen times more strongly at $\mathrm{Re}=5$ and 25 times more
weakly at $\mathrm{Re}=100$.

The last column of the table is the strength of the thermomagnetic
vorticity source, $\mathrm{Mn_F}\,\varepsilon\,\xi\mathcal L'(\xi)$ from
Eq.~\eqref{eq:fhdsource}, evaluated at $H=1$. It should be compared with the
inertial and viscous terms of the vorticity equation, which are of order one
and $1/\mathrm{Re}$. The sources thus range from weak (a single wire, source
$<10^{-3}$) through moderate (bundles of 100--500 ampere-turns, rods of
1--4\,mT, source $0.05$--$1.3$) to strong (the larger rods and the cores,
source $\gg1$). The large magnetic numbers of the strong cases are of the
order that has caused numerical difficulties in BFD computations
\citep{Tzirtzilakis2013cavity}. They are approached by continuation in the
field strength (\S\ref{sec:numerics}).

\begin{table}[t]
\centering\small
\caption{Field sources for EFH1 at $\mathrm{Re}=20$ ($\bar r=5$\,mm,
$\bar\Omega=3.97$\,rad/s, $C=4.5$, $\bar T_0=300$\,K, $\Delta\bar T=10$\,K)
and the corresponding dimensionless parameters. $\bar B_0=\bar\mu_0\bar H_0$ is
the reference magnetic flux density: at the corral centre for one external source, at 17.5\,mm
from the axis for a core. Common values: $\varepsilon=0.033$,
$\mathrm{Pr}=58$, $\mathrm{Ec}\approx2\times10^{-8}$, $N<10^{-14}$.
$\mathrm{Mn_F^{lin}}=\alpha\,\mathrm{Mn_F}/3$ is the magnetic number of the
linear law used by \citet{Tzirtzilakis2013cavity}.}
\label{tab:scenarios}
\footnotesize\setlength{\tabcolsep}{4pt}
\begin{tabular}{llrrrrr}
\toprule
Source & Strength & $\bar B_0$ (mT) & $\alpha$ & $\mathrm{Mn_F}$ & $\mathrm{Mn_F^{lin}}$ & $\mathrm{Mn_F}\varepsilon\,\xi\mathcal L'(\xi)$\\
\midrule
Wire, $d_s=5$          & $NI=10$\,A          & 0.080 & 0.0085 & 5.88  & 0.0167 & $5.6\times10^{-4}$\\
                       & $NI=100$\,A-turns   & 0.80  & 0.085  & 58.8  & 1.67   & 0.056\\
                       & $NI=500$\,A-turns   & 4.0   & 0.43   & 294   & 41.7   & 1.34\\
N42 rod, $d_s=6$       & $\bar R_m=0.59$\,mm & 0.25  & 0.027  & 18.4  & 0.163  & $5.4\times10^{-3}$\\
                       & $\bar R_m=1.18$\,mm & 1.0   & 0.106  & 73.5  & 2.61   & 0.087\\
                       & $\bar R_m=2.35$\,mm & 4.0   & 0.43   & 294   & 41.7   & 1.34\\
                       & $\bar R_m=4.71$\,mm & 16.0  & 1.70   & 1176  & 668    & 13.5\\
N42 core, $d_s=3.5$    & $\bar a_c=2$\,mm    & 8.49  & 0.90   & 624   & 188    & 5.4\\
                       & $\bar a_c=3$\,mm    & 19.1  & 2.03   & 1404  & 951    & 16\\
                       & $\bar a_c=4$\,mm    & 34.0  & 3.61   & 2497  & 3007   & 22\\
\bottomrule
\end{tabular}
\end{table}

\paragraph{Limitations of the mathematical formulation.}
Three assumptions of the model deserve attention for this fluid.

   The first is the \emph{field modified by the magnetisation}. With $\chi_0\approx1.6$,
    EFH1 is not weakly magnetisable, so the field inside the fluid differs
    from the applied field. The BFD assumption that the field is not modified
    by the magnetisation (\S\ref{sec:problem}) is therefore an approximation.
    The strength of the thermomagnetic source depends on it, and a
    quantitative comparison with
    experiment would require solving the magnetostatic problem
    $\nabla\cdot(\bm H+\bm M)=0$, $\nabla\times\bm H=\bm0$, with the
    interface conditions on the rotor and corral boundaries. The size of the
    correction is not small. For a disc of linear susceptibility $\chi_0$ in
    a uniform transverse field the internal field is
    $2\bar H_0/(2+\chi_0)=0.56\,\bar H_0$, so in the Curie regime, where the
    source scales as $\bar H^{2}$, it would be reduced to about one third. The
    field of the localised sources is, in addition, redistributed, and its
    gradient, on which the mechanism depends directly, is changed. The
    dimensional field strengths quoted in this paper are therefore those of
    the prescribed applied field. They are not predictions of the strengths
    at which the effects would appear in a ferrofluid with
    $\chi_0\approx1.6$.

The second is the \emph{temperature dependence of the properties}. The temperature
dependence of the viscosity and of $\bar M_s$ is neglected. For a heating of 10\,K
    this is a first approximation. The omitted part of the pyromagnetic
    response is estimated above.

The third is \emph{gravity}. With $\bar\beta_T=8.6\times10^{-4}$\,K$^{-1}$
    \citep{Lee2019thermomagnetic}, $\bar r=5$\,mm and
    $\Delta\bar T=10$\,K, the Richardson number
    $\mathrm{Ri}=\bar g\bar\beta_T\Delta\bar T/(\bar\Omega^{2}\bar r)$ is
    4.3, 1.07 and 0.27 at $\mathrm{Re}=10$, 20 and 40. Buoyancy in the plane
    of the flow would therefore not be negligible. Its baroclinic source,
    $\mathrm{Ri}\,\hat{\bm z}\cdot(\nabla\theta\times\hat{\bm g})$, can be
    compared with the thermomagnetic source~\eqref{eq:fhdsource}, whose
    strength is $S|\nabla H|$ with $S$ the last column of
    Table~\ref{tab:scenarios}. The ratio
    $S|\nabla H|/\mathrm{Ri}=\bar\mu_0\bar M_s\xi\mathcal L'(\xi)|\bar\nabla\bar H|/(\bar T_0\bar\rho\bar g\bar\beta_T)$
    is independent of $\mathrm{Re}$ and of $\Delta\bar T$. At the reference
    point it is about $0.03$, $0.1$ and $0.4$ for a rod of 1, 2 and 4\,mT
    ($0.06$ and $0.25$ for wires of 2 and 4\,mT), but 2.9, 8.7 and 12 for
    the three cores. With horizontal axes, gravity would therefore dominate
    the thermomagnetic forcing of the external sources in the range where
    they reverse or strengthen the coupling. With vertical axes buoyancy
    drives an axial flow that, in the axially homogeneous model, does not
    act on the planar problem (\S\ref{sec:2D}). Its magnitude is not small: for
    the conduction regime of a vertical slot of the width of the fluid layer
    around $A$ (9--16\,mm), the axial velocity is of order 1--3\,cm/s,
    comparable with the rotor surface speed $\bar\Omega\bar r=1$--4\,cm/s
    for $\mathrm{Re}=10$--40. Near the ends of rotors of height $8\bar r$
    this flow turns round and has in-plane components, which are not
    represented by the planar model. A laboratory test of the present
    predictions therefore requires vertical axes, a large aspect ratio and,
    preferably, sources whose field and gradient are strong enough for the
    thermomagnetic forcing to exceed buoyancy.

\section{Numerical method}\label{sec:numerics}

\subsection{Steady torque-free states}
\citet{Pan2026planar} integrated the planar two-rotor problem in time. In all
the cases they computed up to $\mathrm{Re}=400$, the passive rotor relaxed to a
steady terminal spin. Its value was fixed by the zero-torque condition and was
independent of the rotor inertia. These terminal states are therefore computed
directly from the steady system~\eqref{eq:fcont}--\eqref{eq:ftorque}. The gear
ratio $\Gamma$ is an additional unknown, and the torque-free
condition~\eqref{eq:ftorque} is the additional equation. When the terminal
state of Eq.~\eqref{eq:ndtorque} is steady and unique, the two formulations
give the same $\Gamma$. 

\subsection{Meshes and finite element spaces}
\begin{figure}[t]
\centering
\includegraphics[width=\textwidth]{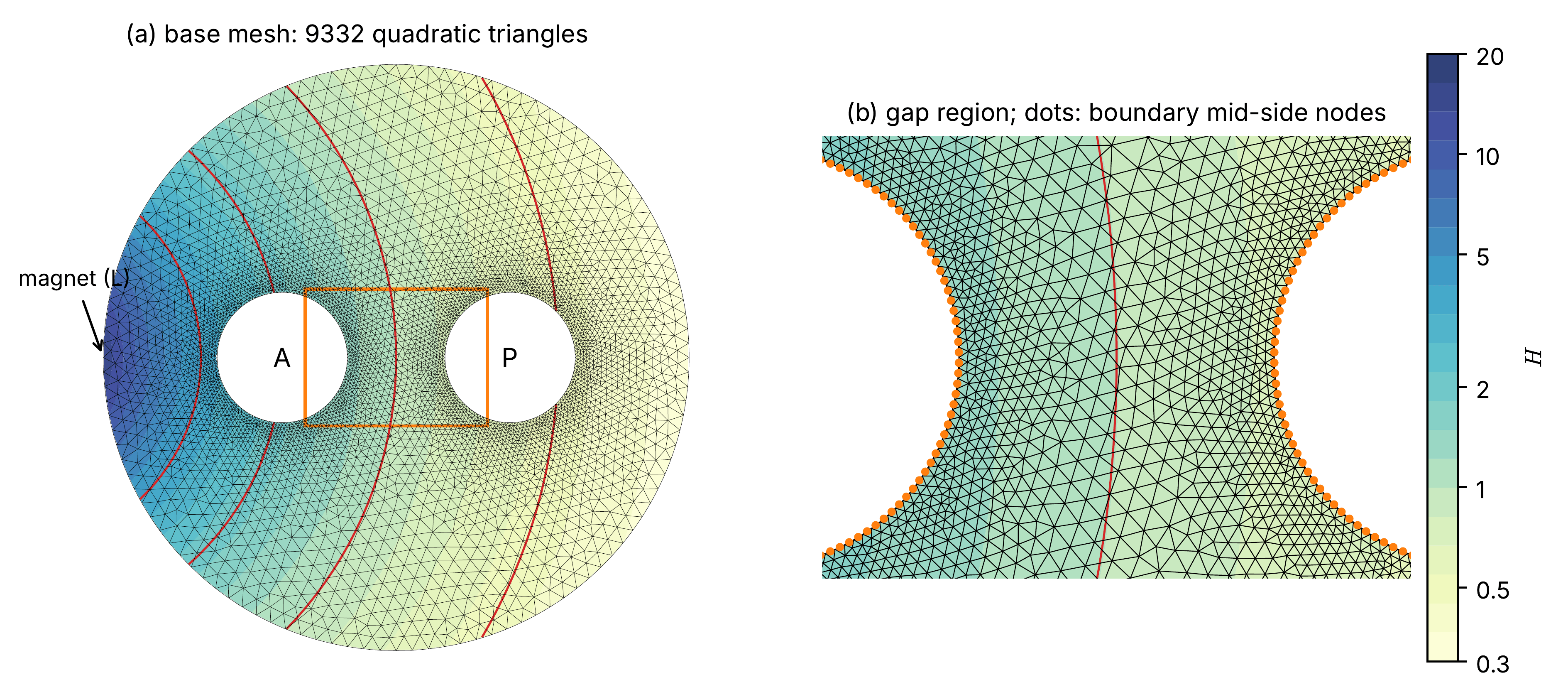}
\caption{Base mesh for $C=4.5$, $G=0.3$ ($h=0.05$ on the rotors, $h_w=0.25$ on
the corral wall), drawn with its curved quadratic edges, over the dimensionless
magnetic field strength intensity $H$ of an NdFeB rod at position L (red lines: $H=0.5$, 1, 2,
4). (a) Whole domain. (b) The gap region marked in (a), with the mid-side
nodes of the boundary edges, which lie on the exact circles.}
\label{fig:mesh}
\end{figure}

The fluid domain $\mathcal F$ is the corral disc with the two rotor discs
removed. It is covered by unstructured triangular meshes generated with Gmsh
\citep{Geuzaine2009gmsh} using the Frontal--Delaunay algorithm. The element
size is $h$ on both rotor surfaces. It grows linearly from $h$ at a distance
$0.15$ from the rotors to $h_w$ at a distance $2$, and $h_w$ is also used on
the corral wall (all lengths in units of $\bar r$). Four meshes are
used, with a refinement ratio of about 1.4 between consecutive ones. On the
\emph{coarse} mesh the element size is $h=0.07$ on the rotor surfaces and $h_w=0.35$
on the corral wall. In the same notation, the \emph{base} mesh has
$(h,h_w)=(0.05,0.25)$, the \emph{fine} mesh $(0.035,0.175)$ and the \emph{finest}
mesh $(0.025,0.125)$. 
For $C=4.5$ and $G=0.3$ the base mesh has 9\,332 triangles, about 126
elements around each rotor and 61\,942 unknowns for $(u,v,p,\theta)$. The
fine and finest meshes have 19\,014 and 36\,824 triangles and 125\,428 and
241\,928 unknowns. For the external sources the part of the corral wall
nearest to each source is refined further: the element size there is $2h$
within a distance 0.8 of the wall point closest to the source and grows
linearly to $h_w$ at a distance 2.5. The field gradient, and with it the
Kelvin force, is largest there. Without this refinement the global energy
balance of \S\ref{sec:torqueflux} was violated by 7\% on the base mesh for
the pair of magnets, against 0.08\% with it. The magnetic results of
\S\ref{sec:results} are computed on the finest mesh, and the other three
meshes are used for the mesh study (\S\ref{sec:meshcheck}). The base mesh for
a characteristic case is shown in Fig.~\ref{fig:mesh}, together with the field
of a magnet at position L.

All elements are six-node quadratic triangles. Each is the image of the
reference triangle $\hat K$ under a quadratic (isoparametric) map
$\bm x=\bm F_K(\hat{\bm x})$. The mid-side nodes of the boundary edges are
placed on the exact circles, so the rotor surfaces and the corral wall are
represented to third order, and $\Gamma$ is mesh-independent to four
significant digits (\S\ref{sec:validation}). The coupling is weak, and $\Gamma$
results from a small imbalance of opposing shear contributions. A small error
in the representation of the rotor surface therefore produces a comparatively
large error in $\Gamma$.

On this mesh three spaces of continuous, piecewise polynomial
functions are used: the velocity space $\bm V_h$, quadratic in each component, the
pressure space $Q_h$, linear, and the temperature space $\Theta_h$,
quadratic.
The pair $(\bm V_h,Q_h)$ is the Taylor--Hood element \citep{Taylor1973hood},
which satisfies the inf--sup condition. $\bm V_{h,0}$ and $\Theta_{h,0}$
denote the subspaces of functions that vanish on the boundaries where the
velocity and the temperature are prescribed. The unknowns are the nodal
values at the vertices and mid-side nodes (velocity, temperature) and at the
vertices (pressure).

\subsection{Weak formulation}
The discrete problem is obtained by multiplying
Eqs.~\eqref{eq:fcont}--\eqref{eq:fT} by test functions and integrating the
viscous, pressure and conduction terms by parts: $(\bm u_h,p_h,\theta_h)\in\bm V_h\times Q_h\times\Theta_h$
and $\Gamma\in\mathbb R$ are sought such that $\bm u_h$ and $\theta_h$ satisfy the
Dirichlet conditions~\eqref{eq:fbc}, and
\begin{align}
&\int_{\mathcal F}(\bm u_h\cdot\nabla\bm u_h)\cdot\bm w\,dA
+\frac{2}{\mathrm{Re}}\int_{\mathcal F}\mathsf D(\bm u_h):\mathsf D(\bm w)\,dA
-\int_{\mathcal F}p_h\,\nabla\cdot\bm w\,dA
-\int_{\mathcal F}\bm f_m(\theta_h)\cdot\bm w\,dA=0, \label{eq:weakmom}\\[2pt]
&\int_{\mathcal F}q\,\nabla\cdot\bm u_h\,dA=0, \label{eq:weakcont}\\[2pt]
&\int_{\mathcal F}(\bm u_h\cdot\nabla\theta_h)\,\phi\,dA
+\frac{1}{\mathrm{Pe}}\int_{\mathcal F}\nabla\theta_h\cdot\nabla\phi\,dA
+\sum_K\int_K\tau_K\,(\bm u_h\cdot\nabla\theta_h)(\bm u_h\cdot\nabla\phi)\,dA=0, \label{eq:weakT}\\[2pt]
&\mathcal T_P(\bm u_h,p_h,\theta_h)=0, \label{eq:weaktorque}
\end{align}
for all $\bm w\in\bm V_{h,0}$, $q\in Q_h$ and $\phi\in\Theta_{h,0}$.
Here $\mathsf D(\bm u)=\tfrac12(\nabla\bm u+\nabla\bm u^{\mathsf T})$
is the rate-of-strain tensor and $\mathrm{Pe}=\mathrm{Re}\,\mathrm{Pr}$. The force
$\bm f_m(\theta)=\mathrm{Mn_F}\,\mathcal L(\xi)\nabla H$, with
$\xi=\alpha H/(1+\varepsilon\theta)$, is the Kelvin force of
Eqs.~\eqref{eq:fx}--\eqref{eq:fy}, and the discrete torque $\mathcal T_P$ is
defined in \S\ref{sec:torqueflux}.
For a divergence-free velocity, the viscous term can be written either with
$\nabla\bm u$ or with $\mathsf D(\bm u)$. The symmetric form is used because
its boundary term is the physical traction, which is needed for the torque.

\paragraph{Stabilisation of the energy equation.} The P\'eclet number is of
order $10^{3}$. The Galerkin form of the energy equation would then give
oscillatory temperature fields on the meshes used. The last term of
Eq.~\eqref{eq:weakT} is a streamline-diffusion stabilisation. It has the
form of the streamline-upwind Petrov--Galerkin (SUPG) term of
\citet{Brooks1982supg}, but only the convective part
$\bm u_h\cdot\nabla\theta_h$ of the element residual is retained. The
diffusive part $-\mathrm{Pe}^{-1}\Delta\theta_h$ is omitted. For quadratic,
and in particular isoparametric, elements this part does not vanish inside
the elements. The stabilisation is therefore not strongly consistent: for the
exact solution of Eq.~\eqref{eq:fT} it leaves the residual
$\sum_K\int_K\tau_K\,\mathrm{Pe}^{-1}\Delta\theta\,(\bm u\cdot\nabla\phi)\,dA$,
which is of the order of $\tau_K$ and vanishes only as the mesh is refined.
It adds diffusion along the streamlines only. The parameter is the usual one,
$\tau_K=\frac{h_K}{2|\bm u_h|}\big(\coth\mathrm{Pe}_K-\mathrm{Pe}_K^{-1}\big)$,
with $\mathrm{Pe}_K=|\bm u_h|h_K\mathrm{Pe}/2$. The length $h_K$ is half the
element diameter, to account for the quadratic interpolation. For
$\mathrm{Pe}_K\gg1$, $\tau_K\to h_K/(2|\bm u_h|)$ decreases in proportion to
$h$, so the stabilisation, and with it the consistency error, is reduced by
about 30\% from the fine to the finest mesh. The mesh study of
\S\ref{sec:meshcheck} therefore includes the change of $\tau_K$. Because
$\bm u_h\cdot\nabla\phi=0$ for $\phi=1$, the stabilisation term drops out of
the sum of the discrete energy equations, and the global energy balance of
\S\ref{sec:torqueflux} is insensitive to its contribution in the interior of
the domain.

\paragraph{Field and quadrature.} All integrals are evaluated with a
six-point Gauss rule on $\hat K$, which is exact for polynomials of degree
four. The field enters only through $H$ and $\nabla H$ in $\bm f_m$. Both are
evaluated analytically at the physical positions $\bm F_K(\hat{\bm x}_q)$ of
the quadrature points, from Eqs.~\eqref{eq:Hnd}--\eqref{eq:Hdipnd} and their
derivatives. For superposed sources, $\bm H=\sum_k\bm H_k$ and its Jacobian
are computed first, and then $\nabla H=(\nabla\bm H)^{\mathsf T}\bm H/H$. The
field is therefore not interpolated, and no discretisation error arises from
its steep gradients near the sources.

\subsection{Newton's method with the gear ratio as unknown}
The vector of nodal values is denoted by $\bm X=(\bm U,\bm P,\bm\Theta)$ and
the residual vector of Eqs.~\eqref{eq:weakmom}--\eqref{eq:weakT}, assembled
for all basis functions, including those of the boundary nodes, by $\bm R(\bm X)$. The
Jacobian $\mathsf J=\partial\bm R/\partial\bm X$ is assembled exactly. It has
the block structure
\begin{equation}
\mathsf J=\begin{pmatrix}
\mathsf A+\mathsf N(\bm U) & \mathsf B^{\mathsf T} & \mathsf C_{u\theta}\\
\mathsf B & 0 & 0\\
\mathsf C_{\theta u}(\bm U,\bm\Theta) & 0 & \mathsf K+\mathsf S(\bm U)
\end{pmatrix}.
\label{eq:jacobian}
\end{equation}
Here $\mathsf A$ is the viscous matrix, $\mathsf N$ the linearised
convection, containing both $(\delta\bm u\cdot\nabla)\bm u_h$ and
$(\bm u_h\cdot\nabla)\delta\bm u$, and $\mathsf B$ the discrete divergence. The
block $\mathsf C_{u\theta}$ is the derivative of the magnetic force with respect to
the temperature,
$\partial\bm f_m/\partial\theta=-\mathrm{Mn_F}\,\varepsilon\,\xi\mathcal L'(\xi)\nabla H/(1+\varepsilon\theta)$,
and $\mathsf C_{\theta u}$ is the derivative of the convective and SUPG terms with
respect to the velocity. Finally, $\mathsf K$ is the conduction matrix and
$\mathsf S$ the convection--SUPG matrix, with $\tau_K$ frozen at its current
value.

\paragraph{Boundary conditions and the bordered system.} The Dirichlet
conditions~\eqref{eq:fbc} are imposed on the nodal values, including the
mid-side nodes. On the corral and on the active rotor the velocity values are
fixed. On the heated rotor and on the corral the temperature values are
fixed. The pressure is fixed at one node, since it is determined only
up to a constant. On the passive rotor the nodal velocities are
$\bm U_P=\Gamma\,\bm g_P$. Here $\bm g_P$ is the nodal vector of the rigid
rotation $\hat{\bm z}\times(\bm x-\bm x_P)$ on that rotor, so the
passive-rotor values are tied to the single unknown $\Gamma$. The extra
equation is the torque-free condition, written with the discrete torque
$\mathcal T_P=-\bm g_P\cdot\bm R_P$ (\S\ref{sec:torqueflux}). The free (non-Dirichlet) degrees of freedom are
denoted by $F$ and those on the passive rotor by $P$. Each Newton correction then solves the bordered system
\begin{equation}
\begin{pmatrix}
\mathsf J_{FF} & \mathsf J_{FP}\,\bm g_P\\
-\bm g_P^{\mathsf T}\mathsf J_{PF} & -\bm g_P^{\mathsf T}\mathsf J_{PP}\,\bm g_P
\end{pmatrix}
\begin{pmatrix}\delta\bm X_F\\ \delta\Gamma\end{pmatrix}
=-\begin{pmatrix}\bm R_F\\ \mathcal T_P\end{pmatrix},
\label{eq:bordered}
\end{equation}
followed by $\bm X_F\leftarrow\bm X_F+\lambda\,\delta\bm X_F$ and
$\Gamma\leftarrow\Gamma+\lambda\,\delta\Gamma$. The step length $\lambda$ is
chosen by backtracking (halving, at most seven times) until the Euclidean
norm of $(\bm R_F,\mathcal T_P)$ decreases. Iterations stop when the maximum
nodal correction falls below $10^{-9}$, and at most 30 iterations are
allowed. At convergence the residual torque on
the passive rotor is at round-off level, $|\mathcal T_P|\sim10^{-17}$.

The velocity, pressure, temperature and gear ratio are thus obtained
simultaneously. In the zero-field computations a secant iteration on
$\Gamma$, with one Navier--Stokes solve per step, was used initially. Both
procedures solve the same discrete equations, and they give the same $\Gamma$
to seven digits. The linear systems~\eqref{eq:bordered} are solved with a
sparse direct LU factorisation: PARDISO \citep{Schenk2004pardiso} in the
present computations, SuperLU (SciPy, \citealp{Virtanen2020scipy}) in the
earlier zero-field runs. On a single core of an otherwise idle machine, one
Newton step takes about 1, 2, 4 and 9\,s on the coarse, base, fine and
finest meshes. With all cores of the server busy it takes several times
longer. Each continuation
path (see below) runs as an independent single-threaded process. This uses a
many-core machine better than a multithreaded direct solver, which scales
poorly at these problem sizes. The computations of \S\ref{sec:results} were
run in this way on a four-socket server with 128 cores. A path on the
finest mesh took between 43 minutes and 17 hours.

\paragraph{Continuation.} The magnetic solutions are reached by
continuation in the reference magnetic field strength intensity $\bar H_0$.
The field is scaled physically, so that
$\mathrm{Mn_F}\propto\bar H_0$ and $\alpha\propto\bar H_0$ increase together.
The zero-field solution is computed first, and the temperature field is
then added with a negligible field ($\mathrm{Mn_F}=10^{-9}$). The field strength is
finally increased through the levels reported in \S\ref{sec:results}, each solution
serving as the initial guess for the next level.
If Newton's method fails between two levels, intermediate levels at the
geometric mean are inserted recursively, up to four times. When even this
fails, the path stops, and the last converged level is recorded. A
stopped path thus marks a failure of the Newton continuation. By itself it
does not distinguish a turning point of the steady branch, the loss of a
steady solution, or an ordinary failure of the nonlinear solver. For 31 of
the paths that stopped, the continuation was repeated with at most 120
Newton iterations and the tolerance $2\times10^{-9}$. Their results replace
the original ones (\S\ref{sec:limitations}).
Solutions at higher $\mathrm{Re}$ in the zero-field benchmark are obtained by
continuation in $\mathrm{Re}$.

\subsection{Torque and heat flux}\label{sec:torqueflux}
Boundary fluxes are computed from the discrete residual (consistent reactions,
\citealp{Gresho1987consistent}) rather than by differentiating the discrete
solution at the wall. For the torque on rotor $k\in\{A,P\}$,
\begin{equation}
\mathcal T_k=-\,\mathrm{Re}\;\bm g_k\cdot\bm R_{u}(\bm X),
\label{eq:residualtorque}
\end{equation}
where $\bm R_u$ is the momentum residual~\eqref{eq:weakmom}, including the
Kelvin force, and $\bm g_k$ is the rigid rotation about $\bm x_k$ at
the velocity nodes of that rotor. The factor $\mathrm{Re}$ converts to the
scale $\bar\mu\bar\Omega\bar r^{2}$ of Eq.~\eqref{eq:ftorque}. Because of the
symmetric viscous form, Eq.~\eqref{eq:residualtorque} equals the torque of the
traction $\bm\sigma\cdot\bm n$, with $\bm\sigma=-p\,\mathsf I+(2/\mathrm{Re})\mathsf D(\bm u)$.
The pressure exerts no torque on a circular rotor. In the same way,
the heat transferred from the heated rotor is the sum of the
Galerkin energy residual over the temperature nodes of that rotor:
\begin{equation}
\mathrm{Nu}=\frac{\mathrm{Pe}}{2\pi}\sum_{i\in\text{heated rotor}}
\Big[\int_{\mathcal F}(\bm u_h\cdot\nabla\theta_h)\phi_i\,dA
+\frac{1}{\mathrm{Pe}}\int_{\mathcal F}\nabla\theta_h\cdot\nabla\phi_i\,dA\Big].
\label{eq:residualNu}
\end{equation}
The same sum over the corral nodes gives the heat leaving through the wall.
Since the passive rotor is adiabatic, the two must be equal. Their relative
difference, $|\mathrm{Nu}+\mathrm{Nu}_{\mathrm{corral}}|/\mathrm{Nu}$ with
the sign convention of the residual, is a check of the global energy
balance. On the finest mesh it is of order $10^{-5}$ for weak fields (at most
$1.1\times10^{-4}$ at the first strength level of every path). It grows
for strong sources, whose thermomagnetic flow produces thin thermal layers at
the corral wall, and in \S\ref{sec:results} states with an error above 1\%
are treated as unresolved. The balance is a necessary condition, used
here to exclude unresolved states. It is not a sufficient verification of the
coupled magnetic solution. It reflects mainly the discrete mass conservation
of the Taylor--Hood velocity, weighted by the temperature, which deteriorates
where the thermal layers are thin, and it is insensitive to the interior
contribution of the stabilisation term. The accuracy of the accepted states
is assessed by the comparison of meshes (\S\ref{sec:meshcheck}), and near a
zero of $\Gamma$ only the absolute differences are meaningful.

Finally, the solver is written in Python. Assembly uses scikit-fem
\citep{Gustafsson2020skfem}, and the sparse linear algebra uses SciPy
\citep{Virtanen2020scipy} and PARDISO.

\section{Results and discussion}\label{sec:results}

\subsection{Validation and grid independence}\label{sec:validation}
The solver is first validated on circular Couette flow. A rotor of unit radius
spins at the centre of a corral of radius $C=4$. The azimuthal flow
$u_\theta=A\rho+B/\rho$ is an exact solution of the Navier--Stokes equations
for every Reynolds number. The torque on the rotor per unit length is
$\bar{\mathcal T}_{\mathrm{ex}}=-4\pi\bar\mu\bar\Omega\bar r^{2}C^{2}/(C^{2}-1)$. The viscous and convective
terms, the pressure--centripetal balance and the residual
torque~\eqref{eq:residualtorque} are therefore tested by this flow.
The flow is computed on the same unstructured isoparametric meshes as
the two-rotor problem, generated with the element size $h$ on the rotor surface and
$5h$ on the corral wall, so the circular boundaries are represented by the curved
quadratic elements and are not fitted by the grid. The computed torque
$\bar{\mathcal T}_h$ is the residual torque~\eqref{eq:residualtorque} evaluated
from the discrete solution, and the relative error reported in
Table~\ref{tab:couette} is defined as
$|\bar{\mathcal T}_h-\bar{\mathcal T}_{\mathrm{ex}}|/|\bar{\mathcal T}_{\mathrm{ex}}|$.
The results are presented in Table~\ref{tab:couette}. On the reference mesh the relative error of the torque is about
$10^{-6}$, and it decreases at roughly fourth order in $h$, which is consistent with
the superconvergence of the residual-based torque for quadratic velocity
elements.

\begin{table}[h]
\centering\small
\caption{Circular Couette flow ($C=4$): relative error of the computed torque on the rotor.}
\label{tab:couette}
\begin{tabular}{ccccc}
\toprule
$h$ & unknowns & $\mathrm{Re}=1$ & $\mathrm{Re}=20$ & $\mathrm{Re}=200$\\
\midrule
0.080 & 10\,390 & $7.6\times10^{-6}$ & $5.8\times10^{-6}$ & $7.0\times10^{-5}$\\
0.050 & 25\,745 & $1.2\times10^{-6}$ & $1.1\times10^{-6}$ & $8.4\times10^{-6}$\\
0.035 & 51\,318 & $3.0\times10^{-7}$ & $2.8\times10^{-7}$ & $1.6\times10^{-6}$\\
\bottomrule
\end{tabular}
\end{table}

The grid study for the two-rotor problem at $C=4.5$, $G=0.3$ and
$\mathrm{Re}=20$ and $100$ is reported in Table~\ref{tab:grid}. When the
element size is reduced from $h=0.08$ to $h=0.025$, which increases the number
of unknowns tenfold, $\Gamma$ changes by less than $0.002\%$ at
$\mathrm{Re}=20$ and by $0.04\%$ at $\mathrm{Re}=100$. The torque on the
active rotor changes by less than $0.01\%$. At $\mathrm{Re}=400$, the base, fine and finest meshes give the same
$\Gamma$ to three significant digits. The zero-field results of
\S\ref{sec:benchmark} were afterwards recomputed on all four meshes of
\S\ref{sec:numerics} (216 continuation paths). Every value quoted in
Table~\ref{tab:benchmark} is reproduced, and on the fine and finest meshes
$\Gamma$ differs by at most $1.4\times10^{-6}$. The magnetic results of
\S\ref{sec:magnetic} are computed on the finest mesh, and their mesh
independence is examined in \S\ref{sec:meshcheck}.

\begin{table}[h]
\centering\small
\caption{Grid study for $C=4.5$, $G=0.3$: gear ratio and torque on the active
rotor, $\bar{\mathcal T}_A/(\bar\mu\bar\Omega\bar r^{2})$, per unit length.
The residual torque on the passive rotor is below $2\times10^{-15}$ in all
cases. The unknowns are those of $(u,v,p)$. These zero-field meshes were
generated independently of the meshes of \S\ref{sec:numerics}, whose element
counts differ slightly.} \label{tab:grid}
\begin{tabular}{rrrcccc}
\toprule
 & & & \multicolumn{2}{c}{$\mathrm{Re}=20$} & \multicolumn{2}{c}{$\mathrm{Re}=100$}\\
\cmidrule(lr){4-5}\cmidrule(lr){6-7}
$h$ & elements & unknowns & $\Gamma$ & $\bar{\mathcal T}_A$ & $\Gamma$ & $\bar{\mathcal T}_A$\\
\midrule
0.080 & 3\,729 & 17\,355 & $-0.0236229$ & $-14.1975$ & $-0.0099410$ & $-14.2847$\\
0.050 & 9\,332 & 42\,911 & $-0.0236227$ & $-14.1975$ & $-0.0099374$ & $-14.2849$\\
0.035 & 18\,978 & 86\,713 & $-0.0236226$ & $-14.1974$ & $-0.0099370$ & $-14.2849$\\
0.025 & 36\,824 & 167\,545 & $-0.0236225$ & $-14.1974$ & $-0.0099369$ & $-14.2849$\\
\bottomrule
\end{tabular}
\end{table}

\subsection{Zero-field benchmark: comparison with Pan and He}\label{sec:benchmark}
\begin{figure}[t]
\centering
\includegraphics[width=\textwidth]{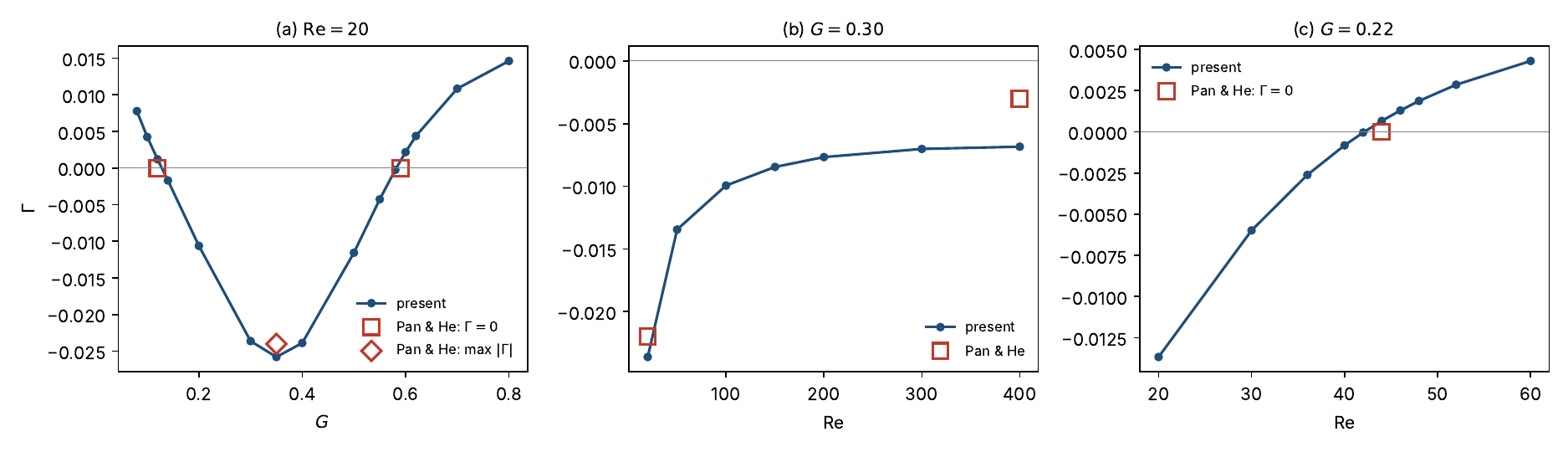}
\caption{Gear ratio for $C=4.5$ without magnetic field: present results
(lines with symbols) and values reported by \citet{Pan2026planar} (open red
symbols). (a) Gap transect at $\mathrm{Re}=20$. (b) Reynolds-number transect at
$G=0.30$. (c) Reynolds-number transect at $G=0.22$.}
\label{fig:benchmark}
\end{figure}

Without a magnetic field ($N=\mathrm{Mn_F}=0$) the problem reduces to the
planar problem of \citet{Pan2026planar}. Their solution was obtained with a
different method: a fictitious-domain method on a Cartesian grid, with time
integration to the terminal state, whereas the present method is a
body-fitted steady formulation. A comparison is made with their results along
the three transects they discuss in detail and with their corral-size trend.
The comparison is summarised in Fig.~\ref{fig:benchmark} and
Table~\ref{tab:benchmark}. The values
attributed to \citet{Pan2026planar} are those stated in their text, not
digitised from their figures.

\begin{table}[t]
\centering\small
\caption{Zero-field benchmark ($C=4.5$ unless stated). Zero crossings are
obtained by linear interpolation between computed points. Pan \& He: values
as stated in the text of \citet{Pan2026planar}.}
\label{tab:benchmark}
\begin{tabular}{lcc}
\toprule
Quantity & \citet{Pan2026planar} & present\\
\midrule
$\mathrm{Re}=20$: corotation at small gap for & $G\lesssim0.12$ & $G<0.128$\\
$\mathrm{Re}=20$: corotation at wide gap for & $G\gtrsim0.59$ & $G>0.582$\\
$\mathrm{Re}=20$: strongest counter-rotation & $|\Gamma|\approx0.024$ near $G\approx0.35$ & $\Gamma=-0.0258$ at $G=0.35$\\
$G=0.30$, $\mathrm{Re}=20$ & $\Gamma\approx-0.022$ & $\Gamma=-0.0236$\\
$G=0.30$, $\mathrm{Re}=400$ & $\Gamma\approx-0.003$ & $\Gamma=-0.0068$\\
$G=0.30$: sign change for $\mathrm{Re}\le400$ & none & none\\
$G=0.22$: spin reversal & $\mathrm{Re}\approx44$ & $\mathrm{Re}=42.1$\\
$\max|\Gamma|$, $C=3$ & $\lesssim0.062$ & $0.0622$ ($\mathrm{Re}=10$, $G=0.30$)\\
$\max|\Gamma|$, $C=4.5$ & $\lesssim0.029$ & $0.0308$ ($\mathrm{Re}=10$, $G=0.30$)\\
$\max|\Gamma|$, $C=6$ & $\lesssim0.016$ & $0.0166$ ($\mathrm{Re}=10$, $G=0.30$)\\
\bottomrule
\end{tabular}
\end{table}

The gap transect at $\mathrm{Re}=20$ is the transect that
\citet{Pan2026planar} regard as the cleanest benchmark of the gap-controlled
route, and it shows the clearest agreement
(Fig.~\ref{fig:benchmark}a). The re-entrant sequence is recovered by the
present computations: corotation at small gaps, a broad band of
counter-rotation, and
corotation again at wide gaps. The two sign changes occur at $G=0.128$ and
$G=0.582$, against $G\approx0.12$ and $G\approx0.59$ in
\citet{Pan2026planar}. The strongest counter-rotation occurs at the same gap,
$G\approx0.35$, with $|\Gamma|=0.0258$ against $0.024$.

The agreement along the Reynolds-number transects is
similar. At the experimental gap $G=0.30$ (Fig.~\ref{fig:benchmark}b) a
counter-rotating passive rotor whose $|\Gamma|$ decreases monotonically with
$\mathrm{Re}$ is obtained in both computations. Neither shows a change of sign up to
$\mathrm{Re}=400$. The values differ by about 7\% at $\mathrm{Re}=20$
($-0.0236$ against $-0.022$). At higher $\mathrm{Re}$ the curves separate:
the present $\Gamma$ levels off at about $-0.007$ for $\mathrm{Re}\gtrsim300$, whereas
\citet{Pan2026planar} report $\Gamma\approx-0.003$ at $\mathrm{Re}=400$. At
the narrower gap $G=0.22$ (Fig.~\ref{fig:benchmark}c), the Reynolds-driven
spin reversal reported by \citet{Pan2026planar} near $\mathrm{Re}\approx44$
is reproduced at $\mathrm{Re}=42.1$.

\begin{figure}[t]
\centering
\includegraphics[width=0.6\textwidth]{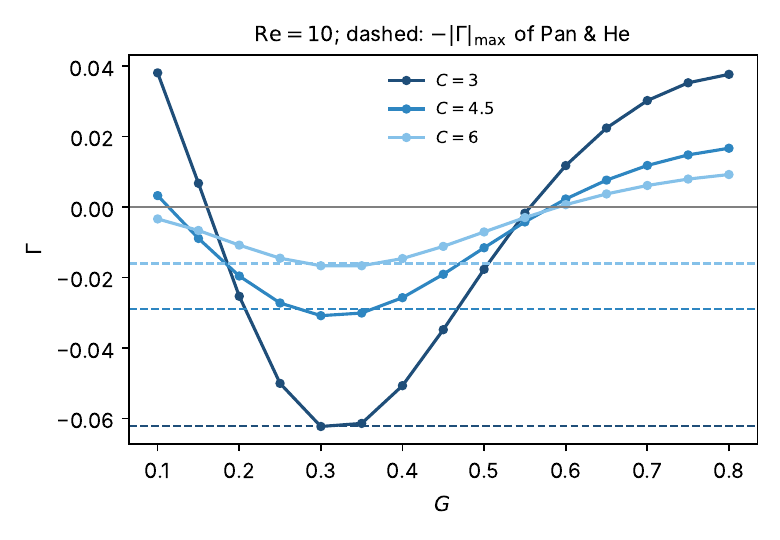}
\caption{Effect of corral size on the gap transect at $\mathrm{Re}=10$.
Dashed lines: $-|\Gamma|_{\max}$ reported by \citet{Pan2026planar} for each
$C$.}
\label{fig:corral}
\end{figure}

The effect of the corral size is compared next. Gap
transects at $\mathrm{Re}=10$ for $C=3$, $4.5$ and $6$ are shown in
Fig.~\ref{fig:corral}. The coupling weakens rapidly as the corral grows. The largest
counter-rotation, at $G\approx0.3$ for all three corrals, is $|\Gamma|=0.0622$,
$0.0308$ and $0.0166$. \citet{Pan2026planar} give the upper bounds
$|\Gamma|\lesssim0.062$, $0.029$ and $0.016$ over their phase maps. The trend
that tighter corrals produce stronger coupling, which is also observed
experimentally \citep{Smith2026spin}, is thus reproduced quantitatively.

\begin{figure}[t]
\centering
\includegraphics[width=\textwidth]{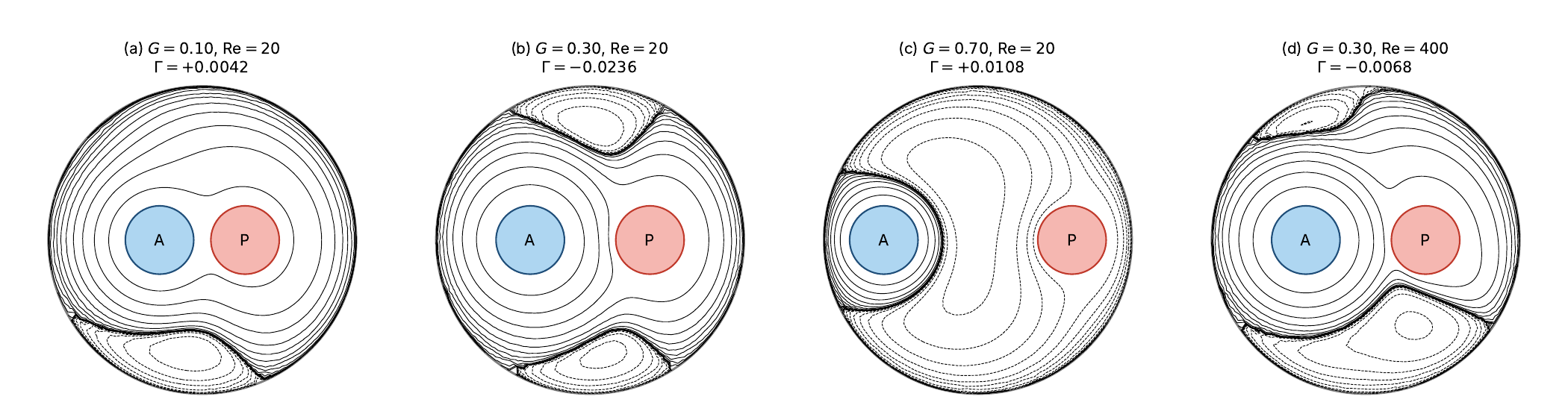}
\caption{Streamlines for $C=4.5$ without magnetic field (solid: positive
stream function relative to the corral wall, dashed: negative). $A$: active
rotor (counter-clockwise). $P$: passive rotor.}
\label{fig:streamlines}
\end{figure}

\begin{figure}[t]
\centering
\includegraphics[width=0.62\textwidth]{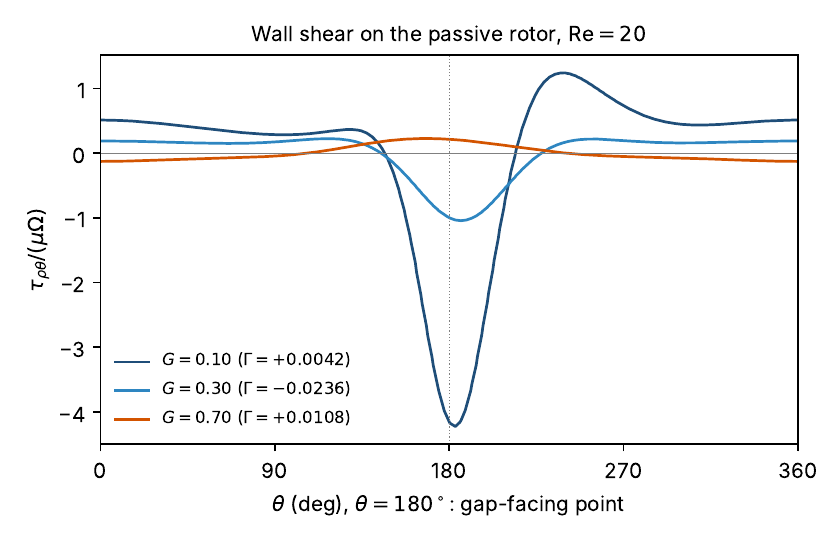}
\caption{Shear stress on the passive rotor at $\mathrm{Re}=20$, $C=4.5$.
Positive values drive the rotor in the sense of the active rotor (corotation).
$\theta$ is measured from the line of centres, and $\theta=180^\circ$ is the
point facing the active rotor.}
\label{fig:wallshear}
\end{figure}

The comparison closes with the flow fields and the wall
shear. Streamlines for representative states are shown in
Fig.~\ref{fig:streamlines}, and the shear stress $\tau_{\rho\theta}$ on the
passive rotor at $\mathrm{Re}=20$ in Fig.~\ref{fig:wallshear}. The computed $\Gamma$ is thereby
connected with the shear-competition picture of \citet{Smith2026spin}.

At $G=0.30$ (counter-rotation, Fig.~\ref{fig:streamlines}b) a corral-scale
circulation encloses both rotors, and two counter-rotating eddies are attached
to the corral wall. The passive rotor is sheared negatively, i.e.\ in the
counter-rotating sense, on a gap-facing arc
$145^\circ\lesssim\theta\lesssim228^\circ$, where $\theta=180^\circ$ is the
point closest to the active rotor. It is sheared weakly positively on the
rest of its perimeter. The inner arc dominates, and $\Gamma<0$.

At $G=0.10$ (Fig.~\ref{fig:streamlines}a) the negative shear at the gap is
four times stronger. It is, however, confined to a narrower arc
($147^\circ$--$214^\circ$) and flanked by strong positive peaks, and the outer
perimeter is sheared positively. The balance tips towards corotation
($\Gamma=+0.0042$). This is the planar counterpart of the geometric route,
in which ``the gears grind and lose purchase and the belt wins out''
\citep{Smith2026spin}.

At $G=0.70$ (Fig.~\ref{fig:streamlines}c) a separatrix isolates the active
rotor near the wall. The passive rotor lies inside a large recirculation of
opposite sense, and the shear on its gap-facing side is now positive. This is
the topological route, in which the inner shear still dominates but has
reversed sign.

At $G=0.30$ and $\mathrm{Re}=400$ (Fig.~\ref{fig:streamlines}d) the wall eddies
are displaced and strengthened, and the negative arc narrows to
$155^\circ$--$214^\circ$. It does not collapse, so the passive rotor still
counter-rotates. This agrees with the observation of \citet{Pan2026planar}
that, in the planar model, the gap-facing counter-rotating arc narrows with
$\mathrm{Re}$ but does not disappear.

\subsection{Discussion on  the zero-field benchmark}
Two independent planar formulations agree on every qualitative feature of the
hydrodynamic problem: the re-entrant gap route and the positions of its
two sign changes, the location of the strongest coupling, the Reynolds-driven
reversal at $G=0.22$, the absence of a reversal at $G=0.30$, and the corral-size
scaling of $|\Gamma|$.
Quantitatively the differences are about 1--8\% up to
$\mathrm{Re}\approx50$ (Table~\ref{tab:benchmark}). The fictitious-domain method of \citet{Pan2026planar}
is first-order accurate in the maximum norm, as shown by their own Wannier
validation, and it represents the circular boundaries through collocated
Lagrange multipliers. Since $\Gamma$ results from a small imbalance of
opposing shear contributions, it is sensitive to the representation of the
rotor surface. A difference of this size between the two methods is
therefore to be expected.

The one substantive difference is the magnitude of $\Gamma$ at high Reynolds
number along $G=0.30$: $-0.0068$ here against $-0.003$ in
\citet{Pan2026planar} at $\mathrm{Re}=400$. The value computed here does not change under
mesh refinement, so it is not a resolution artefact of the present
computation. Two explanations remain open. First, the first-order method may
lose accuracy at high $\mathrm{Re}$, where the wall eddies strengthen and the
shear layers thin. Second, the steady Newton iteration may converge to a
steady branch that differs from the attractor reached by time integration.
A time-dependent computation from rest, or a linear stability analysis of the
steady branch, would discriminate between the two explanations and is left for
future work.
Both computations nevertheless agree that the strictly planar flow does not
reproduce the experimental counter-to-corotation transition at
$\mathrm{Re}\approx55$ for $G=0.3$ \citep{Smith2026spin}. This independently
supports the conclusion of \citet{Pan2026planar} that finite-depth effects,
namely secondary motion and end-wall stresses, are responsible for that
transition. This is relevant for the magnetic problem, since an axial field
acts precisely through the end-wall (Hartmann) layers (\S\ref{sec:uniform}).

Since it has been established that the solver reproduces the hydrodynamic
benchmark, the zero-field results of this section are used as the reference
against which the magnetic effects are measured.

\subsection{Effect of the magnetic field: overview of the computations}\label{sec:magnetic}
The computations of this work form a database of 1\,526 continuation paths and
16\,744 converged steady states, obtained in about 3\,650 core-hours. Each path
starts from the zero-field solution at fixed geometry, Reynolds number and
heating, and along it the field strength is raised through a prescribed
sequence of levels (\S\ref{sec:numerics}). The paths cover 34 source configurations in
four groups (Table~\ref{tab:campaigns}). The first group holds the
eight single sources, a wire or an NdFeB rod at one of the positions B, T, L and R of
Fig.~\ref{fig:geometry}. The second holds the eight pairs, with the two sources at T
and B or at L and R, carrying opposite or equal currents (wires) or magnetised
parallel or antiparallel (rods). The third holds the sixteen crosses of four sources
at T, B, L and R, eight of wires and eight of rods, one for each arrangement of the
relative current senses or magnetisations that is distinct up to a global sign. The
fourth holds the two stationary magnet cores R1 and R2.
For every configuration the gap was varied over $G=0.1,0.2,\dots,0.7$ and the
Reynolds number over $\mathrm{Re}=10$, 20 and 40, with
$\Delta\bar T=10$\,K. At $G=0.3$ and $\mathrm{Re}=20$ the heating was also set
to 5 and 20\,K. The strength of each source was raised in 14 levels for the
wires ($NI=1$--500 ampere-turns, $\bar B_0=0.008$--4\,mT), 15 levels for the
rods ($\bar B_0=0.25$--16\,mT) and 21 levels for the cores (0.05--34\,mT at
17.5\,mm from the core). In the arrangements with several sources, each
source alone has this strength, and the fields are superposed. The
782 paths of the production and heating campaigns, which cover all 34
configurations, were computed on the finest mesh and reached 10\,515 of the
11\,638 requested levels. A further
528 paths for the eleven configurations of the first campaign (wire and rod at B, L and R, the three pairs T+B of Fig.~\ref{fig:Hconfigs} and the two cores), computed on the coarse, base, fine and finest
meshes, form the mesh study
(\S\ref{sec:meshcheck}), and the 216 zero-field paths repeat the benchmark of
\S\ref{sec:benchmark} on all four meshes. In 265 of these 782 paths the continuation stopped before the strongest level. These paths are
discussed in \S\ref{sec:limitations}. Below, each of them contributes the
levels it reached. Such a volume of results cannot be presented case by case.
The response of all configurations is therefore first summarised
(\S\ref{sec:map}), and representative cases, chosen for the clarity or
the size of their effect, are then presented: the single sources, which isolate the
role of the source position, and the flow fields that explain it
(\S\ref{sec:single}), the pairs and crosses, which show how sources combine
(\S\ref{sec:pairs}), the dependence on the gap and the Reynolds number
(\S\ref{sec:gapre}), the torque on the active rotor and the heat transfer
(\S\ref{sec:torquenu}), and the heating, together with the parameter that combines
it with the field strength (\S\ref{sec:heating}).

\begin{table}[t]
\centering\small
\caption{Magnetic computations (EFH1, $C=4.5$). $\bar B_0$: magnetic flux
density of one source at the reference point (corral centre, or 17.5\,mm from the axis for a
core). Crosses: senses of the currents or magnetisations at (T, B, L, R).
Last column: number of configurations, of strength levels and of
continuation paths.}
\label{tab:campaigns}
\footnotesize\setlength{\tabcolsep}{4pt}
\begin{tabular}{p{2.3cm}p{10.6cm}r}
\toprule
\multicolumn{3}{l}{\textit{Source configurations}}\\
Single sources & wire or rod at B, T, L or R & 8\\
Pairs & wires at T+B or L+R with opposite or equal currents, and rods at T+B or L+R magnetised parallel or antiparallel & 8\\
Crosses & four wires or four rods at T, B, L and R with senses (T, B, L, R) $=(+,+,+,+)$, $(-,+,+,+)$, $(+,-,+,+)$, $(+,+,-,+)$, $(+,+,+,-)$, $(+,+,-,-)$, $(+,-,+,-)$, $(+,-,-,+)$ & 16\\
Cores & R1 (core in $A$, $P$ heated) and R2 (core in $P$, $A$ heated) & 2\\
\midrule
\multicolumn{3}{l}{\textit{Strength levels}}\\
Wires & $NI=1$--500 ampere-turns, $\bar B_0=0.008$--4\,mT & 14\\
Rods & $\bar B_0=0.25$--16\,mT & 15\\
Cores & 0.05--34\,mT at 17.5\,mm from the core & 21\\
\midrule
\multicolumn{2}{l}{\textit{Campaigns}} & \textit{paths}\\
Production & all 34 configurations at $G=0.1$--0.7 and $\mathrm{Re}=10$, 20 and 40, with $\Delta\bar T=10$\,K, on the finest mesh & 714\\
Heating & all 34 configurations at $G=0.3$ and $\mathrm{Re}=20$, with $\Delta\bar T=5$ and 20\,K, on the finest mesh & 68\\
Mesh study & the 11 configurations of the first campaign\textsuperscript{a} at $G=0.1$--0.7 and $\mathrm{Re}=10$, 20 and 40, on the base and fine meshes & 462\\
 & the same at $G=0.3$, $\mathrm{Re}=20$ on the coarse and finest meshes, and with $\Delta\bar T=5$, 20\,K on the base and fine meshes & 66\\
Zero field & the 54 cases of \S\ref{sec:benchmark} on four meshes & 216\\
\midrule
Total & 16\,744 converged steady states & 1\,526\\
\bottomrule
\multicolumn{3}{@{}p{13.5cm}@{}}{\footnotesize\textsuperscript{a}\,wire and rod at B, L and R, the wire pairs T+B (opposite and equal currents), the parallel rod pair T+B, and R1, R2.}\\
\end{tabular}
\end{table}

\paragraph{Quantities and accuracy criterion.} Three quantities are reported
for every steady state: the gear ratio $\Gamma$, which measures the
transmission of rotation to the passive rotor, the torque $\mathcal T_A$ on the
active rotor, whose magnitude is the power needed to drive it, and the Nusselt
number $\mathrm{Nu}$ of the heated rotor.
The magnetic effect is measured against the zero-field values $\Gamma_0$,
$\mathcal T_{A,0}$ and $\mathrm{Nu}_0$ of the same path. The coupling is termed
\emph{strengthened} when $|\Gamma|$ increases at unchanged sign,
\emph{weakened} when $|\Gamma|$ decreases, and \emph{reversed} when $\Gamma$
changes sign. A sign change that leaves $|\Gamma|<0.002$ is counted as
weakened, not as a reversal. Each state is also checked against the global energy balance of
\S\ref{sec:torqueflux}: the heat released by the heated rotor must leave
through the corral wall. In strong fields the thermomagnetic flow produces
very thin thermal layers at the part of the corral wall nearest to the
sources. On the finest mesh the balance error exceeds 1\% in 2\,063 of the
10\,515 states: 1\,270 states of the crosses of four rods (from 2\,mT), 655
of the other rod arrangements (from 2.5\,mT, mostly above 4\,mT) and 138 of
the cores (from 5\,mT). For the wires it stays below 1\% throughout. Where the error is below 1\%, the fine and finest meshes give the same
$\Gamma$ to within $10^{-3}$ in 99\% of 2\,966 comparisons (median difference
$2\times10^{-6}$, \S\ref{sec:meshcheck}). The large differences between meshes
occur only where the balance fails. For all quantitative statements below,
the 8\,452 states with a balance error below 1\% are therefore used. In the
figures, states that fail the criterion are shown as open symbols. The
criterion serves to exclude unresolved states. It is not by itself a proof of
accuracy (\S\ref{sec:torqueflux}). The accuracy of the accepted states rests
on the comparison of the fine and finest meshes as demonstrated at \S\ref{sec:meshcheck}.

\subsection{Response of all configurations}\label{sec:map}
\begin{figure}[p]
\centering
\includegraphics[width=\textwidth]{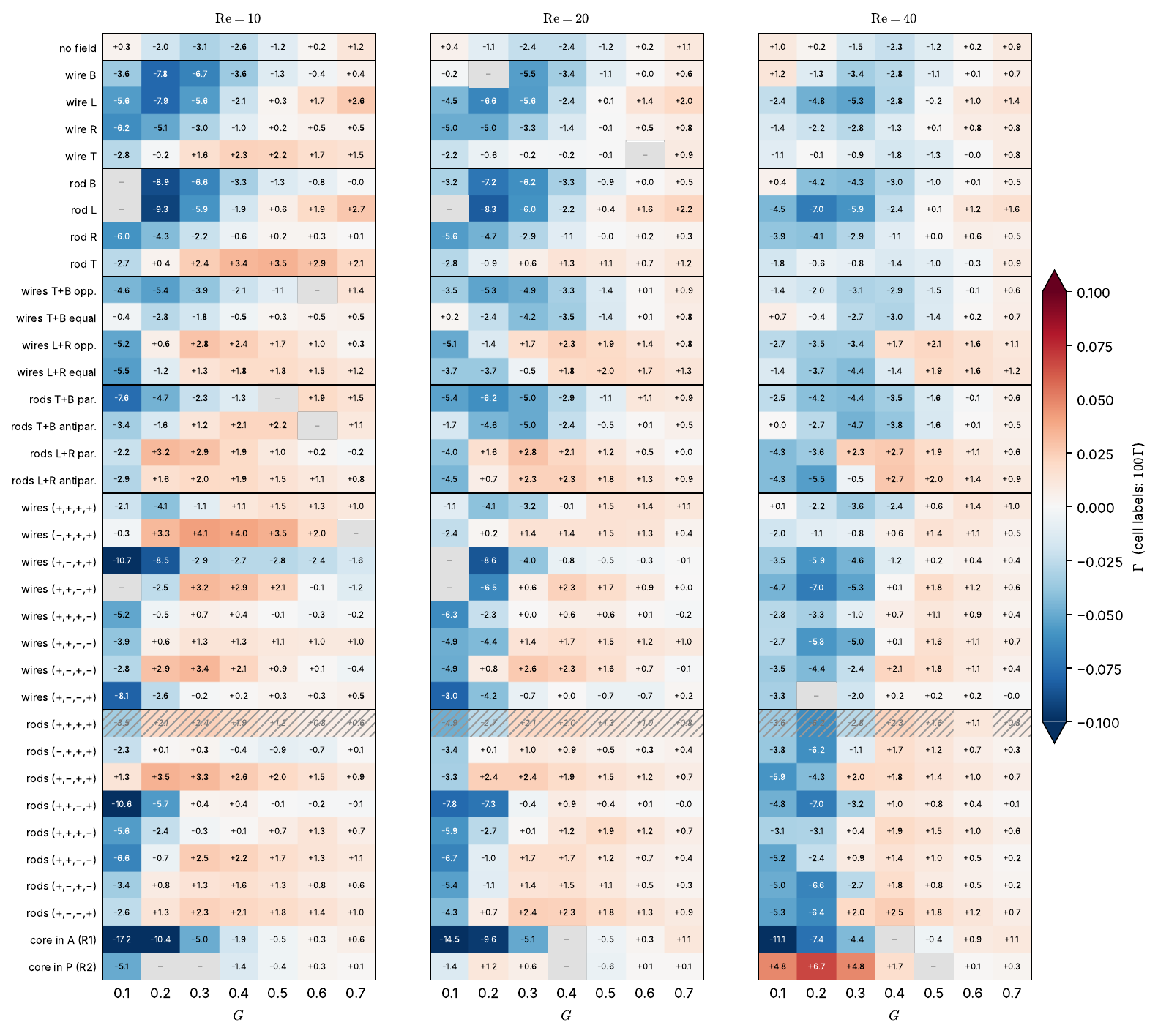}
\caption{Gear ratio at $\bar B_0=2$\,mT for all 34 configurations, the seven
gaps and the three Reynolds numbers ($\Delta\bar T=10$\,K, finest mesh). Cell
labels: $100\,\Gamma$. Top row: zero field. Blue: counter-rotation. Red:
corotation. Hatched, grey labels: energy-balance error above 1\%. Grey cells:
level not reached by the continuation. Crosses: senses at (T, B, L, R).}
\label{fig:overview}
\end{figure}

The values of $\Gamma$ at $\bar B_0=2$\,mT for all configurations, gaps and
Reynolds numbers are presented in Fig.~\ref{fig:overview}, together with the
zero-field values. The strength 2\,mT is reached by almost every path, and the energy
balance is satisfied there for all configurations except the cross of four
parallel rods. Of the 714 cases (configuration, $G$, $\mathrm{Re}$) with
$\Delta\bar T=10$\,K, 675 have a valid state at this strength. With respect to
the zero-field value, the coupling is reversed in 290 of them, strengthened in
171 and weakened in 152, and $\Gamma$ changes by less than 0.002 in 62.
The response of every configuration at the experimental gap $G=0.3$ and
$\mathrm{Re}=20$, where $\Gamma_0=-0.0236$, is listed in Table~\ref{tab:summary}.

The following general features are apparent from the map. The
response is set mainly by the arrangement of the sources. Wires and rods at the same
positions give the same trends, the rods acting at somewhat lower $\bar B_0$, since
their field decays faster and its gradient is steeper near the wall. Reversal is the
most frequent response. In 162 of the 290 reversed cases counter-rotation becomes
corotation. In the other 128 the weak corotation at the smallest or the widest gaps
becomes counter-rotation. At the experimental gap (Table~\ref{tab:summary})
corotation results from a single source at T, the pairs L+R, most crosses and R2,
whereas single sources at B and L, the pairs T+B (up to about 2\,mT) and R1
strengthen the counter-rotation.

The strongest counter-rotation is produced by the core in $A$ (R1) at small gaps:
$\Gamma=-0.145$ at $G=0.1$, $\mathrm{Re}=20$ and 2\,mT, and $\Gamma=-0.221$ at
$\mathrm{Re}=10$ and 6.5\,mT. The latter is seven times the largest hydrodynamic
gear ratio, $|\Gamma_0|=0.031$, found at $G=0.3$ and $\mathrm{Re}=10$. At the same
gap the zero-field rotors corotate weakly ($\Gamma_0=+0.003$). The strongest
corotation is produced by a rod at T at $\mathrm{Re}=10$ and 14\,mT:
$\Gamma=+0.118$ at $G=0.1$ and $+0.107$ at $G=0.3$, against $\Gamma_0=+0.003$ and
$-0.031$. At fixed field the effect decreases with $\mathrm{Re}$, since
$\mathrm{Mn_F}\propto\mathrm{Re}^{-2}$ (\S\ref{sec:parameters}). At wide gaps,
$G\ge0.5$, the change in $\Gamma$ at 2\,mT is on average 2.5 to 7 times smaller
than at $G=0.3$ (mean $|\Gamma-\Gamma_0|$ of 0.005--0.012 for the four groups,
against 0.025--0.034), although individual configurations still change $\Gamma$
there by up to 0.047.

\begin{table}[p]
\centering\small
\caption{Response at the experimental gap $G=0.3$, $\mathrm{Re}=20$,
$\Delta\bar T=10$\,K ($\Gamma_0=-0.0236$). $\Gamma$ at $\bar B_0\approx1$
(0.8\,mT for wires), 2 and 4\,mT. Most negative and most positive $\Gamma$
over the valid states, with $\bar B_0$ in mT in parentheses. Torque on the
active rotor and Nusselt number relative to zero field at 2\,mT. Italics:
energy-balance error above 1\%. --: level not reached.}
\label{tab:summary}
\footnotesize\setlength{\tabcolsep}{3.5pt}
\begin{tabular}{lrrrrrrr}
\toprule
Configuration & $\Gamma$ (1\,mT) & $\Gamma$ (2\,mT) & $\Gamma$ (4\,mT) & $\Gamma_{\min}$ ($\bar B_0$) & $\Gamma_{\max}$ ($\bar B_0$) & $\mathcal T_A/\mathcal T_{A,0}$ & $\mathrm{Nu}/\mathrm{Nu}_0$\\
\midrule
wire B & $-$0.034 & $-$0.055 & $-$0.059 & $-$0.061\,(3.2) & $-$0.024\,(0.008) & 1.03 & 1.22\\
wire L & $-$0.036 & $-$0.056 & $-$0.062 & $-$0.062\,(4) & $-$0.024\,(0.008) & 1.21 & 2.07\\
wire R & $-$0.030 & $-$0.033 & $-$0.023 & $-$0.033\,(1.6) & $-$0.023\,(4) & 1.01 & 1.06\\
wire T & $-$0.019 & $-$0.002 & +0.019 & $-$0.024\,(0.008) & +0.019\,(4) & 1.02 & 1.07\\
\midrule
rod B & $-$0.047 & $-$0.062 & $-$0.052 & $-$0.062\,(2.5) & $-$0.002\,(10) & 1.07 & 1.38\\
rod L & $-$0.053 & $-$0.060 & $-$0.063 & $-$0.063\,(4) & $-$0.026\,(0.25) & 1.38 & 2.97\\
rod R & $-$0.032 & $-$0.029 & $-$0.012 & $-$0.032\,(1) & +0.005\,(8) & 1.01 & 1.08\\
rod T & $-$0.013 & +0.006 & +0.025 & $-$0.023\,(0.25) & +0.068\,(12) & 1.07 & 1.21\\
\midrule
wires T+B opp. & $-$0.030 & $-$0.049 & $-$0.025 & $-$0.050\,(2.4) & $-$0.024\,(0.008) & 1.06 & 1.25\\
wires T+B equal & $-$0.028 & $-$0.042 & $-$0.020 & $-$0.049\,(2.8) & $-$0.020\,(4) & 1.01 & 1.08\\
wires L+R opp. & $-$0.026 & +0.017 & +0.025 & $-$0.026\,(0.4) & +0.027\,(3.2) & 1.26 & 2.07\\
wires L+R equal & $-$0.031 & $-$0.005 & +0.022 & $-$0.032\,(1.2) & +0.022\,(4) & 1.18 & 1.70\\
\midrule
rods T+B par. & $-$0.043 & $-$0.050 & -- & $-$0.051\,(1.5) & $-$0.025\,(0.25) & 1.14 & 1.47\\
rods T+B antipar. & $-$0.040 & $-$0.050 & +0.024 & $-$0.051\,(1.5) & +0.024\,(4) & 1.07 & 1.27\\
rods L+R par. & +0.008 & +0.028 & \textit{+0.019} & $-$0.028\,(0.25) & +0.028\,(2) & 1.44 & 2.91\\
rods L+R antipar. & $-$0.004 & +0.023 & \textit{+0.023} & $-$0.030\,(0.5) & +0.024\,(2.5) & 1.38 & 2.51\\
\midrule
wires $(+,+,+,+)$ & $-$0.030 & $-$0.032 & $-$0.014 & $-$0.032\,(1.6) & $-$0.014\,(4) & 1.09 & 1.33\\
wires $(-,+,+,+)$ & $-$0.014 & +0.014 & +0.038 & $-$0.024\,(0.008) & +0.038\,(4) & 1.26 & 1.58\\
wires $(+,-,+,+)$ & $-$0.047 & $-$0.040 & $-$0.021 & $-$0.052\,(1.2) & $-$0.018\,(3.2) & 1.24 & 2.04\\
wires $(+,+,-,+)$ & $-$0.045 & +0.006 & +0.037 & $-$0.047\,(1.2) & +0.037\,(4) & 1.35 & 2.28\\
wires $(+,+,+,-)$ & $-$0.027 & +0.000 & +0.004 & $-$0.027\,(0.8) & +0.005\,(3.2) & 1.21 & 1.73\\
wires $(+,+,-,-)$ & $-$0.034 & +0.014 & +0.020 & $-$0.038\,(1.2) & +0.020\,(4) & 1.29 & 1.85\\
wires $(+,-,+,-)$ & $-$0.030 & +0.026 & +0.025 & $-$0.030\,(0.8) & +0.029\,(2.8) & 1.29 & 2.23\\
wires $(+,-,-,+)$ & $-$0.032 & $-$0.007 & +0.001 & $-$0.032\,(0.8) & +0.001\,(4) & 1.30 & 1.72\\
\midrule
rods $(+,+,+,+)$ & $-$0.023 & \textit{+0.021} & \textit{+0.032} & $-$0.031\,(0.5) & +0.014\,(1.5) & \textit{1.40} & \textit{2.24}\\
rods $(-,+,+,+)$ & $-$0.001 & +0.010 & \textit{+0.008} & $-$0.028\,(0.5) & +0.010\,(2) & 1.49 & 2.85\\
rods $(+,-,+,+)$ & +0.013 & +0.024 & \textit{+0.030} & $-$0.030\,(0.25) & +0.024\,(2) & 1.51 & 2.70\\
rods $(+,+,-,+)$ & $-$0.035 & $-$0.004 & \textit{+0.003} & $-$0.036\,(0.5) & $-$0.003\,(10) & 1.44 & 2.58\\
rods $(+,+,+,-)$ & $-$0.010 & +0.001 & \textit{$-$0.007} & $-$0.029\,(0.5) & +0.001\,(2) & 1.25 & 1.66\\
rods $(+,+,-,-)$ & $-$0.001 & +0.017 & \textit{+0.023} & $-$0.027\,(0.25) & +0.025\,(10) & 1.49 & 2.64\\
rods $(+,-,+,-)$ & $-$0.028 & +0.014 & \textit{+0.017} & $-$0.031\,(0.5) & +0.014\,(2) & 1.46 & 2.45\\
rods $(+,-,-,+)$ & +0.004 & +0.024 & \textit{+0.023} & $-$0.034\,(0.5) & +0.024\,(2) & 1.44 & 2.55\\
\midrule
core in $A$ (R1) & $-$0.046 & $-$0.051 & $-$0.046 & $-$0.051\,(2.5) & $-$0.024\,(0.05) & 1.00 & 1.55\\
core in $P$ (R2) & +0.022 & +0.006 & -- & $-$0.023\,(0.05) & +0.022\,(1) & 1.28 & 1.93\\
\bottomrule
\end{tabular}
\end{table}

\subsection{Single sources: the role of the source position}\label{sec:single}
\begin{figure}[t]
\centering
\includegraphics[width=\textwidth]{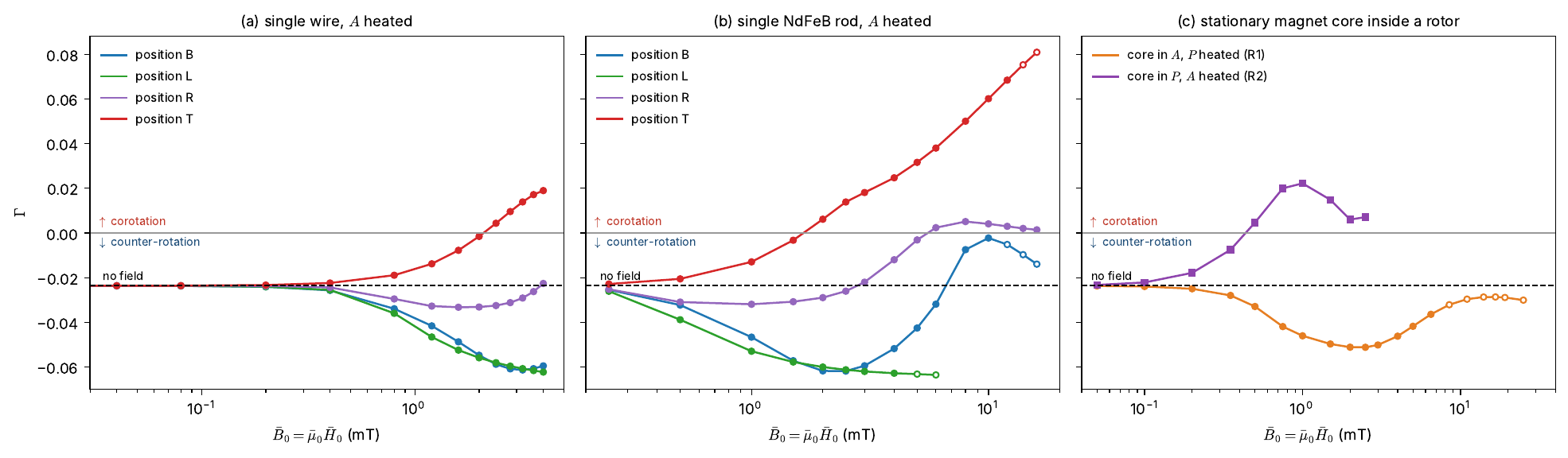}
\caption{Gear ratio at $G=0.3$, $\mathrm{Re}=20$, $\Delta\bar T=10$\,K as a
function of the magnetic flux density $\bar B_0$ at the reference point. (a) Single wire and
(b) single NdFeB rod at the positions B, L, R and T of
Fig.~\ref{fig:geometry}, $A$ heated. (c) Stationary core inside a rotor.
Dashed: zero field. Open symbols: energy-balance error above 1\%.}
\label{fig:single}
\end{figure}

In Fig.~\ref{fig:single}, $\Gamma$ at $G=0.3$, $\mathrm{Re}=20$ is plotted as
the strength of a single source is increased. For weak fields the change is
small. For the wires at B, L and T it grows as $\bar B_0^{2}$ up to about
0.2\,mT, as expected from the linear (Curie) part of the Langevin law. For
stronger fields the four positions behave differently.
At position B, below the corral, counter-rotation is strengthened. The
wire gives $\Gamma=-0.061$ at 3.2\,mT and the rod $-0.062$ at 2.5\,mT, 2.6 times
$\Gamma_0$. For the rod the response then reverses its trend: $|\Gamma|$ falls to
0.002 at 10\,mT, so that the passive rotor almost stops. At position L, behind $A$,
counter-rotation is strengthened monotonically, to $\Gamma=-0.062$ for the wire and
$-0.063$ for the rod at 4\,mT. At position R, behind $P$, the effect is weaker.
$|\Gamma|$ rises to 0.033 for the wire (1.6\,mT) and 0.032 for the rod (1\,mT) and
then decreases. For the rod, $\Gamma$ changes sign at 5.6\,mT and stays
small and positive, at most $0.005$, up to 16\,mT. At position T, above the corral,
the coupling is reversed. $\Gamma$ crosses zero at 2.1\,mT for the wire and at
1.7\,mT for the rod, and the rod gives $\Gamma=+0.068$ at 12\,mT, i.e.\ corotation
three times as strong as the zero-field counter-rotation.
The cores (Fig.~\ref{fig:single}c) show the two opposite responses within one
geometry. With the core in $A$ and $P$ heated (R1), counter-rotation is
strengthened to $\Gamma=-0.051$ at 2--2.5\,mT, and $|\Gamma|$ then decreases
to 0.036 at 6.5\,mT. With the core in $P$ and $A$ heated (R2), $\Gamma$ changes sign at
0.44\,mT and reaches $+0.022$ at 1\,mT. The R2 path stopped at 2.5\,mT.

\begin{figure}[p]
\centering
\includegraphics[width=0.95\textwidth]{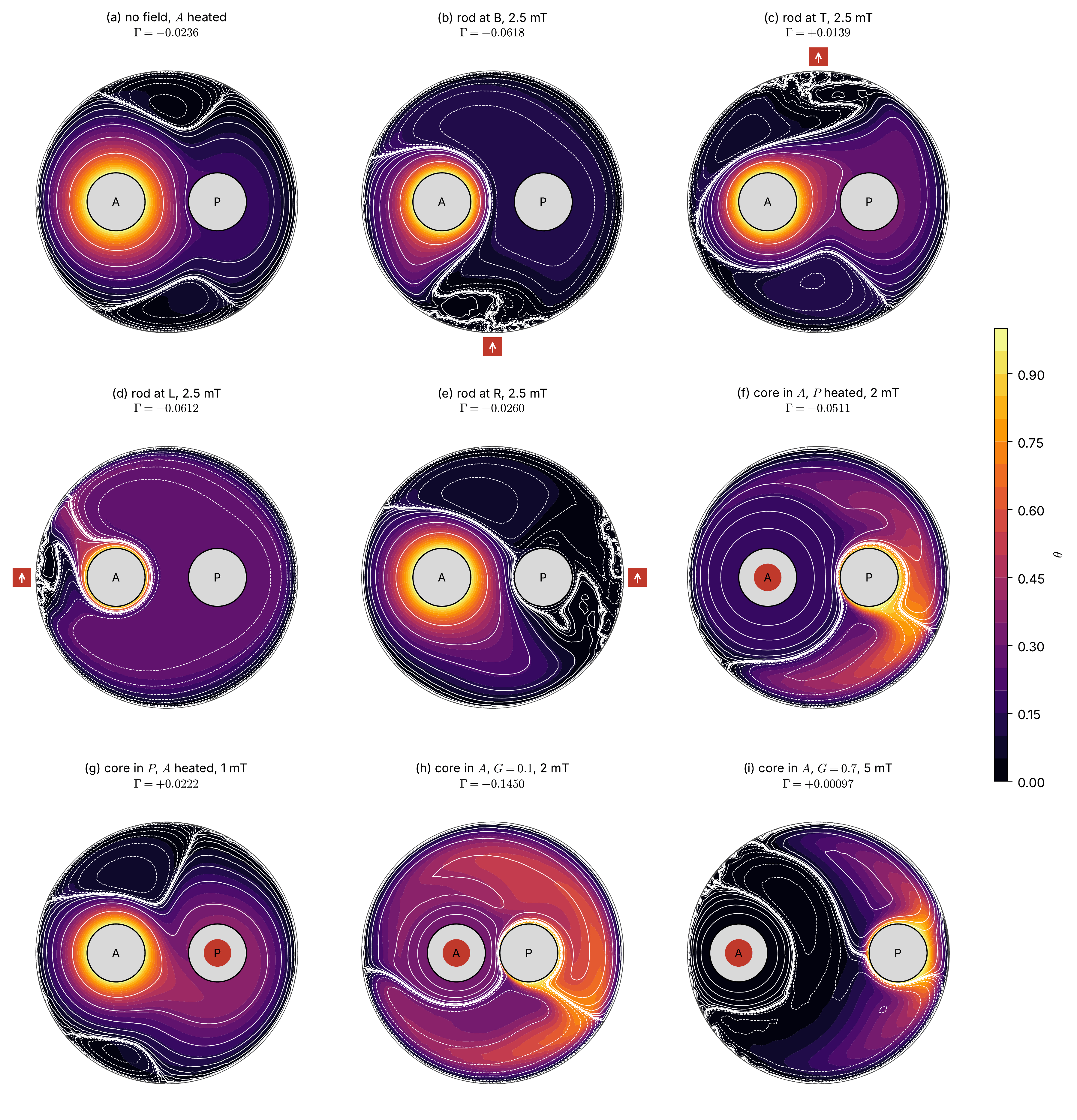}
\caption{Temperature $\theta$ (colour) and streamlines (white, solid for the
counter-clockwise circulation, the sense of $A$, and dashed for clockwise),
$\mathrm{Re}=20$, $\Delta\bar T=10$\,K, $G=0.3$ unless stated. Red squares:
rods, with the arrow along the magnetisation (drawn at the corral wall, while the
rods lie at $d_s=6$). Red discs: cores. (a)~No field. (b)--(e)~Rod at B, T,
L and R, 2.5\,mT. (f)~Core in $A$, $P$ heated, 2\,mT. (g)~Core in $P$, $A$
heated, 1\,mT. (h),~(i)~Core in $A$ at $G=0.1$ (2\,mT) and $G=0.7$ (5\,mT).
Fields from the base mesh. $\Gamma$ from the finest mesh (the two differ by
less than $7\times10^{-4}$).}
\label{fig:magfields}
\end{figure}

\paragraph{Why the position matters.} As noted in \S\ref{sec:final}, the
vorticity source~\eqref{eq:fhdsource} pushes fluid that is hotter, and
therefore less magnetised, away from the field source. Its effect on the
passive rotor is governed by symmetry. The field is symmetric about the line
of centres for a single source at L or R, for all the pairs, and for the
crosses with equal senses at T and B. If the temperature field were also
symmetric about that line, as it is for conduction from the heated rotor
alone, the source~\eqref{eq:fhdsource} would be antisymmetric and would drive
a pair of counter-rotating cells that exert no net torque on the passive
rotor. In the present problem, however, the thermal boundary layer of the
heated rotor is swept around it by the rotation of the active rotor
($\mathrm{Pe}=\mathrm{Re}\,\mathrm{Pr}\approx10^{3}$), and the temperature
field is asymmetric. A net torque is therefore possible also for symmetric
fields. The symmetry argument only predicts that the effect must vanish as
$\mathrm{Pe}\to0$. The computations show that symmetric fields act as
strongly as asymmetric ones: at the experimental gap the source at L gives
the largest strengthening of all single sources (Table~\ref{tab:summary}).
For a source at B or T the field has no such symmetry. Reflection about the
line of centres maps position B onto T, but it also reverses the sense of
rotation of $A$.
The two problems are therefore not equivalent, and the difference between
them is the clearest indication of how the field acts. The active rotor turns
counter-clockwise, so on its side facing the gap the fluid moves upwards, from
the B side towards the T side, and drags the facing side of $P$ in the same
direction. This is the gear-like counter-rotation. From the temperature and
streamline fields of Fig.~\ref{fig:magfields} the following is observed.
 Without a field (Fig.~\ref{fig:magfields}a), the heat is held in an
annulus around $A$ by the closed streamlines that surround it. A counter-clockwise
circulation encloses both rotors, and two clockwise eddies are attached to the corral
wall above and below the gap. With a rod at B (Fig.~\ref{fig:magfields}b), the warm
fluid around $A$ is displaced upwards, away from the rod, which reinforces the upward
stream on the gap side of $A$. A single clockwise cell now fills the right half and
the bottom of the corral and encloses $P$, which is driven in the counter-rotating
sense: $\Gamma=-0.062$. With a rod at T (Fig.~\ref{fig:magfields}c), the warm fluid
is pushed downwards, against the upward stream in the gap. The counter-clockwise
circulation that encloses both rotors expands, carries warm fluid from $A$ across to
$P$, and confines the clockwise eddies to the corral wall. The passive rotor is now
driven by this belt-like outer circulation and corotates: $\Gamma=+0.014$ at the
same strength. With a rod at L (Fig.~\ref{fig:magfields}d), the thermal layer of $A$
becomes very thin and the heat leaves as a plume along the wall near the rod. As at
B, one clockwise cell encloses $P$, and $\Gamma=-0.061$. With a rod at R
(Fig.~\ref{fig:magfields}e), the fluid next to the rod, behind $P$, stays cold, and
the warm fluid from $A$ is turned downwards before it reaches $P$. The passive rotor
lies on the boundary between the circulation around $A$ and a clockwise cell on its
far side, and $\Gamma$ changes little ($-0.026$). 
In every case the change in $\Gamma$ goes together with a change in the
topology of the streamlines around $P$: a field that encloses $P$ in a cell of
its own strengthens the gear-like coupling, and a field that lets the outer
circulation enclose both rotors produces belt-like corotation. This is the
competition between inner and outer shear identified by
\citet{Smith2026spin}, now controlled by the position of the source.

The cores act in the same way (Fig.~\ref{fig:magfields}f--i). With R1 the heat
leaves $P$ as a plume carried along the corral wall, away from the core in
$A$, and $A$ is surrounded by closed streamlines. $P$ counter-rotates
($\Gamma=-0.051$). With R2 the warm fluid from $A$ is carried across the gap
towards $P$ inside a circulation that encloses both rotors, and $P$ corotates
($\Gamma=+0.022$). At $G=0.1$ the plume from $P$ fills most of the corral and
drives $P$ against $A$ ($\Gamma=-0.145$). At $G=0.7$ the passive rotor lies
near the wall, on the boundary between the cold closed region around $A$ and a
clockwise cell, and the two shears on it nearly cancel: $\Gamma$ falls from
$0.011$ to $0.001$ at 5\,mT and to $-0.0004$ at 25\,mT. The field then
practically decouples the passive rotor. No magnetic torque acts on the rotors
in any of these cases: the cores are stationary inside non-magnetic sleeves,
and $H$ is axisymmetric about each core.

\subsection{Pairs and crosses: how sources combine}\label{sec:pairs}
\begin{figure}[t]
\centering
\includegraphics[width=\textwidth]{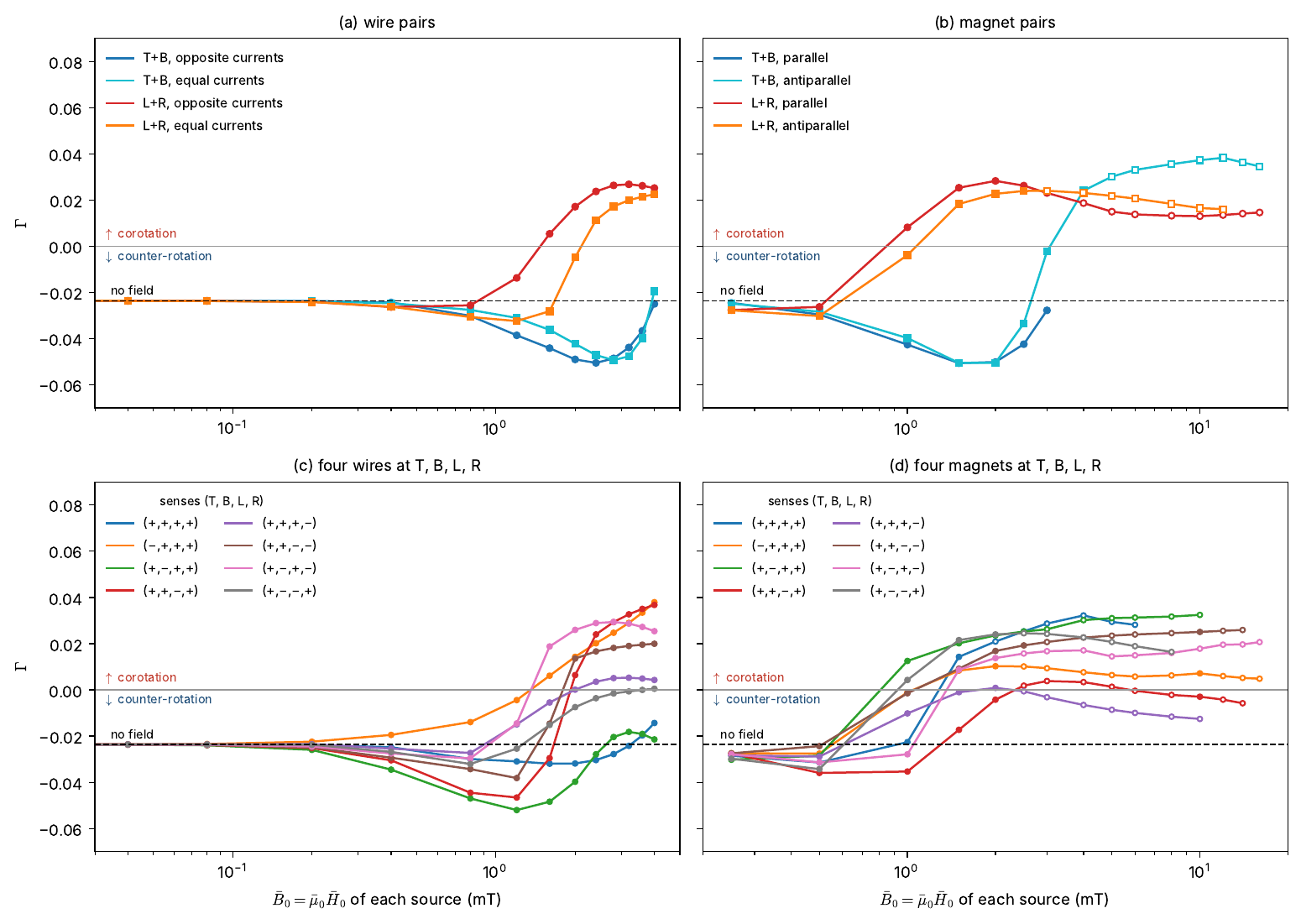}
\caption{Gear ratio at $G=0.3$, $\mathrm{Re}=20$, $\Delta\bar T=10$\,K for
(a)~wire pairs, (b)~rod pairs, (c)~crosses of four wires and (d)~crosses of
four rods. $\bar B_0$ is the magnetic flux density of one source at the
corral centre. Senses
of the currents or magnetisations at (T, B, L, R) as in
Table~\ref{tab:campaigns}. Dashed: zero field. Open symbols: energy-balance
error above 1\%.}
\label{fig:pairs}
\end{figure}

The pairs and the crosses at $G=0.3$, $\mathrm{Re}=20$ are shown in
Fig.~\ref{fig:pairs}. Their magnetic field strength intensity is shown in
Figs.~\ref{fig:Hconfigs} and~\ref{fig:Hcrosses}. The main result is that the effects of the sources do not add.
The Kelvin force depends on the magnitude $H$ of the superposed magnetic field
strength intensity, and the
source~\eqref{eq:fhdsource} on its gradient, so two sources that each
strengthen the coupling can reverse it together.
For the pairs T+B, both wire pairs and both rod pairs strengthen the
counter-rotation, to $\Gamma\approx-0.05$ at 1.5--2.8\,mT, i.e.\ about twice
$\Gamma_0$, and then weaken it. The antiparallel rods reverse the coupling at
3.1\,mT and reach $+0.024$ at 4\,mT. For the pairs L+R, each of the single sources
at L and at R strengthens the counter-rotation, yet every L+R pair reverses it: at
1.5 and 2.1\,mT for the wires with opposite and equal currents, and at 0.88 and
1.1\,mT for the parallel and antiparallel rods. The corotation that results,
$\Gamma=+0.022$ to $+0.028$, is as strong as the zero-field counter-rotation.
Of the crosses, five of the eight wire crosses and six of the eight rod crosses
reverse the coupling, between 0.8 and 2\,mT. (The rod cross $(+,+,+,-)$ also crosses
zero, near 1.8\,mT, but $|\Gamma|$ stays below 0.001 and the sign reverts above
2\,mT. By the convention above it is a decoupling, not a reversal.) The wire cross
$(+,-,+,+)$ strengthens the counter-rotation to $\Gamma=-0.052$ at 1.2\,mT. The
wire cross $(+,-,-,+)$ drives $\Gamma$ to zero: $|\Gamma|<0.002$ between 350 and
500\,A-turns (2.8--4\,mT). The rod cross $(+,+,+,-)$ does the same at 1.5--2\,mT.
For most crosses of four rods the energy balance fails above about 2\,mT
(Fig.~\ref{fig:pairs}d).

\begin{figure}[p]
\centering
\includegraphics[width=\textwidth]{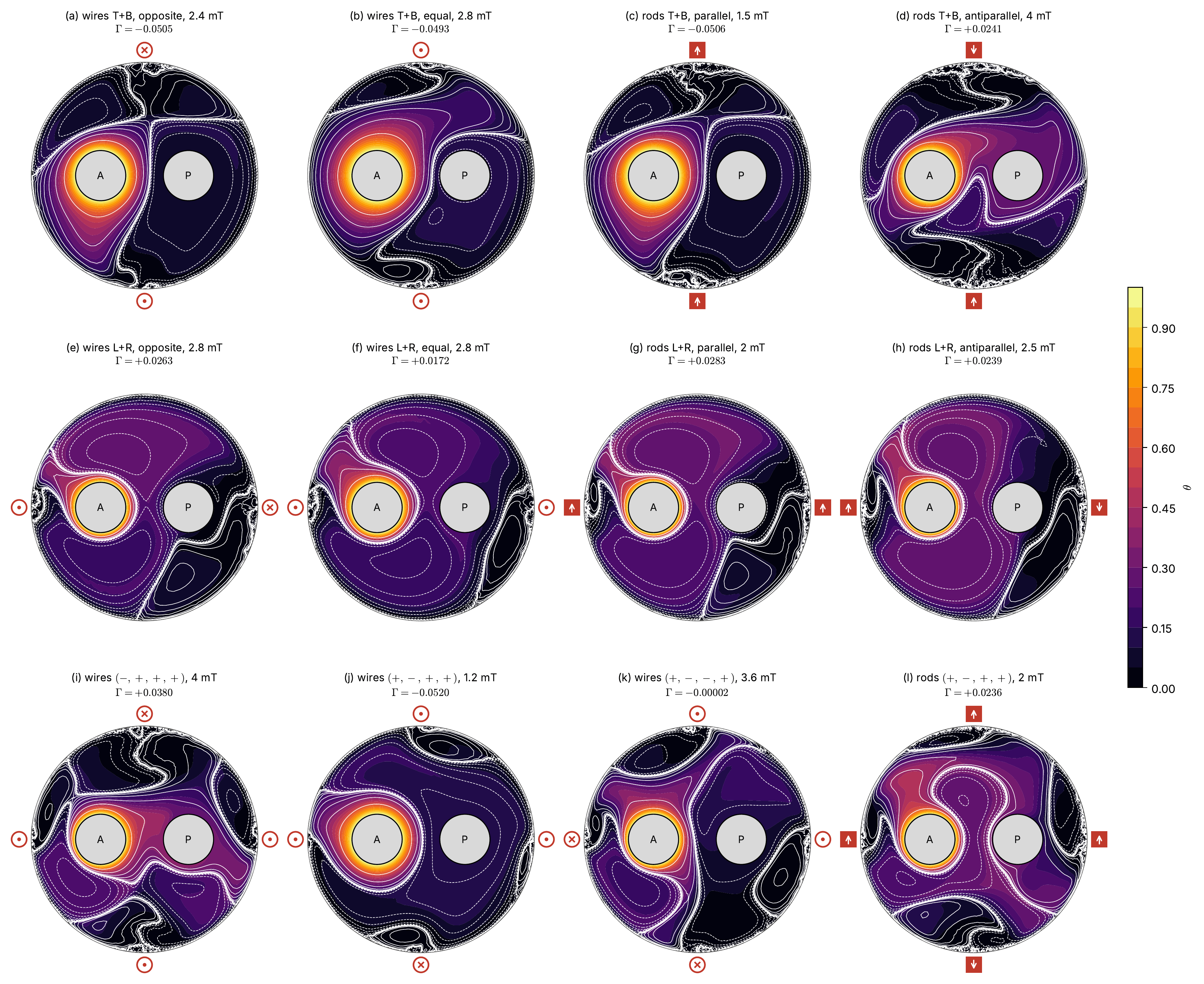}
\caption{Temperature $\theta$ (colour) and streamlines (white, solid for the
counter-clockwise circulation, dashed for clockwise) for the pairs and crosses,
$G=0.3$, $\mathrm{Re}=20$, $\Delta\bar T=10$\,K. Wires: $\odot$ positive and
$\otimes$ negative current. Rods: red squares with the arrow along the
magnetisation, drawn at the corral wall (wires lie at $d_s=5$, rods at
$d_s=6$). Top row: pairs T+B, (a)~wires with opposite and (b)~equal
currents, (c)~parallel and (d)~antiparallel rods. Middle row: the same for
the pairs L+R, (e)--(h). Bottom row: crosses with senses (T, B, L, R),
(i)--(k)~wires, (l)~rods. $\bar B_0$ is the magnetic flux density of one
source. Fields from
the base mesh, except (b), (d), (g)--(i), (k), (l) from the fine mesh.
$\Gamma$ from the finest mesh (the two differ by less than
$4\times10^{-4}$).}
\label{fig:magfieldsmulti}
\end{figure}

The temperature and the streamlines for all eight
pairs and for four crosses, at $G=0.3$, $\mathrm{Re}=20$ and at a strength
where each effect is well developed, are presented in
Fig.~\ref{fig:magfieldsmulti}. Together with Fig.~\ref{fig:magfields}, the patterns fall into
four classes.
\begin{itemize}
  \item \emph{Isolation of $P$ (gear-like).} A separatrix between the rotors
    confines the heated annulus of $A$ to the left half of the corral, and
    $P$ turns inside one large clockwise cell. This is the pattern of the
    source at B or L, of the three pairs T+B that strengthen the coupling
    (Fig.~\ref{fig:magfieldsmulti}a--c, $\Gamma\approx-0.05$), and of the wire
    cross $(+,-,+,+)$ (j, $\Gamma=-0.052$).
  \item \emph{Enclosure of both rotors (belt-like).} A counter-clockwise
    circulation carries warm fluid from $A$ past $P$ and encloses both
    rotors, while the clockwise eddies are pushed to the corral wall. This is
    the pattern of the source at T, of the antiparallel rods T+B at 4\,mT
    (d, $\Gamma=+0.024$), of the wire cross $(-,+,+,+)$ (i, $\Gamma=+0.038$)
    and of the core R2.
  \item \emph{Reversal of the gap flow (topological).} For all four pairs L+R
    (e--h) and for the rod cross $(+,-,+,+)$ (l), the thermal layer of $A$
    becomes thin, its heat leaves as a plume towards the upper wall on the
    side of $A$, and a large clockwise cell forms above the line of centres.
    This cell drives the fluid between the rotors downwards, against the
    stream of $A$. The gap-facing side of $P$ is therefore swept in the
    reversed direction, and $P$ corotates ($\Gamma=+0.017$ to $+0.028$). This
    is the field-induced counterpart of the topological route of
    \citet{Smith2026spin}, in which an intervening recirculation reverses
    the inner shear. It is not the belt-like mode of the source at T.
  \item \emph{Balance (decoupling).} With the wire cross $(+,-,-,+)$ at
    3.6\,mT (k) the circulation of $A$ is confined to the left of a separatrix
    that runs from the upper to the lower wall. Cells of opposite sense
    adjoin $P$, the shears on its surface cancel, and $\Gamma=-0.00002$. The
    core R1 at $G=0.7$ (Fig.~\ref{fig:magfields}i) gives the same result.
\end{itemize}
The two corotating classes therefore reverse the coupling by different
mechanisms. The field reproduces, at fixed geometry and Reynolds number, both
the belt-like and the topological route to corotation that
\citet{Smith2026spin} found by changing the Reynolds number or the gap.

\subsection{Effect of the gap and of the Reynolds number}\label{sec:gapre}
\begin{figure}[t]
\centering
\includegraphics[width=\textwidth]{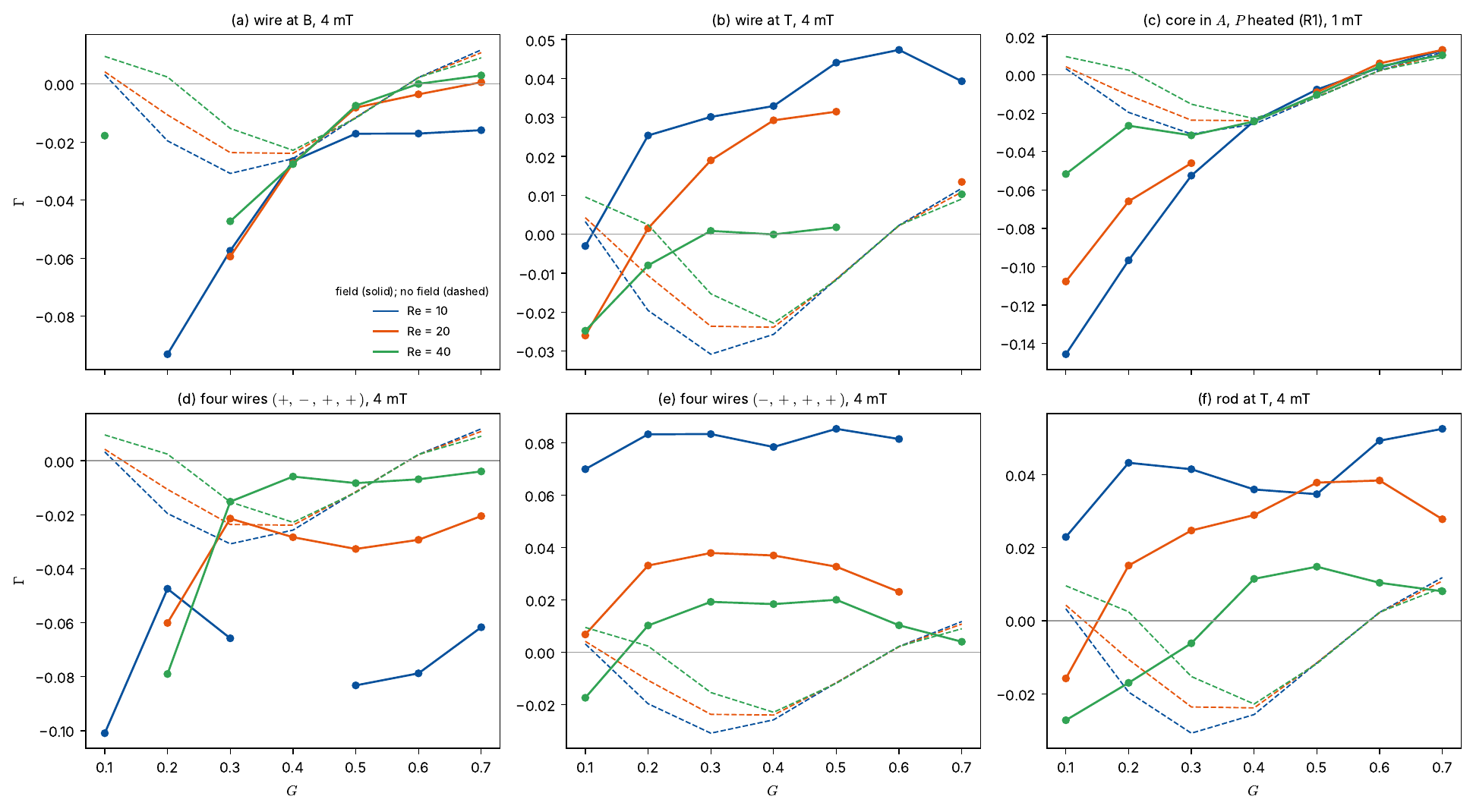}
\caption{Gap transects at $\mathrm{Re}=10$, 20 and 40 ($\Delta\bar T=10$\,K)
for six configurations at the strengths indicated. Solid lines with symbols:
with field. Dashed: zero field (same colour). Open symbols: energy-balance
error above 1\%. Missing points: level not reached by the continuation.}
\label{fig:gap}
\end{figure}

Gap transects at the three Reynolds numbers are shown in Fig.~\ref{fig:gap}
for six representative configurations: two single wires (B, T), the core R1, the two
crosses with the largest effects of opposite sign, and the rod at T.
For the wire at B at 4\,mT (Fig.~\ref{fig:gap}a), the
counter-rotating band is deepened, most strongly at $G=0.2$, where $\Gamma=-0.093$
at $\mathrm{Re}=10$ (against $\Gamma_0=-0.020$). At $\mathrm{Re}=10$ the wide-gap
corotation at $G=0.6$ and 0.7 is also replaced by counter-rotation, and
$\Gamma\approx-0.017$ for $G=0.5$--0.7. At $\mathrm{Re}=20$ and 40 the wide-gap
values move towards zero, and at $G=0.6$, $\mathrm{Re}=20$ the sign changes
($\Gamma=-0.0035$).

For the wire at T at 4\,mT (Fig.~\ref{fig:gap}b), the re-entrant gap route of
\S\ref{sec:benchmark} is inverted. At $\mathrm{Re}=10$ the passive rotor corotates
at every gap from 0.2 to 0.7, with $\Gamma$ up to $+0.047$. At the smallest gap, by
contrast, the zero-field corotation becomes counter-rotation ($\Gamma=-0.026$ at
$\mathrm{Re}=20$, against $\Gamma_0=+0.004$). At $\mathrm{Re}=40$ the field brings
$\Gamma$ close to zero, $|\Gamma|<0.002$, for $G=0.3$--0.5.

For the core in $A$ at 1\,mT (Fig.~\ref{fig:gap}c), the effect is confined to small
gaps, where the heated passive rotor is close to the core: $\Gamma=-0.146$, $-0.108$
and $-0.052$ at $G=0.1$ for $\mathrm{Re}=10$, 20 and 40. For $G\ge0.4$ they differ
from the zero-field transects by less than 0.004.

For the crosses $(+,-,+,+)$ and $(-,+,+,+)$ at 4\,mT (Fig.~\ref{fig:gap}d,~e), the
two arrangements, mirror images of each other in the line of centres, impose a sense
of rotation that hardly depends on the gap. At $\mathrm{Re}=10$ the first gives
counter-rotation at every gap, $\Gamma=-0.047$ to $-0.101$ (at $G=0.4$, where the
path stopped at 3.6\,mT, $\Gamma=-0.068$), including the wide gaps where the
zero-field rotors corotate. The second gives corotation, $\Gamma=+0.070$ to $+0.085$
for $G=0.1$--0.6. The field then overrides both the geometric and the topological
route of \citet{Smith2026spin}.

For the rod at T at 4\,mT (Fig.~\ref{fig:gap}f), at $\mathrm{Re}=10$ the passive
rotor corotates at all gaps ($\Gamma=+0.023$ to $+0.053$). At $\mathrm{Re}=40$ it
counter-rotates for $G\le0.3$ and corotates for $G\ge0.4$. 
The Reynolds number acts mainly through the strength of the field relative to
inertia: at a given $\bar B_0$, $\mathrm{Mn_F}$ is four times larger at
$\mathrm{Re}=10$ than at $\mathrm{Re}=20$ and sixteen times larger than at
$\mathrm{Re}=40$.

\subsection{Torque on the active rotor and heat transfer}\label{sec:torquenu}
\begin{figure}[t]
\centering
\includegraphics[width=\textwidth]{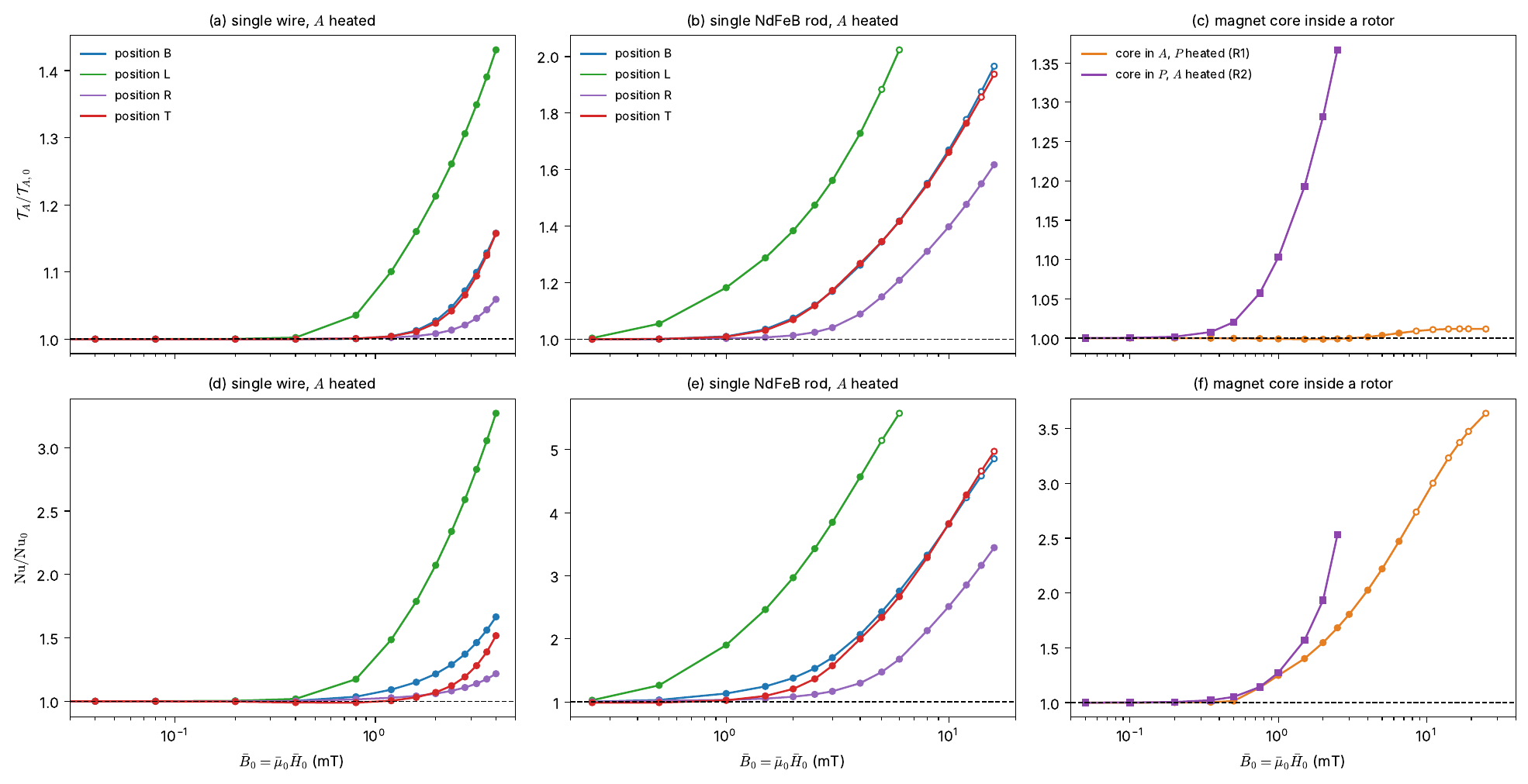}
\caption{Torque on the active rotor (top) and Nusselt number of the heated
rotor (bottom) relative to zero field, at $G=0.3$, $\mathrm{Re}=20$,
$\Delta\bar T=10$\,K, for the single sources and the cores of
Fig.~\ref{fig:single}. Open symbols: energy-balance error above 1\%.}
\label{fig:torquenu}
\end{figure}

The torque on the active rotor and the Nusselt number of the heated rotor for
the single sources and the cores at $G=0.3$, $\mathrm{Re}=20$ are plotted in
Fig.~\ref{fig:torquenu}.
With $A$ heated, the field increases $|\mathcal T_A|$, i.e.\ the power
needed to drive $A$, in almost every state. The few decreases are below 0.7\%. At
4\,mT the increase is 16\%, 43\%, 6\% and 16\% for the wires at B, L, R and T,
and 26\%, 73\%, 9\% and 27\% for the rods. Of the 7\,720 valid states with
$\Delta\bar T=10$\,K, $|\mathcal T_A|$ increases by more than 1\% in 4\,757, and
by up to a factor of 3.6 for a cross of four rods. With the core in $A$ and $P$
heated (R1), the torque changes by less than 1\% up to 6.5\,mT, although $|\Gamma|$
is doubled. The strengthened coupling then costs no additional driving power.

For the positions B and T the torque on $A$ is the same to within 0.6\%, and the
Nusselt number to within 15\%, although $\Gamma$ has opposite signs. The
thermomagnetic flow near $A$ is equally strong in the two cases, and only its
direction relative to the gap differs.

As far as the heat transfer is concerned, above about 1\,mT $\mathrm{Nu}$ grows
with the field for all positions, most for position L: by factors of 3.3 (wire) and
4.6 (rod) at 4\,mT. Over all valid states with $\Delta\bar T=10$\,K, $\mathrm{Nu}$
increases by up to a factor of 6.4, and the largest decrease, 18\%, occurs for R1 at
$G=0.7$.

\subsection{Effect of the heating}\label{sec:heating}
\begin{figure}[t]
\centering
\includegraphics[width=\textwidth]{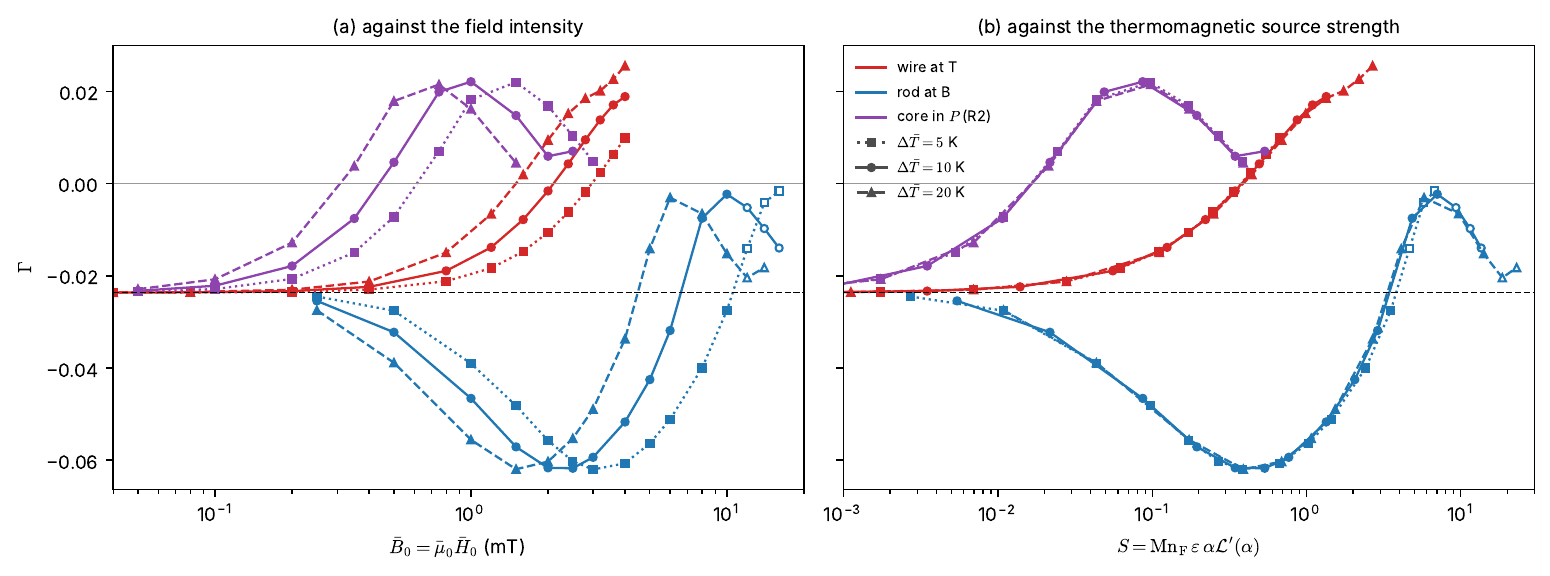}
\caption{Gear ratio at $G=0.3$, $\mathrm{Re}=20$ for the wire at T, the rod at B
and the core in $P$ (R2), with $\Delta\bar T=5$, 10 and 20\,K. (a)~Against the
magnetic flux density $\bar B_0$. (b)~Against the source strength $S$ of
Eq.~\eqref{eq:Ssource}. Open symbols: energy-balance error above 1\%.}
\label{fig:dT}
\end{figure}

At $G=0.3$ and $\mathrm{Re}=20$ every configuration was also computed with
$\Delta\bar T=5$ and 20\,K. Plotted against $\bar B_0$, the curves for the
three heatings differ: stronger heating moves the response to weaker fields
(Fig.~\ref{fig:dT}a). Plotted against the strength of the vorticity
source~\eqref{eq:fhdsource} at the reference point,
\begin{equation}
S=\mathrm{Mn_F}\,\varepsilon\,\xi\mathcal L'(\xi)\big|_{\xi=\alpha},
\label{eq:Ssource}
\end{equation}
the last column of Table~\ref{tab:scenarios}, they collapse onto one curve
(Fig.~\ref{fig:dT}b). For the single wires, the single rods and the cores the
largest deviation from the $\Delta\bar T=10$\,K curve is 0.0006--0.005,
against 0.009--0.033 when the curves are compared at equal $\bar B_0$. The
collapse is close also for most pairs and crosses, but less so where $\Gamma$
changes steeply. For fixed geometry and Reynolds number, the heating and the
field strength thus act approximately through the single parameter $S$. A field of
strength $\bar B_0$ with heating $\Delta\bar T$ is equivalent, to this
accuracy, to the field that gives the same $S$ with any other heating. In the
Curie regime $S\propto\Delta\bar T\,\bar B_0^{2}$. The collapse is a
numerical observation, not an exact reduction. $S$ is evaluated at the
reference point, whereas the source~\eqref{eq:fhdsource} depends on $\alpha$
and $\varepsilon$ separately through $\xi=\alpha H/(1+\varepsilon\theta)$,
which varies across the fluid. Equal $S$ therefore does not imply an equal
source throughout the domain. Deviations of up to 0.005 are small compared
with the changes of $\Gamma$ produced by the field, but they are comparable
with $\Gamma$ itself near a decoupling or a reversal, where the collapse is
only qualitative.

\subsection{Limitations}\label{sec:limitations}
Six limitations of these results should be kept in mind.

\emph{Steady solutions.} All magnetic results are steady torque-free solutions. Their stability with respect to time-dependent
    perturbations, with the torque balance~\eqref{eq:ndtorque} of the passive rotor, has not been examined. The hydrodynamic evidence of
    \citet{Pan2026planar} does not extend to the heated magnetic problem. The reversals and amplifications reported above are therefore properties of
    steady solution branches, and it is not established that a passive rotor starting from rest would reach them. In 283 of the 782 finest-mesh paths the
    continuation first stopped before the strongest level, even after four levels of geometric sub-steps. When 31 of them were repeated with a larger
    iteration limit (120 Newton iterations, tolerance $2\times10^{-9}$), 18 were carried to the last level, so that 265 paths remain stopped: 193 of
    the 368 rod paths, 29 of the 46 core paths and 43 of the 368 wire paths.
    This may indicate the loss of a steady state, i.e.\ time-dependent thermomagnetic
    convection, a turning point of the steady branch, or an ordinary
    failure of the Newton iteration. The present computations do not
    distinguish between these possibilities. A time-dependent computation or a
    linear stability analysis of the steady states, and a continuation that
    can follow turning points, are required. These paths
    will be examined separately.

\emph{Thin thermal layers.} For the strongest sources the energy
    balance shows that the thermal layers at the corral wall are not resolved
    even on the finest mesh. Those states have been excluded from the
    discussion.

 \emph{Stabilisation.} The streamline-diffusion term of
    Eq.~\eqref{eq:weakT} is not strongly consistent (\S\ref{sec:numerics}).
    Its effect is contained in the mesh comparison of \S\ref{sec:meshcheck},
    but it has not been compared directly with a consistent (full-residual)
    formulation.

\emph{Self-field.} The magnetostatic self-field of the fluid is
    neglected (\S\ref{sec:parameters}). With $\chi_0\approx1.6$ it would reduce
    and redistribute the field inside the fluid. For a uniformly
    magnetised disc the internal field would be reduced to 0.56 of the applied
    field. The threshold strengths quoted in this section, e.g.\ the reversal
    by a rod at T at 1.7\,mT, refer to the prescribed applied field and are
    not quantitative predictions for EFH1. Whether the dependence on the
    source position is preserved when the field is computed self-consistently
    has not been demonstrated.

\emph{Constant properties.} The temperature dependence of $\bar M_s$
    and of the viscosity is neglected. The thermal dilution of the
    particles alone would raise the pyromagnetic coefficient by about 26\% for
    the external sources and by 29--70\% at the reference points of the
    cores (\S\ref{sec:parameters}). The sensitivity of $\Gamma$ to this
    omitted part of the pyromagnetic response has not been computed.

 \emph{Gravity.} The planar, gravity-free problem is exact only in
    the axially homogeneous limit with vertical axes (\S\ref{sec:2D}). The
    Richardson number of the laboratory scaling is $\mathrm{Ri}=0.27$--4.3,
    and the axial buoyant flow is comparable with the rotor speed
    (\S\ref{sec:parameters}). End-wall effects of this flow are not included.

\subsection{Mesh independence}\label{sec:meshcheck}
For the eleven configurations of the first campaign (wire and rod at B, L and R, the three pairs T+B of Fig.~\ref{fig:Hconfigs} and the two cores) every path was
computed on the base and fine meshes over the whole range of $G$ and
$\mathrm{Re}$, and at $G=0.3$, $\mathrm{Re}=20$ also on the coarse and finest
meshes. Representative values at $G=0.3$, $\mathrm{Re}=20$ are given in
Table~\ref{tab:meshmag}. With the wall refinement of \S\ref{sec:numerics} the finest
mesh has 34\,549--43\,378 triangles and 0.23--0.28 million unknowns for
$(u,v,p,\theta)$, depending on the configuration and the gap. The comparison
of the fine and finest meshes over all states that satisfy the energy-balance
criterion on the finest mesh (2\,966 states) gives the following agreement.
The differences are reported for $\Gamma$ in absolute terms, because
$\Gamma$ passes through zero, and for $\mathcal T_A$ and $\mathrm{Nu}$ as relative
differences, $|x_{\mathrm{fine}}-x_{\mathrm{finest}}|/|x_{\mathrm{finest}}|$, with
the finest-mesh value as the reference. For $\Gamma$ the median difference is
$2.4\times10^{-6}$, 95\% of the
differences are below $1.7\times10^{-4}$, and 99\% are below $10^{-3}$. For
$\mathcal T_A$ the median relative difference is $2.3\times10^{-6}$ and 99\% are
below $1.7\times10^{-3}$. For $\mathrm{Nu}$ the median relative difference is
$5.7\times10^{-5}$, with 95\% below $3\times10^{-3}$ and 99\% below
$9\times10^{-3}$.
Where both meshes satisfy the criterion (2\,659 states), the two values of
$\Gamma$ differ by more than $2\times10^{-3}$ only once. The few large
differences, up to 0.06, all involve a fine-mesh state whose balance error
exceeds 1\%. In some of them $\Gamma$ even has the opposite sign, which
suggests that the under-resolved continuation has followed a different
solution branch. Near a zero of $\Gamma$ the relative difference can
be large although the absolute one is small (rod at R in
Table~\ref{tab:meshmag}). 

\begin{table}[t]
\centering\small
\caption{Gear ratio on the four meshes at $G=0.3$, $\mathrm{Re}=20$,
$\Delta\bar T=10$\,K. Last columns: difference between the fine and finest
meshes and energy-balance error on the finest mesh. --: level not reached on
that mesh.}
\label{tab:meshmag}
\footnotesize\setlength{\tabcolsep}{4pt}
\begin{tabular}{lrrrrrrr}
\toprule
Configuration & $\bar B_0$ (mT) & coarse & base & fine & finest & $|\Delta\Gamma|$ fine--finest & balance (\%)\\
\midrule
wire B & 2.4 & -- & $-0.05874$ & $-0.05875$ & $-0.05875$ & $4.5\times10^{-7}$ & 0.03\\
wire L & 4 & -- & $-0.06269$ & $-0.06223$ & $-0.06224$ & $5.4\times10^{-6}$ & 0.10\\
wires T+B opp. & 2.4 & -- & $-0.05014$ & $-0.05049$ & $-0.05049$ & $3.9\times10^{-6}$ & 0.04\\
rod B & 2 & $-0.06293$ & $-0.06163$ & $-0.06171$ & $-0.06172$ & $1.0\times10^{-5}$ & 0.14\\
rod L & 4 & -- & $-0.06266$ & $-0.06277$ & $-0.06276$ & $1.1\times10^{-5}$ & 0.82\\
rod R & 6 & -- & -- & $+0.00210$ & $+0.00236$ & $2.7\times10^{-4}$ & 0.18\\
core R1 & 2 & $-0.05105$ & $-0.05111$ & $-0.05114$ & $-0.05111$ & $2.5\times10^{-5}$ & 0.02\\
core R2 & 1 & $+0.02253$ & $+0.02218$ & $+0.02220$ & $+0.02220$ & $7.0\times10^{-6}$ & 0.01\\
\bottomrule
\end{tabular}
\end{table}

\clearpage

\section{Conclusions}\label{sec:conclusions}

The two-rotor problem of \citet{Smith2026spin} has been extended to a magnetic
fluid. An active rotor drives a torque-free passive rotor inside a corral
filled with the oil-based ferrofluid EFH1, one rotor is heated, and the
field of wires, NdFeB rods or a stationary NdFeB core acts on the fluid
through the full Kelvin force with a temperature-dependent Langevin
magnetisation. The gear ratio $\Gamma$ was
computed directly from
the steady equations, with $\Gamma$ as an additional unknown fixed by the
torque-free condition. The solver was first validated against the planar
simulations of \citet{Pan2026planar}. A database of 1\,526 continuation paths
and 16\,744 steady states was then obtained for 34 arrangements of the field
sources, gaps $G=0.1$--0.7, $\mathrm{Re}=10$--40 and heatings of 5--20\,K.
All quantitative statements rest on states that satisfy the global energy
balance to within 1\% on the finest mesh. 

The main conclusions are the following.
\begin{enumerate}
  \item \emph{Benchmark.} Without a field, the body-fitted steady
    formulation reproduces every qualitative feature of the planar problem
    reported by \citet{Pan2026planar}: the re-entrant gap route with sign
    changes at $G=0.128$ and 0.582 for $\mathrm{Re}=20$, the location of the
    strongest coupling, the Reynolds-driven reversal at $G=0.22$
    ($\mathrm{Re}=42.1$), the absence of a reversal at $G=0.3$ up to
    $\mathrm{Re}=400$, and the corral-size scaling of $|\Gamma|$. The
    differences are 1--8\% up to $\mathrm{Re}\approx50$. At $G=0.3$ and
    $\mathrm{Re}=400$, $\Gamma=-0.0068$ is obtained, against $-0.003$ in
    \citet{Pan2026planar}. This value does not change under mesh
    refinement, and the cause of the difference remains open.
  \item \emph{Sensitivity.} Because the hydrodynamic coupling is weak
    ($|\Gamma_0|\le0.031$ for $C=4.5$), modest fields change it by
    an amount of the order of $\Gamma_0$ itself. At $\bar B_0=2$\,mT, which
    corresponds to a coil side of 250 ampere-turns at 25\,mm or an N42 rod of
    1.7\,mm radius at 30\,mm, the gear ratio is reversed in 290 of the 675
    cases with a valid state, strengthened in 171 and weakened in 152, and it
    changes by less than 0.002 in only 62. The field acts through the
    Kelvin force alone, which drives thermomagnetic convection from the
    heated rotor wherever the field gradient is not parallel to the
    temperature gradient. No magnetic
    torque acts on the rotors.
  \item \emph{The position of the source selects the mode of coupling.} At the
    experimental gap $G=0.3$ and $\mathrm{Re}=20$, a single source below the
    corral (B) or behind the active rotor (L) strengthens the gear-like
    counter-rotation by a factor 2.6--2.7, to $\Gamma\approx-0.062$. The same
    source above the corral (T) reverses it into belt-like corotation from
    1.7--2.1\,mT on, up to $\Gamma=+0.068$ at $\mathrm{Re}=20$ and $+0.107$
    at $\mathrm{Re}=10$. The magnetic field therefore controls the
    inner--outer shear competition identified by \citet{Smith2026spin}. It
    adds a thermomagnetic route between the two modes, alongside the
    geometric, topological and inertial routes. Unlike those three routes, it
    can be followed in both directions at fixed geometry and Reynolds number:
    of the 290 reversals at 2\,mT, 162 turn counter-rotation into corotation
    and 128 turn corotation into counter-rotation.
  \item \emph{Four flow patterns.} The temperature and streamline fields of
    all groups of sources fall into four classes.
    \begin{itemize}
      \item \emph{Isolation of $P$} in a clockwise cell gives gear-like
        counter-rotation: sources at B and L, pairs T+B at moderate fields,
        and the wire cross $(+,-,+,+)$.
      \item \emph{Enclosure of both rotors} by a counter-clockwise
        circulation gives belt-like corotation: source at T, antiparallel
        rods T+B at 4\,mT, the wire cross $(-,+,+,+)$, and R2.
      \item \emph{A field-induced cell between the rotors} reverses the gap
        flow, as in the topological route of \citet{Smith2026spin}: all pairs
        L+R and the rod cross $(+,-,+,+)$.
      \item \emph{A balance} of shears of opposite sense on $P$ decouples it.
    \end{itemize}
    The field thus reproduces, at fixed geometry and Reynolds number, both
    the belt-like and the topological route to corotation.
  \item \emph{Symmetry does not suppress the effect.} Fields that are
    symmetric about the line of centres (position L or R, and all pairs) act
    as strongly as asymmetric ones. At $G=0.3$ the source at L gives the
    largest strengthening of all single sources. At $\mathrm{Pe}\approx10^{3}$ the
    rotation of the active rotor sweeps its thermal layer around it, and this
    breaks the symmetry that would otherwise cancel the net torque on the
    passive rotor.
  \item \emph{Sources do not add.} The single sources at L and at R both
    strengthen the counter-rotation, yet every L+R pair reverses it, at
    0.85--2.1\,mT, to $\Gamma=+0.022$ to $+0.028$, through the cell that
    forms between the rotors. Five of the eight crosses
    of four wires and six of the eight crosses of four rods reverse the
    coupling. At $\mathrm{Re}=10$ and 4\,mT, two crosses of wires that differ
    only in the senses of the currents at T and B impose a sense of rotation
    that hardly depends on the gap: corotation with $\Gamma=+0.070$ to
    $+0.085$, or counter-rotation with $\Gamma=-0.047$ to $-0.101$. Here the
    field overrides both the geometric and the topological route, and the
    sense of the passive rotor can be switched electrically, on the
    steady branches computed here.
  \item \emph{Magnet cores: strong amplification and decoupling.} A
    stationary core inside the active rotor, with the passive rotor heated
    (R1), produces the strongest counter-rotation of the study:
    $\Gamma=-0.221$ at $G=0.1$ and $\mathrm{Re}=10$, seven times the largest
    hydrodynamic gear ratio. At $G=0.3$ it doubles $|\Gamma|$ while changing
    the torque needed to drive the active rotor by less than 1\%. At wide
    gaps the same core decouples the rotors: at $G=0.7$ and $\mathrm{Re}=20$,
    $\Gamma$ falls from
    0.011 to 0.001 at 5\,mT and stays near zero up to 25\,mT. With the core
    inside the passive rotor and the active rotor heated (R2), the coupling is
    reversed above 0.44\,mT.
  \item \emph{Cost and benefit.} With the active rotor heated, the field
    increases the torque needed to drive it in almost every state: at 4\,mT
    and $G=0.3$ by 6--73\% for single sources, and up to a factor of 3.6 for
    crosses of rods. It also increases the heat transfer from the heated
    rotor, $\mathrm{Nu}$ by up to a factor of 6.4. Sources at B and T need the
    same driving torque to within 0.6\% but give gear ratios of opposite
    sign, so the choice between the two modes costs nothing in driving
    power.
  \item \emph{One governing parameter.} For fixed geometry and Reynolds
    number, $\Gamma$ depends on the field strength and the heating almost
    only through the thermomagnetic source strength
    $S=\mathrm{Mn_F}\,\varepsilon\,\alpha\mathcal L'(\alpha)$
    (Eq.~\eqref{eq:Ssource}). Curves for $\Delta\bar T=5$, 10 and 20\,K
    collapse when plotted against $S$, and a weaker field with stronger
    heating is equivalent to a stronger field with weaker heating. In the
    Curie regime $S\propto\Delta\bar T\,\bar B_0^{2}$. The collapse is
    approximate, with deviations of up to 0.005, which are comparable with
    $\Gamma$ near a decoupling or a reversal. Since
    $\mathrm{Mn_F}\propto\mathrm{Re}^{-2}$, a given field acts more strongly
    on slower rotors.
  \item \emph{Numerical requirements.} With the full Kelvin force, the
    corral wall nearest to an external source must be refined locally.
    Without this refinement the global energy balance was violated by 7\%
    on the base mesh. The energy balance proved auseful filter for
    unresolved states, although it is not by itself a verification of the
    coupled solution. Where it holds to within 1\%, the fine and finest meshes give
    the same $\Gamma$ to within $10^{-3}$ in 99\% of the comparisons. Where it
    fails, which happens for rods above about 2--2.5\,mT and cores above
    5\,mT,
    the thermal layers at the wall are not resolved, and the mesh-to-mesh
    differences can reach the size of $\Gamma$ itself.
\end{enumerate}

Within the planar, prescribed-field model, these results show that a
magnetic field turns the weak and passively
determined hydrodynamic spin coupling into a controllable one. With a fixed
geometry and a fixed rotation rate, the field can strengthen the coupling
severalfold, reverse its sense, or cancel it, and the outcome is selected by
the position and arrangement of the sources and by the current senses rather
than by the mechanical layout of the rotors. Fluid gears transmit rotation
without contact, and their sense of rotation is selected by the flow rather
than by a fixed mechanism \citep{Smith2026spin}. In the model, a field
of a few millitesla adds to them the functions of a gearbox and a clutch,
without moving parts and without wear.

Several limitations point to further work. The limitations of the model and
of the computations, namely the unexamined stability of the steady branches, the
stopped continuation paths, the inconsistent streamline-diffusion stabilisation, and
the prescribed field and constant properties of the ferrofluid, were discussed in
\S\ref{sec:limitations} and are not repeated here. They call for time-dependent
computations with the torque balance of the passive rotor, a linear stability
analysis of representative states, continuation methods that can follow turning
points, a direct comparison with full-residual SUPG, and self-consistent
magnetostatic computations for representative cases. Two further steps go beyond
the planar model.

The first concerns three-dimensional effects. Finite-depth
secondary motion and end-wall stresses, which \citet{Pan2026planar}
propose as plausible contributors to the experimental high-$\mathrm{Re}$
transition, will also modify the thermomagnetic convection. For
electrically conducting fluids an axial field acts only through the
Hartmann layers on the end walls (\S\ref{sec:uniform}), and a
three-dimensional or quasi-two-dimensional computation of that case is a
natural next step.

The second is a laboratory test. The predictions can be tested qualitatively
in the apparatus of \citet{Smith2026spin} filled with a ferrofluid: one
heated rotor and a permanent magnet or a coil side moved from below to
above the corral should reverse the passive rotor. In the present
model this occurs at about 2\,mT of applied field for $\mathrm{Re}=20$ and
$\Delta\bar T=10$\,K. For an experiment this value is only an order of
magnitude, for three reasons. The self-field and the omitted pyromagnetic
contributions change it. The rotor axes must be vertical, since with
horizontal axes buoyancy ($\mathrm{Ri}=0.27$--4.3) would exceed the
thermomagnetic forcing of the external sources. Finally, with vertical axes
the axial buoyant flow, of the order of the rotor speed, is affected by
the end walls, so that a large aspect ratio is required.
Beyond the two-rotor problem, the same mechanism should apply to arrays of
rotors and to freely translating or non-circular passive bodies, the
extensions proposed by \citet{Smith2026spin} and \citet{Pan2026planar}. In
such systems, field sources placed outside the container could organise the
collective spin states.

\bibliographystyle{unsrtnat}
\bibliography{references}

\end{document}